\documentclass[
    aps,pre,
    twocolumn,
    reprint,
    superscriptaddress,
    floatfix,
    amssymb,
    long bibliography
]{revtex4-1}

\usepackage{amsmath}
\usepackage{amsfonts}
\usepackage{graphicx}
\usepackage{bm}
\usepackage{dcolumn}
\usepackage{overpic}
\usepackage{mathtools}
\usepackage{dsfont}
\usepackage{nicefrac}
\usepackage[normalem]{ulem}
\usepackage{xcolor}
\definecolor{linkColor}{rgb}{0,0.3,0.7}
\usepackage{hyperref}
\hypersetup{colorlinks=true,
            allcolors=linkColor,
            pdfborder={0 0 0},
            pdfencoding = auto
            }

\begin{document}

\title{Capillary wave formation in conserved active emulsions}

\author{Florian Raßhofer}
\thanks{FR and SB contributed equally.}
\affiliation{Arnold Sommerfeld Center for Theoretical Physics and Center for NanoScience, Department of Physics, Ludwig-Maximilians-Universit\"at M\"unchen, Theresienstra\ss e 37, D-80333 Munich, Germany}
\author{Simon Bauer}
\thanks{FR and SB contributed equally.}
\affiliation{Arnold Sommerfeld Center for Theoretical Physics and Center for NanoScience, Department of Physics, Ludwig-Maximilians-Universit\"at M\"unchen, Theresienstra\ss e 37, D-80333 Munich, Germany}
\author{Alexander Ziepke}
\affiliation{Arnold Sommerfeld Center for Theoretical Physics and Center for NanoScience, Department of Physics, Ludwig-Maximilians-Universit\"at M\"unchen, Theresienstra\ss e 37, D-80333 Munich, Germany}
\author{Ivan Maryshev}
\affiliation{Arnold Sommerfeld Center for Theoretical Physics and Center for NanoScience, Department of Physics, Ludwig-Maximilians-Universit\"at M\"unchen, Theresienstra\ss e 37, D-80333 Munich, Germany}
\author{Erwin Frey}
\email[Corresponding author: ]{frey@lmu.de}
\affiliation{Arnold Sommerfeld Center for Theoretical Physics and Center for NanoScience, Department of Physics, Ludwig-Maximilians-Universit\"at M\"unchen, Theresienstra\ss e 37, D-80333 Munich, Germany}
\affiliation{Max Planck School Matter to Life, Hofgartenstraße 8, 80539 Munich, Germany}

\date{\today}

\begin{abstract}
The dynamics of phase-separated interfaces shape the behavior of both passive and active condensates.
While surface tension in equilibrium systems minimizes interface length, non-equilibrium fluxes can destabilize flat or constantly curved interfaces, giving rise to complex interface morphologies.
Starting from a minimal model that couples a conserved, phase-separating species to a self-generated chemical field, we identify the conditions under which interfacial instabilities may emerge.
Specifically, we show that non-reciprocal chemotactic interactions induce two distinct types of instabilities: a stationary (non-oscillatory) instability that promotes interface deformations, and an oscillatory instability that can give rise to persistent capillary waves propagating along the boundaries of phase-separated domains.
To characterize these phenomena, we develop a perturbative framework that predicts the onset, wavelength, and velocity of capillary waves, and quantitatively validate these predictions through numerical simulations.
Beyond the linear regime, our simulations reveal that capillary waves undergo a secondary instability, leading to either stationary or dynamically evolving superpositions of different wave modes.
Finally, we investigate whether capillary waves can facilitate directed mass transport, either along phase boundaries (conveyor belts) or through self-sustained liquid gears crawling along a solid wall.
Taken together, our results establish a general framework for interfacial dynamics in active phase-separating systems and suggest new strategies for controlling mass transport in soft matter and biological condensates.
\end{abstract}

\maketitle
%
%
Phase separation is a fundamental mechanism driving the organization of liquid and soft matter across a wide range of physical\@~\cite{Kim.2013}, chemical\@~\cite{Lee.2020}, culinary\@~\cite{Benabdelhalim.2022,Bartolucci.2024}, and biological\@~\cite{Hyman.2014,Banani.2017,Siteur.2023} systems.
In thermodynamic equilibrium, phase separation is dictated by thermodynamic constraints, leading to the formation of coexisting condensed and dilute phases. 
These condensates emerge due to microscopic pairwise interactions between different constituents, which drive the system toward a state that minimizes free energy.
Depending on the system’s quench conditions, phase separation proceeds via either spinodal decomposition or nucleation and growth\@~\cite{Weber.2019}. 
In the spinodal regime, small density fluctuations are amplified due to negative effective diffusivity, leading to the rapid formation of phase-separated domains.
Outside this regime, the surface tension of phase-separated domains induces an energy barrier, so that only rare, sufficiently large fluctuations can form stable nuclei, which then grow by further accumulating material.
Over time, these domains coarsen through Ostwald ripening\@~\cite{Ostwald.1897,Wagner.1961,Lifshitz.1961}, with larger droplets growing at the expense of smaller ones, until a single macroscopically large domain remains.

While these equilibrium phase separation processes are well understood, many biological cells and synthetic systems operate far from equilibrium, where phase separation, or more generally self-organization into spatiotemporal structures, is coupled to energy-consuming processes such as chemical reactions\@~\cite{Glotzer.1994,Zwicker.2022, Halatek.2018b,Demarchi.2023,Goychuk.2024} or externally maintained gradients\@~\cite{Doan.2024, Burkart.2022}.
These systems evade classical thermodynamic constraints, often exhibiting complex patterns or persistent dynamical steady states rather than passive coarsening.

One important example is that of biomolecular condensates, which have been argued to play a crucial role in organizing intracellular processes such as mid-cell identification\@~\cite{Schumacher.2017,Bergeler.2018}, germ-cell specification\@~\cite{Brangwynne.2009}, cell division\@~\cite{Bornens.2012,Zwicker.2014}, and memory formation\@~\cite{Zeng.2016}.
Beyond their role in cellular organization, biomolecular condensates also modulate biochemical reactions by spatially localizing enzymes and their substrates\@~\cite{Smokers.2024,O'Flynn.2021,Prouteau.2018}.
For instance, pyrenoids and carboxysomes enhance carbon fixation by increasing the local carbon dioxide concentration\@~\cite{Wunder.2018,Wang.2019}; multi-layered nucleoli coordinate ribosome assembly\@~\cite{Feric.2016}; and transcriptional condensates facilitate gene expression by concentrating transcriptional machinery at specific genomic loci\@~\cite{Hnisz.2017,Henninger.2021}.

Although biomolecular condensates are believed to form via passive liquid-liquid phase separation\@~\cite{Bracha.2018,Fritsch.2021}, their biological function typically relies on active processes, which require the consumption of chemical energy, often in the form of Adenosintriphosphat (ATP).
This continuous energy input results in emergent material fluxes and chemical gradients, leading to phenomena that cannot be explained within an equilibrium framework.
If the condensate hosts a chemical reaction, inhomogeneous reactant distributions induce chemophoretic flows, which can counteract Ostwald ripening and drive droplet self-propulsion\@~\cite{Demarchi.2023,Goychuk.2024,Jambon-Puillet.2024}.
Similarly, previous studies have demonstrated that continuous turnover of phase-separating proteins can stabilize multiple equally sized droplets\@~\cite{Zwicker.2015,Wurtz.2018,Kirschbaum.2021,Donau.2023}, promote the formation of liquid shells\@~\cite{Bauermann.2023}, enable diffusiophoretic transport of droplets\@~\cite{Haefner.2024}, and even facilitate droplet division\@~\cite{Zwicker.2017}. 

Another prominent example of non-equilibrium phase separation that has been receiving considerable attention is motility-induced phase separation (MIPS)\@~\cite{Tailleur.2008, Fily.2012,Redner.2013,Cates.2014}, where self-propelled particles spontaneously cluster due to a feedback loop between local density and motility suppression.
In an ecological context, such a phenomenon was discovered in mussel bed aggregation, where density-dependent movement drives spatial self-organization\@~\cite{Liu.2013}.

Unlike equilibrium phase separation, which relies on energetically favorable interactions, MIPS arises purely from self-propulsion and density-dependent slowdown, making it an intrinsically non-equilibrium process. 
This distinguishes it from biomolecular condensates, where passive phase separation is merely modified by additional active processes.
Nevertheless, under certain conditions, the large-scale behavior of MIPS can be mapped onto an equilibrium-like description, providing a thermodynamic interpretation of phase coexistence\@~\cite{Cates.2014,Tjhung.2018,Takatori.2015,Solon.2018}.

Similar clustering effects arise in chemotactic systems, where biological cells\@~\cite{Singer.2019}, enzymes\@~\cite{Feng.2020}, or synthetic particles\@~\cite{Jin.2017,Liebchen.2018} migrate in response to self-generated chemical gradients. 
Unlike MIPS, which is driven by velocity-dependent aggregation, chemotactic clustering relies on long-range signaling interactions that coordinate collective motion.
These systems are commonly described by coarse-grained Keller–Segel models\@~\cite{Keller.1970,Keller.1971}, which capture a diverse range of behaviors, including aggregate coarsening, stable cluster formation, and chaotic aggregate dynamics\@~\cite{Painter.2011,Hillen.2008,Tindall.2008}.
Despite their distinct microscopic origins, recent work has revealed formal links between chemotactic aggregation and equilibrium phase separation, highlighting similarities in their underlying dynamics\@~\cite{Weyer.2024,Weyer.2024c}.

Chemical signaling is further known to modify MIPS, enabling new dynamical states such as traveling stripes and self-propelled droplets\@~\cite{Zhao.2023}, which are absent in classical MIPS.
This raises the intriguing question of whether similar phenomena could emerge if chemotactic bacteria undergo passive phase separation through purely physical interactions, possibly induced by the presence of a crowding species\@~\cite{Schwarz-Linek.2012}.

The above discussion illustrates that effective phase-separation dynamics can emerge in a broad class of non-equilibrium systems, even in the absence of energetically favorable short-range interactions.
Remarkably, such behavior has been observed even in systems of non-interacting particles:
Recent theoretical work has established a deep formal analogy between equilibrium phase separation and mass-conserving reaction-diffusion systems \cite{Brauns.2020,Brauns.2021,Frey.2020}, which describe many pattern-forming processes in reconstituted biochemical systems \cite{Würthner.2022} and living cells \cite{Ren.2025}.
This analogy extends to the existence of an effective surface tension, allowing concepts from equilibrium thermodynamics to be used in analyzing the dynamics and stability of phase-separated states\@~\cite{Weyer.2024b}.

A central organizing principle across all these systems---from passive condensates to chemotactic clusters---is the pivotal role of interfaces between coexisting phases, which typically serve as the primary slow variables that govern long-term system evolution.
In thermodynamic equilibrium, interfaces evolve through diffusive mass transfer between domains of different curvature, with surface tension driving the system towards a state of minimal interface length\@~\cite{Bray.2002}.
In contrast, active and chemically driven systems can violate this thermodynamic constraint, giving rise to complex behaviors such as non-equilibrium interface fluctuations\@~\cite{Cho.2023, Fausti.2021}, interface instabilities\@~\cite{Mullins.1963,Zwicker.2017,Alert.2022,Demarchi.2023}, and persistent capillary waves or fluxes along the interface\@~\cite{Brauns.2024,Junco.2019}. 
Another important commonality is the presence of a conserved quantity whose phase-separation dynamics can often be described by an effective chemical potential. 
In systems such as enzymatically active condensates or chemotactic aggregates, this conserved field is typically coupled to at least one non-conserved species, representing reactants or signaling molecules.

Building on these insights, we explore the dynamics of conserved active emulsions, where self-generated chemical gradients influence phase separation.
To this end, we propose a minimal model, in which a phase-separating species migrates along gradients of a self-produced chemical field.
The diffusive nature of the chemical field, combined with the active transport of the phase-separating species, generates long-range non-reciprocal interactions. 
These interactions lead to a rich spectrum of non-equilibrium behaviors, including arrested coarsening, self-propelled domains, and dynamically evolving interfaces.

In this study, we focus primarily on the latter, highlighting the emergence of interfacial instabilities and persistent capillary waves propagating along the boundaries of phase-separated domains. 
These waves result from a two-fold instability, where chemical interactions destabilize an otherwise static interface, whereas the slow spreading of chemical signals generates wave motion.
Unlike their passive counterparts, which result from thermal fluctuations and relax diffusively\@~\cite{Bedeaux.1985,Bray.2001}, we show that the capillary waves observed in our system can be sustained indefinitely.
In the steady state, they manifest as \textit{liquid conveyor belts} and \textit{rotating gears}, structures that facilitate highly localized and directed material movement along interfaces.

To quantify these processes, we develop a perturbative theoretical framework that predicts the onset and wavelength of general interface instabilities, as well as the velocity of capillary waves. 
Our analytical predictions are validated through extensive numerical simulations.
Thereby, we gain key insights into the dynamic behavior of interfaces in active phase-separated systems. 
Specifically, we identify two distinct types of interface instabilities in conserved active emulsions: 
A stationary instability that resembles the well-known Mullins-Sekerka instability during solidification and an oscillatory instability that can give rise to capillary wave formation.
Surprisingly, our simulations reveal that once formed, these waves undergo a secondary instability, giving rise to even more complex interfacial dynamics characterized by the superposition (and coupling) of multiple wave modes. 

Finally, we explore how the system’s non-equilibrium nature can be harnessed to extract work, converting chemical energy into directed motion along interfaces.
In particular, we consider an isolated liquid gear rotating near a system boundary.
We find that its chiral motion generates slow, continuous propulsion parallel to the boundary--effectively causing the gear to ``crawl'' along the wall.

Our findings provide a unifying perspective on interfacial instabilities in non-reciprocal systems, linking classical phase separation with active matter physics and chemically driven pattern formation.
More broadly, they offer a framework for understanding how self-generated gradients and long-ranged chemical interactions govern the behavior of active emulsions and biomolecular condensates, with potential applications in synthetic active materials and mesoscale transport mechanisms.

This paper is organized as follows:
In Sec.\@~\ref{sec:model}, we introduce our theoretical framework for a conserved phase-separating species that facilitates the local turnover of additional chemical components. 
Section\@~\ref{sec:overview} provides an overview of the system’s behavior across different parameter regimes, illustrated through finite element simulations. 
This is complemented by a comprehensive phase diagram, obtained via numerical simulations and linear stability analysis, which is presented in Apps.\@~\ref{app:phaseDiagramLSA}--\ref{app:phaseDiagramNumerics}.

Focusing on capillary wave formation, Sec.\@~\ref{sec:stationaryInterfaces} discusses stationary interface solutions, further detailed in App.\@~\ref{app:stationaryInterfaces}. 
Numerical simulations reveal that these regular interfaces are subject to interfacial instabilities, which we rationalize through a heuristic argument (App.\@~\ref{app:InstabilityCriterion}).
Our central contribution is presented in Sec.\@~\ref{sec:stabilityAnalysis}, where we develop a perturbative framework to quantitatively analyze capillary wave formation. 
By linearizing our model equations around the stationary interface solutions, we derive a self-consistency relation for the interface position, providing the dispersion relation of small interface undulations. 
The details of this derivation are provided in Apps.\@~\ref{app:StabilityAnalysis}–\ref{app:modelBDispersion}. 
From this, we infer a bifurcation diagram of interfacial instabilities and extract predictions for the wavelength and velocity of capillary waves.
Our analytical results are tested in Sec.\@~\ref{sec:DispersionFlat} against extensive numerical simulations for an initially flat interface, with the complementary analysis for curved interfaces presented in App.\@~\ref{app:CurvedInterfaces}. 
Additional methodological details are outlined in Apps.\@~\ref{app:fastestMode}–\ref{app:phaseVelocities}.

Having characterized the onset of instabilities, in Sec.\@~\ref{sec:LongTimeStability} we turn to their long-term behavior, where we show that capillary waves of a fixed wavelength undergo a secondary instability, leading to complex superpositions of multiple wave modes.
Additionally, in Sec.\@~\ref{sec:liquidMechanics}, we explore whether capillary waves can be harnessed for localized mass transport along phase- or system boundaries. 
The methods and concepts underlying this analysis are presented in Apps.\@~\ref{app:conveyorTransport}--\ref{app:translationGears}.
Finally, in Sec.\@~\ref{sec:Discussion}, we summarize our findings, relate them to existing work, and discuss their broader implications. 
In particular, we highlight how our model bridges an important gap in active matter field theories and provides a foundation for future research.
The remaining appendices further revise certain key concepts, including the surface tension of phase-separated domains (App.\@~\ref{app:SurfaceTension}), the local Gibbs-Thomson relation for general interfaces (App.\@~\ref{app:GibbsThomsonSection}), and the details of our numerical simulations (App.\@~\ref{app:Numerics}).
\section{Phase separation with chemorepulsive interactions}
\label{sec:model}
In this section, we present a theoretical framework to describe the chemotactic motion of phase-separated droplets in self-generated chemical gradients.
Our goal is to derive a minimal model that is capable of reproducing the qualitative features for this class of systems.
Whenever possible, we reduce the complexity of the governing equations while retaining the essential physical mechanisms that drive the observed behavior.

As a representative example, we consider an enzymatically enriched condensate that catalyzes the conversion of a substrate into a product species.
Recent theoretical work has shown that such enzymatic reactions can drive self-organization, positioning, and propulsion of condensates via chemically generated gradients\@~\@~\cite{Demarchi.2023,Goychuk.2024}.
Specifically, for different substrate-condensate and product-condensate interactions, the imbalance in concentrations creates an effective driving force acting on the condensate.
To model this, we follow the derivation given in Refs.\@~\cite{Demarchi.2023,Goychuk.2024}, highlighting key steps and introducing simplifying assumptions where appropriate.

As a starting point, we outline how phase separation can be described through an effective chemical potential.
Noting that the total amount of substrate and product is conserved, we assume a time-scale separation that allows us to eliminate one species by imposing a constant total chemical concentration.
The resulting chemical dynamics are further reduced to a mass-action kinetics that captures the essential qualitative features of the underlying, more complex reactions.
Finally, the chemotactic motion of the enzymes is incorporated via a gradient coupling, analogous to that in the classical Keller–Segel model\@~\cite{Keller.1970, Keller.1971}.
\subsection{Phase separation}
%
\begin{figure}[b]
    \centering
    \includegraphics{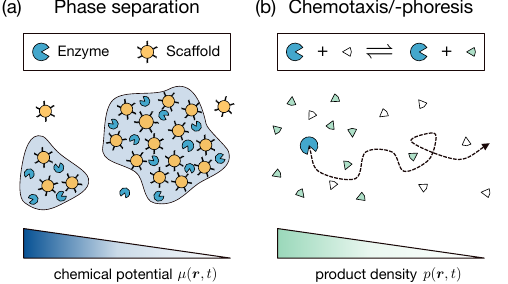}
    \caption{\textbf{Phase separation in chemically active systems}.
    (a)~\@Scaffolds (orange/light gray) and enzymes (blue/gray) co-phase separate into high- and low-density domains. Gradients in chemical potential drive coarsening by redistributing mass from smaller to larger droplets. 
    (b)~\@Enzymes catalyze reactions by transforming substrates (white triangles) into products (green/light gray triangles). Emergent gradients in chemical concentrations result in chemotactic/-phoretic motion.}
    \label{fig:modelOverview}
\end{figure}
In general, biomolecular condensates consist of more than one constituent\@~\cite{Jain.2016,Youn.2019}.
In particular, the enzymes responsible for catalyzing reactions typically do not phase separate on their own.
Instead, to accumulate within condensates, they associate with a phase-separating scaffold species\@~\cite{Banani.2016,Ditlev.2018} [Fig.\@~\ref{fig:modelOverview}(a)].
Modeling such a multi-component mixture, even with a single enzyme (client) species can be technically involved.
However, if scaffold–client interactions are significantly weaker than the scaffold–scaffold interactions driving phase separation, the local scaffold concentration acts as an effective potential for the enzyme distribution $\rho(\boldsymbol{r},t)$.
In this case, the typical enzyme concentrations $\rho_\pm$ inside and outside the condensate follow a Boltzmann distribution, ${\rho_\pm\sim\exp(-\varepsilon s_\pm)}$, where $\varepsilon$ is the interaction strength and $s_\pm$ are the scaffold concentrations inside and outside the condensate, respectively.
We aim to develop a coarse-grained description that qualitatively captures the emergence of enzyme-rich and enzyme-poor domains characterized by densities $\rho_\pm$.

These considerations motivate a thermodynamic description in terms of a binary mixture of enzymes and solvent that undergoes phase separation into domains with characteristic concentrations $\rho_\pm$.
We model the phase separation process via an effective chemical potential
\begin{align}
    \mu[\rho]
    = 
    -\kappa\boldsymbol{\nabla}^2\rho 
    - r \, (\rho-\rho_c) 
    + u \, (\rho-\rho_c)^3
    \, ,
    \label{eq:chemicalPotential}
\end{align}
given by the gradient ${\mu=\delta \mathcal{F}/\delta\rho}$ of a \textit{Ginzburg-Landau} free energy functional $\mathcal{F}[\rho]$\@~\cite{Doi.2013} which is, for simplicity, assumed to be symmetric around $\rho_c$.
The control parameter ${r > 0}$ reflects attractive enzyme-enzyme interactions---mediated through the scaffolding species---that drive phase separation. 
It quantifies the distance from the critical point ${r \sim \chi - \chi_c}$ with\@~$\chi$ denoting the Flory-Huggins parameter.
The parameter ${u > 0}$ is essential for thermodynamic stability, resulting in a free-energy density that takes the form of a double-well potential with a maximum located at the critical concentration $\rho_c$ and the minima at ${\rho_\pm = \rho_c\pm\sqrt{r/u}}$.
Finally, a positive stiffness parameter, ${\kappa > 0}$, penalizes concentration gradients.

The dynamics of phase separation are governed by the Cahn-Hilliard equation\@~\cite{Cahn.1958}, which, due to particle number conservation, takes the form of a continuity equation
\begin{align}
    \partial_t \rho(\boldsymbol{r},t) 
    = 
    - 
    \boldsymbol{\nabla} 
    \boldsymbol{J}_\rho [\rho]
    \, .
    \label{eq:CahnHilliard}
\end{align}
Following the principles of non-equilibrium thermodynamics\@~\cite{Groot.1984,Onsager.1930,*Onsager.1931}, the current $\boldsymbol{J}_\rho [\rho]$ is due to gradients in the chemical potential [Fig.\@~\ref{fig:modelOverview}(a)].
\begin{align}
   \boldsymbol{J}_\rho [\rho] 
   =
   -M(\rho) \,  \boldsymbol{\nabla}\mu[\rho] 
   \, .
   \label{eq:passiveCurrent}
\end{align}
The Onsager coefficient (mobility) $M(\rho)$ is in general a function of the density. 
However, for simplicity, we adopt the common approximation of treating it as constant.  
We determine this constant such that the Cahn-Hilliard equation reduces to an effective diffusion equation whenever the density only weakly deviates from the equilibrium densities\@~$\rho_\pm$.
Expanding Eq.\@~\eqref{eq:CahnHilliard} around \@~$\rho_\pm$ and neglecting terms of order ${\mathcal{O}\left(\boldsymbol{\nabla}^4\right)}$, one obtains 
\begin{align}
    \partial_t \rho(\boldsymbol{r},t) = \partial_\rho \mu\left(\rho_\pm\right) M\left(\rho_\pm\right) \boldsymbol{\nabla}^2 \rho(\boldsymbol{r},t)\,.
\end{align}
For a symmetric free energy and a constant Onsager coefficient, the effective diffusivity reduces to ${D_\rho= 2rM}$\@~\footnote{If we had chosen an asymmetric free energy or a more complex mobility function, the diffusion constants generally differ in the high- and low-density phases.}.

The nonlinear dynamics of the Cahn-Hilliard equation\@~\eqref{eq:CahnHilliard} is characterized by the formation and growth of phase-separated domains \cite{Weber.2019, Doi.2013, Bray.2002}.
In particular, mass is always redistributed from smaller to larger domains, a process known as coarsening or \textit{Ostwald} ripening \cite{Ostwald.1897,Wagner.1961,Lifshitz.1961}. 
The coarsening process continues until the system becomes fully phase-separated and reaches a steady state characterized by a spatially uniform chemical potential.
For a symmetric mixture, as assumed here, the steady state ${\mu=0}$ is characterized by a single flat interface
\begin{align}  
    \rho(z)
    = 
    \rho_- + \frac{1}{2}(\rho_+-\rho_-)\, 
    \big[
    1+ \tanh(z/w)
    \big]
    \,,
\label{eq:1dInterface}
\end{align}
interpolating between the minima of the associated free energy\@~\cite{Krapivsky.2010}. 
Here, $z$ denotes the direction perpendicular to the (equilibrium) interface and the interface width is given by 
\begin{equation}
    w= \sqrt{\frac{2\kappa}{r}} \, .
\end{equation} 
Calculating the free energy of such a configuration, one obtains the associated surface tension \@~\cite{Bray.2002,Weber.2019} (see App.\@~\ref{app:SurfaceTension})
\begin{align}
    \gamma = \frac{2}{3} \, \frac{w \, r^2}{u}\,,
\end{align}
which will be of great importance later on.
Note that the parameters $\kappa$, $r$, $u$, and $\rho_c$ are phenomenological and need to be matched using experimentally accessible quantities like the plateau densities $\rho_\pm$, the interface width $w$, and the effective surface tension $\gamma$. 
In thermodynamic systems, the surface tension $\gamma$ may be inferred from the statistics of interface fluctuations\@~\cite{Bray.2001,Bedeaux.1985}.
\subsection{Chemical dynamics}
Enzymes that catalyze chemical reactions can locally modulate the concentrations of substrate and product species, thereby generating spatial inhomogeneities.
This enzymatic activity gives rise to persistent concentration gradients and dynamic chemical patterns.
We aim to understand how such gradients influence the behavior of condensates composed of enzymes and scaffold proteins.
In this subsection, we focus on the chemical reaction dynamics in isolation, deferring the discussion of feedback mechanisms to the next section.
Specifically, we consider a situation where an enzyme catalyzes a reaction of the form
\begin{align}
    \textrm{substrate} + \textrm{fuel} \rightleftharpoons \textrm{product} + \textrm{waste}\,,
\end{align}
which converts a substrate $S$ into a product $P$; cf.\@~Fig.\@~\ref{fig:modelOverview}(b).
The reaction consumes chemical energy in the form of a fuel $F$, which is transformed into a lower-energy waste product $W$.
When the system is in contact with a reservoir that replenishes fuel and removes waste sufficiently fast, their chemical potentials can be treated as constant\@~\cite{Zwicker.2022}, and the reaction is continuously driven by the difference in chemical potentials
\begin{align}       
    \Delta\mu 
    = \mu_{\mathrm{fuel}} - \mu_{\mathrm{waste}} 
    \,.
\label{eq:chemicalPotentialDiff} 
\end{align}
We also consider a second, passive reaction pathway involving the spontaneous interconversion of $S$ and $P$.
Whereas the catalytic cycle operates out of equilibrium, this spontaneous reaction proceeds near equilibrium and is subject to detailed balance of the rates and thermodynamic consistency\@~\cite{Groot.1984,Kondepudi.2014,Zwicker.2022}.
In particular, the forward and backward rates of the passive reaction are determined by the internal chemical energies of the substrate and product.
Before analyzing this in detail, we introduce additional modeling assumptions and describe the time evolution of the chemical species.

When both the substrate and the product species are dilute and freely diffusing, their local densities ${s(\boldsymbol{r},t)}$ and ${p(\boldsymbol{r},t)}$ obey mass-conserving reaction-diffusion equations\@~\cite{Demarchi.2023,Brauns.2020} 
\begin{subequations}
 \begin{align}
    \partial_t s(\boldsymbol{r},t) 
    &= D_s\boldsymbol{\nabla}^2 s(\boldsymbol{r},t) - R(s,p,\rho)\,, \\
    \partial_t p(\boldsymbol{r},t) 
    &= D_p\boldsymbol{\nabla}^2 p(\boldsymbol{r},t) + R(s,p,\rho)\,.
\end{align}
\end{subequations}
All details of the local chemical reactions are encoded in the reactive flux $R(s,p,\rho)$, which depends explicitly on the local concentration of enzymes ${\rho(\boldsymbol{r},t)}$.
As a result, spatial variations in enzyme concentration modulate the local chemical dynamics and shift the effective chemical equilibria.
Importantly, the reactions conserve the total local density ${n(\boldsymbol{r},t) = s(\boldsymbol{r},t) + p(\boldsymbol{r},t)}$,
whose time evolution is governed by the continuity equation
\begin{align}
     \partial_t n(\boldsymbol{r},t) 
    = D_p\boldsymbol{\nabla}^2 \eta(\boldsymbol{r},t)
    \,.
\end{align}
Here, ${\eta = p + (D_s / D_p) \, s}$ defines the \textit{mass-redistribution potential}, a non-equilibrium quantity that captures how differential diffusion of substrate and product leads to spatial mass redistribution\@~\cite{Halatek.2018,Brauns.2020}. 
Unlike a chemical potential derived from a free energy, $\eta$ is not a thermodynamic potential but emerges from the reaction–diffusion dynamics of mass-conserving systems \cite{Frey.2020}. 
Crucially, it obeys its own equation of motion, reflecting the coupled evolution of substrate and product concentrations under the influence of both diffusion and reaction.

Although one could retain the full dynamics of substrates, products, and enzymes—or even introduce more complex reaction networks—our goal here is to motivate a minimal model that still captures the key aspects of the dynamics.
To achieve this, we assume that the substrate and product have identical diffusivities ${D_s = D_p}$.
Then, the mass redistribution potential reduces to the overall chemical density ${\eta=n}$, which, thus, obeys a closed diffusion equation
\begin{align}
    \partial_t n(\boldsymbol{r},t) 
    = D_p\boldsymbol{\nabla}^2 n(\boldsymbol{r},t)\,.
\end{align}
As pointed out in Refs.\@~\cite{Demarchi.2023, Goychuk.2024}, this implies that, in steady state, even if the enzyme density $\rho$ is spatially nonuniform, the total chemical density remains uniform
${n_\text{stat} (\boldsymbol{r}) = n_0}$.
Here, we build on this exact result and use the relation
\begin{align}
    s(\boldsymbol{r},t)
    =
    n_0-p(\boldsymbol{r},t) \, ,
    \label{eq:SubstrateElimination}
\end{align}
to eliminate the substrate concentration as an independent variable.
The latter reduction simplifies the reaction-diffusion dynamics and allows us to express the reactive flux $R$ solely in terms of the product concentration $p$ and the enzyme density $\rho$.
This assumes that diffusion is either much faster than the reaction kinetics—justifying a quasi-steady-state approximation—or that the system is initially homogeneous.
As diffusion is slow on large scales, the former should be regarded as an uncontrolled approximation.
Nonetheless, we will show that it captures the essential qualitative features of the system.

By eliminating the substrate species $s$ using Eq.\@~\eqref{eq:SubstrateElimination}, the reaction-diffusion dynamics reduce to an effective equation for a single chemical species, $p$,
\begin{align}
    \partial_t p(\boldsymbol{r},t) 
    &= D_p\boldsymbol{\nabla}^2 p(\boldsymbol{r},t) + \tilde R(p,\rho) \, ,
\end{align}
with an effective reaction term ${\tilde R(p,\rho):=R(s(p),p,\rho)}$.
This term contains effective production and degradation processes corresponding to chemical transformation between product and substrate.
Deriving a thermodynamically consistent expression for this reaction term is generally nontrivial and requires detailed knowledge of the entire chemical kinetics, including the physical interactions between the chemical species and the internal energy of the transition states\@~\cite{Hänggi.1990};  see also Sec. II in the SI of Ref.\@~\cite{Demarchi.2023}.

Here we envision a situation where the enzymatic pathway is strongly driven, favoring the conversion of substrate to product within phase-separated domains.
When at the same time the product species has a higher internal energy, the passive pathway strongly favors the formation of substrate.
To capture this behavior in minimal form, we choose a linear reaction term ${\tilde R(p,\rho) = \beta \, \rho - \lambda \, p}$, where the product is produced at rate $\beta$ in the presence of enzymes and otherwise degraded at rate $\lambda$.
This expression can be understood as the linearized limit of a more general reaction law linking enzyme density and product concentration.
Notably, the interplay of degradation and diffusion defines a single characteristic length scale—the \textit{diffusion length}, ${l = \sqrt{D_p/\lambda}}$—which sets the typical range of chemical interactions and will be central to the following analysis.
\subsection{Chemotaxis}
As illustrated in Fig.\@~\ref{fig:modelOverview}(b), we consider a scenario involving directed mass transport of phase-separated droplets along self-generated chemical gradients.
To account for these contributions, we supplement the Cahn-Hilliard equation\@~\eqref{eq:CahnHilliard} by a current
\begin{align}
    \boldsymbol{J}[\rho,p]
    =
    -K(\rho,p)\boldsymbol{\nabla}p(\boldsymbol{r},t)
    \, .
    \label{eq:ChemotaxisCurrent}
\end{align}
Starting from the more general model used in Refs.\@~\cite{Demarchi.2023,Goychuk.2024}, the above relation, Eq.\@~\eqref{eq:ChemotaxisCurrent}, is obtained by eliminating the substrate species using Eq.\@~\eqref{eq:SubstrateElimination}.
One finds ${K(\rho,p)=\Delta\chi M(\rho)}$, with $M$, the equilibrium mobility and $\Delta\chi$, the difference in enzyme-product and enzyme-substrate interaction energies.
From a broader perspective, Eq.\@~\eqref{eq:ChemotaxisCurrent} can be understood as the leading-order term in a gradient expansion of the chemical density $p$.
The proposed current takes the same form as in the classical Keller-Segel (KS) model of chemotaxis\@~\cite{Keller.1970, Keller.1971}, with $p$ playing the role of a chemoattractant or chemorepellent.
In this sense, our choice for the current $\boldsymbol{J}$ encompasses a much broader class of systems,  in which the actual microscopic origin of motion along gradients in $p(\boldsymbol{r},t)$ is encoded in the mobility $K(\rho,p)$.

If the emergent fluxes originate from pairwise interactions between the enzymes and their substrates and products, the current can be shown to derive from a chemical potential\@~\cite{Demarchi.2023,Goychuk.2024}.
This constrains the functional form of the mobility ${K(\rho,p) \sim M(\rho)}$, where $M(\rho)$ is the Onsager parameter associated with the passive current, Eq.\@~\eqref{eq:passiveCurrent}.
In particular, there is no explicit dependence of $K$ on $p(\boldsymbol{r},t)$.
The sign of the interactions is a consequence of both the enzyme-product and enzyme-substrate interactions.
When the interaction of enzymes and products is less favorable than that between enzymes and substrates, $K{(\rho)>0}$.
That is, enzymes tend to migrate towards regions of lower product concentrations.
Due to Onsager's reciprocity relations\@~\cite{Onsager.1930, *Onsager.1931}, the pairwise interactions also induce a current in the chemical species, which may, however, be omitted when considering strong reactive fluxes $\tilde R(\rho,p)$, i.e., fast turnover of the chemical species\@~\cite{Goychuk.2024}.

For the more general case, where the mobilities $M(\rho)$ and $K(\rho,p)$ have different functional forms, the dynamics of the density ${\rho(\boldsymbol{r},t)}$ can no longer be derived from a single chemical potential as in Refs.\@~\cite{Demarchi.2023, Goychuk.2024}.
Instead, the emergent current should be understood as describing active motion along chemical gradients, characteristic of classical chemotaxis.
If one omits the phase-separation dynamics, our model falls in the well-established Keller–Segel (KS) class of models\@~\cite{Keller.1970,Keller.1971,Hillen.2008,Tindall.2008}.
These models typically involve a system of two equations governing the time evolution of the density $\rho(\boldsymbol{r},t)$ of auto-chemotactic agents that migrate up or down gradients in a chemical field $p(\boldsymbol{r},t)$:
\begin{subequations}
\begin{align} 
    \partial_t \rho (\boldsymbol{r},t)
    &= 
    \boldsymbol{\nabla}
    \big[ 
    D(\rho,p)\boldsymbol{\nabla}\rho +  K(\rho,p)  \boldsymbol{\nabla} p
    \big]
    \, , \label{eq:ChemoDensity} \\
    \partial_t p(\boldsymbol{r},t)
    &= 
    D_p \boldsymbol{\nabla}^2 p + \beta(\rho,p)\, \rho - \lambda(\rho,p) \, p
    \, .\label{eq:Product}
\end{align}
\end{subequations}
The first term in Eq.\@~\eqref{eq:ChemoDensity} accounts for diffusion, whereas the second term describes the chemotactic drift, with the sensitivity function $K(\rho,p)$ incorporating various effects such as crowding\@~\cite{Painter.2002} or details of the underlying sensing mechanisms\@~\cite{Segel.1977}.
The second equation, Eq.\@~\eqref{eq:Product}, governs the dynamics of the chemical, which diffuses freely with a diffusivity $D_p$ and is actively produced and degraded at rates $\beta(\rho,p)$ and $\lambda(\rho,p)$, respectively.

While most studies of Keller-Segel-like models assume a constant diffusion coefficient $D$\@~\cite{Weyer.2024,Painter.2011,Kolk.2023,Rapp.2019,Narla.2021,Hillen.2008,Tindall.2008}, some generalize it to density-dependent forms such as ${D(\rho,p) \sim \rho^m}$\@~\cite{Arumugam.2021,Baek.2017}.
These approaches, though insightful, often overlook the impact of short-range particle interactions on transport, particularly in systems where such interactions may drive phase separation.
By incorporating the chemical potential, Eq.\@~\eqref{eq:chemicalPotential}, our model explicitly captures the coupling between non-linear diffusion and interaction-driven phase-separating behavior.
This provides a novel perspective on systems where auto-chemotactic agents experience both chemotactic drift and interaction-driven phase separation.
\subsection{Model equations}
\label{sec:modelEquations}
Taken together, we consider the following reduced set of equations
\begin{subequations}
 \begin{align}
    \partial_t \rho(\boldsymbol{r},t) 
    &= 
    \boldsymbol{\nabla}
    \big[ M(\rho)\boldsymbol{\nabla}\mu[\rho]+K (\rho) \boldsymbol{\nabla}p
    \big] 
    \, , \label{eq:DensityUnscaled} \\
    \partial_t p(\boldsymbol{r},t) 
    &= D_p\boldsymbol{\nabla}^2 p + \beta \rho - \lambda p
    \,,\label{eq:ChemicalUnscaled}
\end{align}
\end{subequations}
where we assume a constant mobility ${M(\rho) = M}$ and take the sensitivity function $K(\rho)$ to depend only on the density field.
With the commonly used choice ${K(\rho) = K \rho}$, the model generalizes the minimal Keller–Segel model \cite{Keller.1970,Keller.1971} by incorporating phase-separation dynamics in the density field.

In the limit ${K = 0}$, the two fields decouple, and one recovers the Cahn–Hilliard equation~\eqref{eq:CahnHilliard}, which describes the formation and subsequent coarsening of phase-separated domains.
Conversely, in the non-interacting limit (${\kappa = u = 0}$, ${r < 0}$), the Cahn–Hilliard dynamics reduce to ordinary diffusion, and one recovers the classical Keller–Segel model of chemotaxis.

When both $M$ and $K$ are non-zero, the additional diffusion length scale ${l = \sqrt{D_p / \lambda}}$ and the dynamical interplay between equilibrium and non-equilibrium currents can lead to qualitatively different behavior, beyond what is captured by the Keller–Segel model and the Cahn–Hilliard model.
Depending on the sign of $K$, the chemotactic interaction either drives aggregation ($K < 0$) or dispersal of the auto-chemotactic agents ($K > 0$).
For $K < 0$, attractive chemical interactions enhance the formation of high-density domains and accelerate the coarsening dynamics.
Notably, while the classical KS model exhibits chemotactic collapse\@~\cite{Hillen.2008,Tindall.2008,Arumugam.2021}, this singularity is avoided in our model due to the stabilizing effects of the entropic term in the chemical potential\@~$\mu$.
As a result, one anticipates a final steady state consisting of a single phase-separated droplet with an average density close to the equilibrium value $\rho_+$.
In contrast, for repulsive interactions (${K > 0}$), the equilibrium and non-equilibrium currents oppose each other, giving rise to novel and potentially more complex dynamics.
In the following, we focus exclusively on the case $K > 0$, where this competition leads to rich and nontrivial behavior.

Since Eq.\@~\eqref{eq:ChemicalUnscaled} is linear in the chemical field $p$ with a source term proportional to $\rho$, it allows one to formally integrate out the chemical field $p(\boldsymbol{r},t)$:
\begin{align}
    p(\boldsymbol{r},t)= \int\mathrm{d}t^\prime \, \mathrm{d}^{d}r^\prime \, 
    G (|\boldsymbol{r}-\boldsymbol{r}^\prime|,t-t^\prime ) \,   \rho(\boldsymbol{r}^\prime,t) \, ,
    \label{eq:FullGreensFuncton}
\end{align}
where $G(r,t)$ is the associated Green's function of the problem and $d$ denotes the system's dimension.
In the limit where the chemical dynamics are much faster than the density field, the quasi-static approximation ${\partial_t p(\boldsymbol{r},t) \approx 0}$ applies. 
In this case, Eq.\@~\eqref{eq:ChemicalUnscaled} simplifies to a screened Poisson equation, with the term proportional to $\rho$ acting as the source and the degradation term proportional to $p$ serving as a screening factor. 
Consequently, Eq.\@~\eqref{eq:FullGreensFuncton} reduces to 
\begin{align}
    p(\boldsymbol{r},t)
    = 
    \int \mathrm{d}^{d}r^\prime \, 
    G (|\boldsymbol{r}-\boldsymbol{r}^\prime|) \,
    \rho(\boldsymbol{r}^\prime,t) \, .
    \label{eq:stationaryConvolution}
\end{align}
Here $G(r)$ represents the instantaneous Green's function; for example in an infinite domain in two dimensions one has ${G(r) \sim \ln (r/l)}$, while in three dimensions ${G(r) \sim \frac{1}{r} e^{-r/l}}$.
In this limit, the chemotactic current ${J\sim-K(\rho) \boldsymbol{\nabla}\mu_\mathrm{ch}}$ may be understood as resulting from a chemical potential ${\mu_\mathrm{ch}=\delta \mathcal{F}_\mathrm{ch}/\delta\rho}$, with a free energy functional
\begin{align}
    \mathcal{F}_\mathrm{ch}[\rho]
    =\int \mathrm{d}^d r^\prime  \mathrm{d}^d r^{\prime\prime} \,
    \rho(\boldsymbol{r}^\prime,t) \, G (|\boldsymbol{r}^\prime-\boldsymbol{r}^{\prime\prime}|) \, \rho(\boldsymbol{r}^{\prime\prime},t) \,.
\end{align}
This expression illustrates that the chemical field $p$ acts as an effectively instantaneous, screened potential that mediates long-ranged repulsive interactions.
Such interactions have been recognized in various thermodynamic systems, including diblock copolymers\@~\cite{Liu.1989} and phase-separating fluids with electrostatic interactions\@~\cite{Kumar.2023}, where long-range repulsion stabilizes micro-phase separation.
Related phenomena also arise in systems driven by membrane-mediated interactions\@~\cite{Yu.2025,Winter.2025}, and in chemically active condensates\@~\cite{Glotzer.1994,Zwicker.2015}.

Although one may anticipate similar phenomenology, our model presents a significant challenge:
The unequal mobilities in Eq.\@~\eqref{eq:DensityUnscaled} violate detailed balance and preclude the existence of a single Lyapunov or free energy functional.
As a result, variational approaches cannot be used to determine steady states, and a full dynamical analysis is required instead.

In a parameter regime where the quasi-static approximation is no longer valid, the additional timescale associated with the relaxation of the chemical field introduces memory effects, as seen in Eq.\@~\eqref{eq:FullGreensFuncton}.
As we show in the following, these memory effects can give rise to qualitatively novel dynamical phenomena that are absent in thermodynamic systems.
\subsection{Non-dimensionalized theory}
\label{sec:rescaling}
To simplify the analysis, we rescale our theory into a dimensionless form and express it in terms of a minimal set of independent parameters. 
We choose to measure distances in units of the equilibrium interface width $\boldsymbol{r} \rightarrow w \, \tilde{\boldsymbol{r}}$, and time is measured in terms of the associated diffusive time scale, ${t\rightarrow (w^2/D_p)\, \tilde t}$.
The basic energy scale is set by the typical interaction strength in the high-density phase, ${r\rho_+^2w^d = 1}$, and concentrations are rescaled in terms of their steady-state values in the high-density phase:
\begin{align}
    \rho(\boldsymbol{r},t) \rightarrow \rho_+ \, \phi (\tilde{\boldsymbol{r}},\tilde t)\,, 
    \quad
    p(\boldsymbol{r},t) \rightarrow \frac{\beta\rho_+}{\lambda} \, \psi (\tilde{\boldsymbol{r}},\tilde t)\,.
\end{align}
For notational simplicity, we omit the tildes in the following.
It is important to note that the dimensionless formulation depends on the distance to the critical point through the scaling choices above.
This rescaling allows us to rewrite Eqs.\@~\eqref{eq:DensityUnscaled}--\eqref{eq:ChemicalUnscaled} in a compact, dimensionless form 
\begin{subequations}
\label{eq:DynamicsRescaled}
\begin{align}
    \partial_t \phi(\boldsymbol{r},t) 
    &= M \boldsymbol{\nabla}
    \big[\boldsymbol{\nabla}\mu[\phi]
    +
    Pe \, \phi \, \boldsymbol{\nabla}\psi
    \big] 
    \label{eq:DenistyRescaled}
    \, , \\
    \partial_t \psi(\boldsymbol{r},t) 
    &= 
    \boldsymbol{\nabla}^2\psi + \frac{1}{l^2} \hat R(\phi,\psi)
    \, . 
    \label{eq:ChemicalRescaled}
\end{align}
\end{subequations}
Here, we defined ${\hat{R}(\phi, \psi) = \phi - \psi}$ as the reaction term, and introduced ${l := \sqrt{D_p / (\lambda w^2)}}$ as the dimensionless diffusive length scale of the chemical field.
The constant dimensionless mobility ${M = D_\rho / (2 D_p)}$ compares the diffusion constants of the density field $\phi$ and the chemical field $\psi$.
To streamline notation, we reuse the symbols $l$ and $M$ from the original dimensionful quantities.
The effective \textit{Péclet} number ${Pe = 2 K \rho_+ \beta / (D_\rho \lambda)}$ quantifies the relative strength of directed chemotactic fluxes to diffusive motion.
The non-dimensionalized chemical potential is defined as ${\mu(\boldsymbol{r}, t) = \delta \mathcal{F} / \delta \phi}$, with an effective free energy functional of the form
\begin{align}
    {\cal F} [\phi] 
    = 
    \int \mathrm{d}^d r
    \bigg[
    \dfrac{1}{4} \left(\boldsymbol{\nabla}\phi\right)^2 
    + f(\phi)
    \bigg] \, .
\label{eq:FreeEnergy}
\end{align}
The local free energy density is given by
\begin{equation}
    f(\phi) = - 
    \frac{1}{2} \, \left(\phi-\phi_c\right)^2
    + 
    \frac{1}{(\Delta \phi)^2} \left(\phi-\phi_c\right)^4 \, ,
    \label{eq:freeEnergyDensity}
\end{equation}
where the parameter 
\begin{align}
    \Delta \phi 
    = \dfrac{\rho_+ - \rho_-}{\rho_+}
    = \phi_+ - \phi_-
    \label{eq:DeltaPhi}
\end{align}
denotes the relative difference in the minima of the square-well potential, that is, the contrast in density between the dense and dilute phases.
It relates to the critical density $\phi_c$ via ${\phi_c=\rho_c/\rho_+=(2-\Delta\phi)/2}$, i.e., the free energy is parameterized by a single dimensionless number.
\goodbreak
\section{Arrested coarsening, self-propulsion, and chiral capillary waves}
\label{sec:overview}
\begin{figure*}
    \centering
    \includegraphics{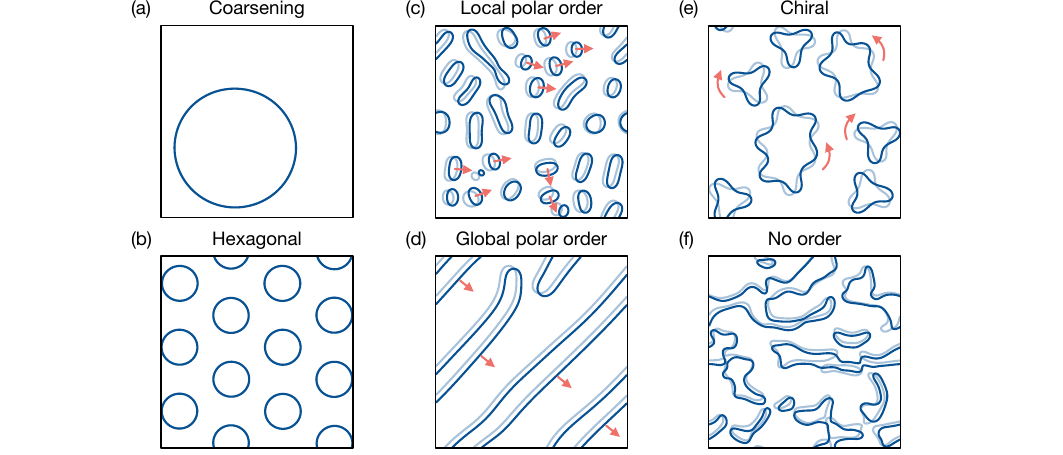}
    \caption{\textbf{Collective dynamic states.} Characteristic collective states of the dynamics described by Eqs.\@~\eqref{eq:DynamicsRescaled}--\eqref{eq:FreeEnergy}, as obtained from finite element simulations. Each figure displays the ${\phi=\phi_c}$ level sets at time $t$ (blue/dark gray lines) and ${t-\Delta t}$ (light blue/gray lines). The time shift $\Delta t$ was chosen in a representative way to properly indicate the dynamical behavior. Depending on the parameters, one observes either static\@~(a--b) or dynamic\@~(c--f) patterns with arrows indicating the direction of propulsion\@~(c--d) or rotation\@~(e). The observed patterns may be further characterized by either hexagonal-order (b), local-\@~(c), or global\@~(d) polar order of the propagation direction or a notion of chirality\@~(e). In the absence of order, the system (transiently) displays irregular behavior\@~(f). All simulations were conducted in a quadratic domain of size ${L=80}$ with periodic boundary conditions, an average density ${\bar \phi=0.59}$, and a critical density ${\phi_c=0.7}$. The remaining parameters ${(M,Pe,l)}$ are (a): ${(1,0.1,5)}$, (b): ${(1,0.5,5)}$, (c): ${(10,1,1)}$, (d): ${(10,1,3.5)}$, (e): ${(15,0.7,1)}$, and (f): ${(20,50,5)}$. The corresponding time evolution for each set of parameters is shown in Video\@~S1\@~\cite{SM}.
}
\label{fig:overview}
\end{figure*}
To build intuition for the dynamics of the model class defined by Eqs.\@~\eqref{eq:DynamicsRescaled}--\eqref{eq:FreeEnergy}, we numerically integrated these dynamical equations using the COMSOL finite element suite\@~\cite{AB.2023,ZN}.
Our results show that the system's dynamics and steady states are strongly influenced by a few key parameters, specifically the relative mobility $M$, the Péclet number $Pe$, and the diffusion length $l$.
While the full phase diagram (App.\@~\ref{app:phaseDiagramLSA}--\ref{app:phaseDiagramNumerics}) reveals complex and intricate behavior, we focus on a specific subset of parameters to illustrate key dynamical regimes and trends rather than attempting an exhaustive analysis.

Figure\@~\ref{fig:overview} showcases snapshots of the characteristic collective states for a representative set of parameters, illustrating the wide range of behaviors observed under different conditions; see also Video\@~S1 of the Supplemental Material\@~\cite{SM}. 
All simulations were initialized from random initial conditions, with ${\Delta\phi = 0.6}$, corresponding to a critical density ${\phi_c = 0.7}$ and an average density ${\bar{\phi} = 0.59}$, i.e., within the spinodal regime.
Periodic boundary conditions were applied to eliminate edge effects and ensure a focus on bulk behavior.
Further information about the numerical procedure is provided in App.\@~\ref{app:Numerics}.

For small P\'eclet numbers (${Pe \ll 1}$), corresponding to weak chemotactic interactions, the system undergoes classical spinodal decomposition followed by Ostwald ripening.
This process eventually leads to a stationary state characterized by the presence of a single phase-separated domain; see Fig.\@~\ref{fig:overview}(a).
In the opposite limit (${Pe \gg 1}$), where long-ranged repulsive interactions dominate, phase separation can be fully suppressed, resulting in a homogeneous steady state.
These two limiting cases correspond to behaviors governed by the classical Cahn-Hilliard equation and the repulsive Keller-Segel model, respectively.

In contrast, for intermediate interactions, ${Pe \geq \mathcal{O}(1)}$, the system displays a rich variety of non-trivial collective states, including static [Fig.\@~\ref{fig:overview}(b)] and dynamic patterns [Fig.\@~\ref{fig:overview}(c--f)].
When the propagation of the chemical field $\psi(\boldsymbol{r}, t)$ is significantly faster than the diffusive mass redistribution of $\phi$, ${M \leq \mathcal{O}(1)}$, the system typically evolves into an arrested state in which equally sized domains arrange into a hexagonal lattice [Fig.\@~\ref{fig:overview}(b)].
This is a classical hallmark of \textit{arrested coarsening}\@~\cite{Zwicker.2015,Brauns.2021,Muratov.2002}.

For slower signal propagation, ${M \geq \mathcal{O}(10)}$, a variety of dynamic states emerge.
These include traveling droplets [Fig.\@~\ref{fig:overview}(c)], stripes [Fig.\@~\ref{fig:overview}(d)], and ruffled domains displaying apparent rotational motion [Fig.\@~\ref{fig:overview}(e)]; see Video\@~S1\@~\cite{SM}.
The stripes and droplets exhibit directed translational motion with a well-defined velocity, as well as varying degrees of local or global polar order.
Here, local order refers to the short-range alignment of velocities among nearby droplets, while global order implies coherent system-wide motion.
In contrast, the ruffled domains [Fig.\@~\ref{fig:overview}(e)] remain essentially immobile but exhibit local chirality, characterized by a distinct handedness in their direction of rotation.
As we will show, this apparent rotation does not correspond to rigid body motion but instead arises from \textit{capillary waves} propagating along the interfaces of the domains.

As a side remark, we note that for very strong interactions, ${Pe\geq\mathcal{O}(10)}$, we observe the onset of behavior that appears irregular and potentially chaotic-like [Fig.\@~\ref{fig:overview}(f)].
The dynamics initially feature highly mobile, irregularly shaped domains that continually merge, split, and mix.
There is no clear notion of order beyond the formation of high- and low-density regions.
Moreover, within the high-density domains, the rescaled enzyme density $\phi$ often exceeds the equilibrium plateau density $\phi_+$; cf.\@~Video\@~S1\@~\cite{SM}.
This is unexpected, as repulsive chemotactic interactions are generally expected to produce a plateau density lower than in the equilibrium case, in line with observations for the other parameter sets in Fig.\@~\ref{fig:overview}.
However, our simulations suggest that this behavior may be transient and should not be considered an independent dynamical phase.
Elucidating the mechanisms that give rise to this irregular regime requires further analysis and remains an open question for future research.

Instead, in this work, we focus on the regime of moderate Péclet numbers, ${Pe=\mathcal{O}(1)}$, where our numerical simulations reveal the critical role of the relative mobility $M$ in shaping the system’s dynamics (App.\@~\ref{app:phaseDiagramNumerics}).
Specifically, $M$ dictates whether the system settles into an arrested state or gives rise to dynamic patterns, including traveling droplets and capillary waves.
Given the extensive prior work on arrested coarsening in systems with long-range interactions\@~\cite{Liu.1989,Glotzer.1994,Kumar.2023,Yu.2025,Winter.2025,Muratov.2002,Zwicker.2015}, we do not further analyze this regime.
Rather, we concentrate on the dynamical regime (${M\geq\mathcal{O}(10)}$), where the emergence of non-equilibrium patterns remains less explored.

Previous studies have examined self-propelled domains in systems driven by enzymatic activity\@~\cite{Demarchi.2023,Goychuk.2024}.
While these systems are more complex in their underlying mechanisms, their insights are largely applicable to our model, Eqs.\@~\eqref{eq:DynamicsRescaled}--\eqref{eq:FreeEnergy}.
Hence, we turn our attention to the unexplored capillary waves that arise in our system.
In what follows, we present a detailed qualitative and quantitative analysis of these waves and their role in shaping the system’s emergent dynamics.
We begin by analyzing how capillary waves emerge and identify the key mechanisms that drive their formation.
\section{Stationary interfaces, gears, and conveyor belts}
\label{sec:stationaryInterfaces}

To analyze the emergence of capillary waves, we first establish stationary interface solutions for our model.
Once these stationary solutions are identified, we analyze their linear stability against small perturbations.
This approach allows us to determine the conditions under which capillary waves form.

In subsection \ref{sec:interfaces}, we focus on constructing stationary interface solutions for both flat and constantly curved interfaces.
To achieve this, we numerically determine the stationary solution to the model equations, Eq.\@~\eqref{eq:DynamicsRescaled}.
Additionally, we employ a sharp interface approximation to derive analytical solutions, which offer complementary insight into the qualitative properties of these interfaces.
In subsection \ref{sec:Gears&Belts}, we numerically probe the stability of the exact solutions and describe the characteristics of the observed instabilities.
Finally, in Sec.~\ref{sec:HeuristicAnalysis}, we develop a heuristic argument to rationalize the onset of interface instabilities.
\subsection{Stationary interfaces}
\label{sec:interfaces}
\begin{figure}
    \centering
    \includegraphics{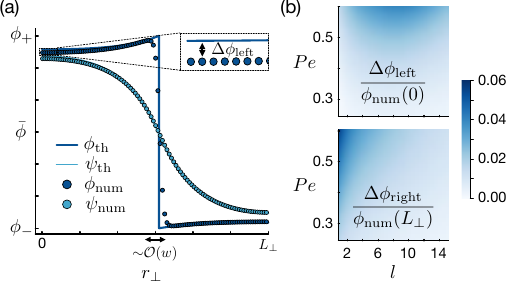}
    \caption{\textbf{Stationary flat interface solutions}. 
    (a) Profiles of the fields $\phi$ (dark blue/gray) and $\psi$ (light blue/gray) obtained from numerical simulations (bullets) and theoretical estimates (solid lines) obtained by solving a piecewise linearized version of Eq.\@~\eqref{eq:stationaryProblem}---as detailed in App.\@~\ref{app:1dInterface}---for a system of size ${L_\perp=60}$ with an average density ${\bar\phi=0.7}$. The deviations between the exact numerical solution and the theoretical estimates at the system boundaries are termed $\Delta\phi_\mathrm{left}$ and $\Delta\phi_\mathrm{right}$, respectively. The remaining parameters are ${M=10}$, ${l=10}$, ${Pe=0.35}$, and ${\phi_c=0.7}$.
    (b) Relative deviations between the numerically and theoretically obtained solutions at the system's boundaries for a range of values $l$ and $Pe$ with all other parameters identical to (a). The value of the deviation at both boundaries is indicated by the colorbar.}
    \label{fig:interfaces}
\end{figure}
For simplicity, we focus on the cases of a flat interface and a circular domain of constant curvature.
This choice enables the use of a global coordinate system with unit vectors aligned parallel and perpendicular to the interface, simplifying the analysis. 
Specifically, for a flat interface, we have ${\boldsymbol{r} = x \boldsymbol{\hat e}_x + z \boldsymbol{\hat e}_z}$, while for a curved interface, ${\boldsymbol{r} = r \boldsymbol{\hat e}_r}$.
Here, $\boldsymbol{\hat e}_x$ and $\boldsymbol{\hat e}_z$ denote the unit vectors in the horizontal and vertical directions, respectively, for the flat interface, whereas $\boldsymbol{\hat e}_r$ is the radial unit vector in the curved case.
To ease the notation, we will jointly denote the vertical coordinate as ${r_\perp = \{ z, r \} }$.

With this choice of the reference frame, stationary interface profiles, denoted as ${\boldsymbol{\Phi}_I(r_\perp)=(\phi_I (r_\perp),\psi_I(r_\perp))}$, are solutions to a one-dimensional problem, which we will refer to as the \textit{profile equations}:
\begin{subequations}   
\label{eq:stationaryProblem}
\begin{align}
    0 &= \partial_\perp\mu + Pe \, \phi \, \partial_\perp\psi \, , 
    \label{eq:stationaryProblemDensity} 
    \\
    0 &= \boldsymbol{\nabla}^2_\perp\psi + \frac{1}{l^2}
    \big( 
    \phi-\psi
    \big) \, .
    \label{eq:stationaryProblemChemical}
\end{align}
\end{subequations}
These equations follow from the stationary limit of the full dynamical model, Eq.\@~\eqref{eq:DynamicsRescaled}, using that for a flat or constantly curved domain there are no variations in the direction parallel to the interface. 
Here ${\partial_\perp=\{\partial_z,\partial_r\}}$ denotes the derivative in the direction of the unit vector\@~$\boldsymbol{\hat r}_\perp$ (pointing from the high- to the low-density phase), while ${\boldsymbol{\nabla}_\perp^2=\{\partial_z^2,\partial_r^2+\partial_r/r\}}$ is the corresponding Laplacian, which accounts for the effects of curvilinear coordinates. 

For both the flat and spherical cases, we consider the system to have a finite extent in the direction perpendicular to the interface, denoted as $L_\perp$.
Consequently, the stationary equations, Eq.\@~\eqref{eq:stationaryProblem}, must be supplemented with appropriate boundary conditions.
In both cases, we require the solution $\boldsymbol{\Phi}_I$ to be differentiable.
In the spherical case, this implies that the derivatives vanish at the origin (${r = 0}$), leading to the boundary conditions ${\partial_\perp \phi(r_\perp) = \partial_\perp \psi(r_\perp) = 0}$.
At the outer boundary, ${r_\perp = L_\perp}$, we impose no-flux conditions, reflecting the absence of mass transport across system boundaries.
For the flat case, equivalent no-flux boundary conditions are applied at both boundaries to ensure mass conservation.

Since the profile equations, Eq.\@~\eqref{eq:stationaryProblem}, are highly non-linear, exact analytical solutions are generally not feasible, and we instead rely on numerical methods. 
Figure\@~\ref{fig:interfaces}(a) illustrates the flat interface solution for a representative parameter set, determined through finite-element simulations.
We highlight two key observations:
(i) The density $\phi$ exhibits distinct regions of low and high density, separated by a sharp interface.
Moreover, within each of these regions, the $\phi$ profile deviates only slightly from the free energy minima, $\phi_\pm$, of the local free energy function, Eq.\@~\eqref{eq:freeEnergyDensity}.
(ii) In contrast, the chemical field $\psi$ varies on a much larger scale, set by the non-dimensional diffusion length.
Within each phase, its value tends to the reactive equilibrium ${\phi=\psi}$.

The first observation suggests that an approximate analytical solution can be constructed using a perturbative approach ${\phi=\phi_\pm+\delta\phi}$, assuming small deviations $\delta\phi$ from the equilibrium interface profile.
The second observation indicates a separation of length scales, meaning that when the diffusive length scale $l$ significantly exceeds the equilibrium interface width $w$, the chemical field $\psi$ is approximately constant across the interface [Fig.\@~\ref{fig:interfaces}(a)].
Under these conditions, chemotactic currents $\sim \boldsymbol{\nabla} \psi$ can be neglected on the scale of $w$, and Eq.\@~\eqref{eq:stationaryProblemDensity} reduces to the equilibrium condition ${\partial_\perp \mu = 0}$.
As a result, the interface profile on the scale of $w$ is governed solely by the chemical potential $\mu$ and matches the equilibrium interface shape [Eq.\@~\eqref{eq:1dInterface}].

Building on these arguments, we adopt the limit of an infinitely sharp interface.
On each side of the interface, we expand the stationarity condition, Eq.\@~\eqref{eq:stationaryProblemDensity}, to linear order in the perturbations $\delta\phi$, retaining only gradient terms up to diffusive order, $\mathcal{O}(\partial_\perp^2)$.
This approximation is consistent with the numerical observation that the profile deviates only weakly from the equilibrium plateau values $\phi_\pm$.
In this sharp interface limit, the only information retained about the interface is the density value on either side:
\begin{align}
    \phi\big\rvert_{r_\perp=X_0^\mp} = \phi_\mathrm{in/out} \, .
    \label{eq:sharpBC}
\end{align}
Here, $X_0$ denotes the interface position, and the notation ${X_0^\mp = \lim_{\varepsilon\rightarrow 0} (X_0 \mp \varepsilon)}$ represents the left and right limits approaching the interface.
For the flat interface, we have ${\phi_{\mathrm{in/out}} = \phi_\pm}$, with the free energy minima $\phi_\pm$ as defined in Eq.\@~\eqref{eq:freeEnergyDensity}.
In contrast, for a curved interface, the densities at the interface $\phi_\mathrm{in/out}$ may generally differ from $\phi_\pm$ (App.\@~\ref{app:GibbsThomsonSection}). 
As illustrated in Fig.\@~\ref{fig:interfaces}(a), the boundary conditions in Eq.\@~\eqref{eq:sharpBC} imply that the density field $\phi$ exhibits a discontinuous jump across the interface.
At the same time, we require the chemical field $\psi$ to be continuous across the interface, which imposes a matching condition on either side:
\begin{align}
    \psi\big\rvert_{X_0^-} = \psi\big\rvert_{X_0^+}, \, \quad 
    \partial_\perp\psi\big\rvert_{X_0^-}=\partial_\perp\psi\big\rvert_{X_0^+}.
    \label{eq:sharpBC2}
\end{align}
That is, within the sharp interface limit, the profile equations, Eq.\@~\eqref{eq:stationaryProblem}, can be solved separately in each of the domains ${r_\perp \in[0,X_0]}$ and ${r_\perp \in[X_0,L_\perp]}$, ensuring that the solutions satisfy the boundary conditions, Eqs.\@~\eqref{eq:sharpBC}--\eqref{eq:sharpBC2}, at ${r_\perp=X_0}$. 
Thereby, we have reduced the non-linear profile equations, Eq.\@~\eqref{eq:stationaryProblem}, to a set of piecewise linear equations, whose solutions are detailed in App.\@~\ref{app:stationaryInterfaces}. 
On each side of the interface, the solution takes the form
\begin{align}
    \phi(r_\perp) 
    = 
    \begin{cases}
        c_+ - \dfrac{Pe}{2} \, \phi_+ \, \psi(r_\perp) \, , \quad &r_\perp < X_0 \, , \\[4mm]
        c_- - \dfrac{Pe}{2} \, \phi_- \, \psi(r_\perp) \, , \quad &r_\perp > X_0 \, .
    \end{cases}
    \label{eq:linearInterfaceRelation}
\end{align}
Here, the constants $c_\pm$ are chosen such that the boundary conditions, Eqs.\@~\eqref{eq:sharpBC}--\eqref{eq:sharpBC2}, are fulfilled.
One finds that the deviations of the density profile $\phi$ from the minima of the free energy are linearly related to the modulations of the chemical field $\psi$ and that their amplitude is proportional to the P\'eclet number $Pe$.
That is, the stronger the chemotactic interactions, the more the density profile deviates from the equilibrium interface profile.

For ${l = 10}$ and the other parameters as specified in Fig.\@~\ref{fig:interfaces}(a), a comparison between the exact numerical solutions of the profile equations, Eq.\@~\eqref{eq:stationaryProblem}, and those obtained via the sharp interface approximation reveals near-perfect agreement [see Fig.\@~\ref{fig:interfaces}(a)].
To further validate our approximations, we compare the theoretical predictions with numerical solutions across a range of $Pe$ and $l$ values.
Specifically, we measure the offsets ${\Delta\phi_\mathrm{left}}$ and ${\Delta\phi_\mathrm{right}}$ between the analytical and numerical solutions at the system boundaries ${r_\perp = 0}$ and ${r_\perp = L_\perp}$.
These results, shown in Fig.\@~\ref{fig:interfaces}(b), demonstrate good overall agreement, with relative deviations of only a few percent.
However, deviations increase for larger Péclet numbers and smaller diffusion lengths $l$:
For small $l$, the sharp interface approximation loses accuracy as the diffusive length scale approaches the interface width.
For large $Pe$, Eq.\@~\eqref{eq:linearInterfaceRelation} predicts that stronger interactions with the chemical field lead to greater deviations from the equilibrium densities.
Both trends are in accordance with the approximations made and illustrate their limitations in extreme parameter regimes. 
\subsection{Gears \& conveyor belts}
\label{sec:Gears&Belts}
\begin{figure*}
\centering
\includegraphics[]{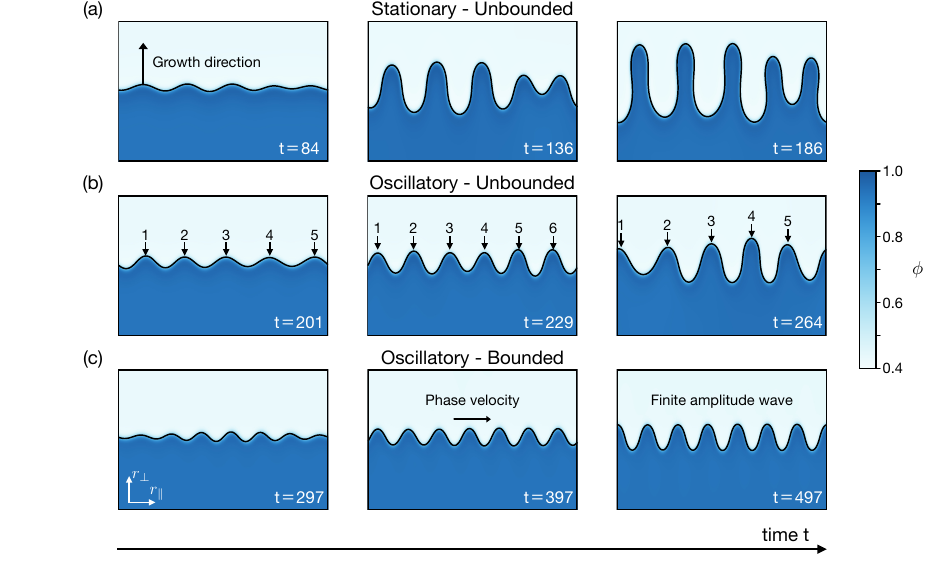}
\caption{\textbf{Instability of a flat interface}.
    (a--c) Time evolution of an initially flat interface, obtained by numerically solving the dynamic equations, Eq.\@~\eqref{eq:DynamicsRescaled}, using a finite element scheme for parameters ${Pe=0.42}$ and ${\bar\phi=\phi_c=0.7}$. The remaining parameters ${(l,M)}$ are (a): ${(3.5,20.5)}$, (b): ${(2.75,20.5)}$, and (c): ${(1.75,22.5)}$. All simulations were initialized with the corresponding exact stationary interface solutions to the profile equations, Eq.\@~\eqref{eq:stationaryProblem}, for a system of size ${L_\parallel=150}$ and ${L_\perp=50}$ with periodic boundary conditions in the direction parallel to the interface and no-flux boundary conditions in the perpendicular direction.
    (a) Stationary fingering-like instability with perturbations growing predominantly in the vertical direction (black arrow).
    (b) Instability oscillating between patterns with ${n=5}$ and ${n=6}$ fingers of increasing amplitude.
    (c) Oscillatory instability saturating in a (transiently) stable finite amplitude capillary wave with a constant phase velocity.
    The full time evolution for each case is displayed in Video\@~S2\@~\cite{SM}.}
\label{fig:categorization}
\end{figure*}
Having established the stationary interface solutions, we now examine their stability in response to small perturbations. 
To that end, we initialize a two-dimensional system with the numerical solutions to the profile equations, Eq.\@~\eqref{eq:stationaryProblem}, for a flat interface.
We numerically evolve this state by employing a finite element scheme to solve the dynamical equations, Eq.\@~\eqref{eq:DynamicsRescaled}, with periodic boundary conditions in the direction parallel to the interface and no-flux conditions in the perpendicular direction.
A detailed description of the numerical implementation and procedures can be found in App.\@~\ref{app:Numerics}.

\begin{figure}
\centering
\includegraphics{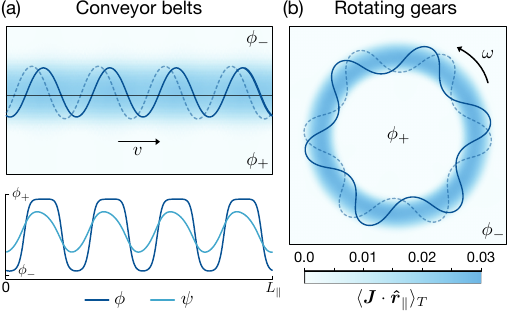}
\caption{\textbf{Capillary waves}. Liquid conveyor belts (a) and rotating gears (b) emerge from capillary waves destabilizing a stationary flat (a) or spherical (b) interface. 
These waves exhibit collective motion (black arrows) characterized by a phase velocity ${v}$ or angular velocity ${\omega}$, driven by a non-zero time-averaged current ${\langle \boldsymbol{J}\cdot\boldsymbol{\hat r}_\parallel\rangle_T}$ [Eq.\@~\eqref{eq:averageCurrent}].
The position of the interface ${\phi=\phi_c}$, separating the high- ($\phi_+$) from the low-density phase ($\phi_-$), at two different times ${t}$ and ${t-\Delta t}$ is shown by solid and dashed blue (gray) lines, respectively.
(a) Density profiles of the fields $\phi$ (dark blue/gray) and $\psi$ (light blue/gray) along the position of the interface at time ${t=0}$ (horizontal black line) reveal a slight offset in their peak positions (bottom panel).
All data were obtained from finite element simulations with no-flux boundary conditions applied perpendicular to the interface and periodic boundary conditions parallel to it in (a), while (b) uses no-flux boundary conditions in all directions.
The chosen parameters are ${\phi_c=0.7}$, ${Pe=0.4}$, ${l=2}$, ${M=21.5}$ and (a) ${\bar \phi=0.7}$, respectively (b) ${\bar \phi=0.64}$. 
The system sizes, in units of the interface width $w$, are (a) $100\times 40$ and (b) $64\times64$.
A corresponding video illustrating the time evolution of gears and conveyor belts is provided in Video\@~S3 of the supplementary material\@~\cite{SM}.}
\label{fig:GearsAndBelts}
\end{figure}

Evolving the system, we observe two distinct outcomes: 
Either the interface remains stable, or it is destabilized by small perturbations that grow over time; see Fig.\@~\ref{fig:categorization} and Video\@~S2\@~\cite{SM}. 
These perturbations manifest in different forms, which can be classified by their growth characteristics. 
Specifically, we distinguish between two types of instabilities: stationary and oscillatory.
In the case of stationary instabilities, perturbations grow predominantly in the vertical (perpendicular) direction, leading to deformations of the interface with negligible lateral motion; see Fig.\@~\ref{fig:categorization}(a). 
The observed behavior is reminiscent of the classical \textit{Mullins-Sekerka} instability\@~\cite{Mullins.1963}, where during solidification small deviations from a flat interface amplify over time, eventually distorting the interface morphology into pronounced, finger-like structures.
In contrast, \textit{oscillatory} instabilities lead to more intricate dynamics. 
In this regime, small perturbations of the interface profile exhibit non-monotonic amplitude growth of individual modes [Fig.\@~\ref{fig:categorization}(b)] or lead to the formation of capillary waves propagating along the interface [Fig.\@~\ref{fig:categorization}(c)]. 
While typically perturbations grow until they reach the system's boundary or destroy the interface structure (cf.\@~Video\@~S2\@~\cite{SM}), the observed capillary waves sometimes saturate at a finite amplitude and cease growing further.
In these cases, they establish a steady pattern of lateral motion, such as traveling waves [Fig.\@~\ref{fig:categorization} and Fig.\@~\ref{fig:GearsAndBelts}(a)] or ``rotating'' domains [Fig.\@~\ref{fig:GearsAndBelts}(b)]; see Video\@~S3\@~\cite{SM}.

To further characterize these dynamical structures, we compute the time-averaged mass flux parallel to the stationary (undeformed) interface\@~\cite{ZN}
\begin{align}
    \langle \boldsymbol{J}\cdot\boldsymbol{\hat r}_\parallel\rangle_T
    = \frac{1}{T} \int_{T_0}^{T_0+T} \mathrm{d}t \,  \boldsymbol{J}\cdot\boldsymbol{\hat r}_\parallel,
    \label{eq:averageCurrent}
\end{align}
where $\boldsymbol{J}$ represents the conserved current of the phase-separating density $\phi$ [cf.\@~Eq.\@~\eqref{eq:DenistyRescaled}] and ${\boldsymbol{\hat r}_\parallel=\{\boldsymbol{\hat e}_x,\boldsymbol{\hat e}_\theta\}}$ is the unit vector parallel to the undeformed interface. 
The times $T_0$ and $T$ are chosen such that the capillary wave pattern has fully developed and has traversed the system several times during the observation period.
Results for a representative parameter set with a flat and spherical interface are displayed in Figs.\@~\ref{fig:GearsAndBelts}(a--b), respectively.
In both cases, one observes a non-vanishing net current transporting mass along the propagation direction of the capillary waves.
This current is sharply localized near the interface and rapidly decays to zero within the high- and low-density phases.
We conclude that capillary waves observed in our model are not merely traveling excitations but instead exhibit finite mass transport confined to the vicinity of the interface.
Because of their directed transport and distinct morphology, we term the emergent dynamical states \textit{conveyor belts}  [Fig.\@~\ref{fig:GearsAndBelts}(a), top panel] and \textit{rotating gears} [Fig.\@~\ref{fig:GearsAndBelts}(b)].
\subsection{Heuristic analysis} 
\label{sec:HeuristicAnalysis}
\begin{figure}
    \centering
    \includegraphics{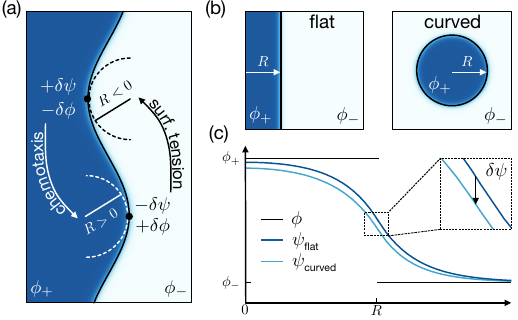}
    \caption{\textbf{Curved interfaces}.
    (a) Weakly curved interface between a high-density (dark blue/gray) and low-density (light blue/gray) domain. Inwards (black circle, ${R<0}$) and outwards bending (white circle, ${R>0}$) areas are described by a local radius of curvature $R$. In the vicinity of the interface, curvature-induced changes in the concentration of the phase-separating ($\delta\phi$) and chemical ($\delta\psi$) densities drive diffusive currents (arrows) in both the high- as well as the low-density phase. Surface tension favors mass distribution from outwards to inwards bending domains, i.e., from positive to negative curvature, whereas chemotactic currents transport mass from negatively to positively curved domains.
    (b) High-density domain of size $R$ confined by an infinite low-density background. 
    (c) Chemical profile along the direction perpendicular to the flat (dark blue/gray) and spherical (light blue/gray) interfaces of panel (b) for a fixed step-like density profile $\phi$ (black). Finite curvature $R^{-1}$ induces a reduction in the chemical concentration ($\delta\psi$) compared to the flat case.}
    \label{fig:intuition}
\end{figure}
The emergence of the observed interface morphologies can be understood through a heuristic argument.
To this end, we consider a flat interface subjected to a small periodic undulation whose wavelength we denote as $2\Lambda$ [cf.\@~Fig.\@~\ref{fig:intuition}(a)]. 
This setup allows us to analyze how local interface curvature induces changes in both the density $\phi$ and the chemical field $\psi$, which drives diffusive fluxes that may either act to flatten the perturbed interface or amplify its shape and thus destabilize it.

In a thermal equilibrium system (${Pe=0}$), the presence of non-vanishing local curvature of the interface induces a \textit{Laplace pressure}, which affects the system's local chemical and mechanical equilibrium\@~\cite{Gennes.2004,Weber.2019}. 
In the vicinity of the interface, this results in a curvature-dependent shift in concentration $\delta \phi$ relative to a flat interface. 
Notably, this shift occurs in both the high- and low-density phases, altering the local balance of diffusive fluxes on either side of the interface.
The dependence of the concentration shift on local curvature is described by the \textit{Gibbs-Thomson} relation\@~\cite{Eyre.1993,Garcke.2000}. 
For a weakly curved interface, it reduces to (App.\@~\ref{app:GibbsThomsonSection}) 
\begin{align}
    \delta \phi
    =
    \frac{1}{2\Delta \phi} 
    \frac{\gamma}{R} \, ,
    \label{eq:GibbsThomson}
\end{align}
where ${\gamma=(\Delta\phi)^2/6}$ denotes the surface tension in dimensionless units (App.\@~\ref{app:SurfaceTension}) and $R$ is the signed local radius of curvature [Fig.\@~\ref{fig:intuition}(a)] with ${R<0}$ (${R>0}$) in inwards (outwards) bending regions.  
Consequently, inwards bending regions (negatively curved) experience a slightly lower concentration of the phase-separating density $\phi$, while regions of outward curvature (positively curved) experience a slightly higher concentration, which we indicate as $\pm\delta\phi$ in Fig.\@~\ref{fig:intuition}(a).
This imbalance in the local concentration of $\phi$ manifests as a corresponding imbalance in the local chemical potential $\mu$, arising from the dependence of $\mu$ on $\phi$ as described by Eq.\@~\eqref{eq:chemicalPotential}. 
Specifically, regions with higher concentrations exhibit a larger $\mu$, while regions with lower concentrations correspond to a smaller\@$\mu$. 
Within both the high- and low-density phases, the emergent gradient in chemical potential, $\boldsymbol{\nabla}\mu$, drives diffusive fluxes ${\boldsymbol{J}_{\mathrm{eq}} = -M \boldsymbol{\nabla}\mu}$, redistributing mass from interface segments with positive curvature (${R>0}$) to those with negative curvature (${R<0}$).
That is, surface tension promotes interface flattening [black arrow in Fig.\@~\ref{fig:intuition}(a)].

For weakly curved interfaces, i.e., when the plateau densities are close to the free energy minima $\phi_\pm$, Eq.\@~\eqref{eq:freeEnergyDensity}, the current can be linearized around these minima yielding ${\boldsymbol{J}_{\mathrm{eq}}=-2M \boldsymbol{\nabla}\phi}$ on both sides of the interface.
To estimate the magnitude of the lateral mass fluxes, we approximate the gradient in the density $\phi$ as $|\boldsymbol{\nabla}\phi| \approx 2|\delta\phi|/\Lambda$, where $\delta\phi$ denotes the curvature-induced shift in concentration. 
Thus, the magnitude of the emergent current on each side of the interface can be approximated as 
\begin{align}
    J_\mathrm{eq}
    = 
    4 M \, 
    \frac{|\delta\phi|}{\Lambda} \,.
\end{align}
Since equilibrium currents act to restore a flat interface, a stationary interface can only be destabilized if out-of-equilibrium chemotactic currents of comparable magnitude redistribute mass in the opposite direction. 
To explore this destabilization process, we analyze how local curvature variations affect the chemical concentration $\psi$ using a similar reasoning as for the density, to estimate the magnitude of the chemotactic currents. 
This allows us to directly compare them to equilibrium currents, yielding a heuristic instability criterion.

As discussed in Sec.\@~\ref{sec:modelEquations}, when the chemical field relaxes sufficiently fast (${M \leq \mathcal{O}(1)}$), memory effects are negligible, and $\psi$ can be considered quasi-static.
In this regime, Eq.\@~\eqref{eq:stationaryConvolution} shows that the local chemical concentration is a weighted spatial average of the phase-separating material in a region of size $l$.
As a result, the chemical concentration near the interface depends on the local interface curvature:
Negatively curved segments, surrounded by more of the high-density phase, exhibit a slightly higher concentration, while positively curved segments, mostly adjacent to the low-density phase, show a lower concentration (cf.\@~Fig.\@~\ref{fig:intuition}). 

To estimate the shifts $\delta\psi$ in chemical concentration, we compare two cases: (i) a high-density domain with radius $R$ in a low-density background [Fig.\@~\ref{fig:intuition}(b), right], and (ii) a flat interface where the high-density region extends over a length $R$ [Fig.\@~\ref{fig:intuition}(b), left].
To isolate curvature effects, we neglect all density modulations and describe the density profile $\phi$ by a step function with plateau values $\phi_\pm$ [Fig.\@~\ref{fig:intuition}(c)].
The stationary chemical profile $\psi$ in both cases is obtained by solving the piecewise profile equation, Eq.\@~\eqref{eq:stationaryProblemChemical}, assuming continuity of the chemical concentration across the interface. 
Detailed calculations are provided in App.\@~\ref{app:InstabilityCriterion}, with a representative solution shown in Fig.\@~\ref{fig:intuition}(c).
As expected from our previous argument, the chemical concentration at the interface ${r=R}$ is found to be smaller than the corresponding value obtained for a flat interface.
To leading order in ${l/R}$, one finds for the shift (App.\@~\ref{app:InstabilityCriterion}) 
\begin{align}
    \delta\psi 
    = 
    - \frac{\Delta \phi}{4}  \, 
    \frac{l}{R}
    \, .
    \label{eq:chemicalShift}
\end{align}

Repeating the same calculation for a spherical low-density domain immersed in a high-density background one obtains the shift in chemical concentration for the inward bending region in Fig.\@~\ref{fig:intuition}(a).
In the above derivation, this amounts to exchanging the high- and low-density phases (${\Delta\phi\rightarrow-\Delta\phi}$), while keeping ${R>0}$ fixed.
That is, the shift in chemical concentration is always \textit{opposite} to shift in the phase-separating density as predicted by the linearized Gibbs-Thomson relation, Eq.\@~\eqref{eq:GibbsThomson}.
As a consequence, the resulting flux ${\boldsymbol{J}_\mathrm{ch}=-M\, Pe\,\phi\boldsymbol{\nabla}\psi}$ redistributes mass from inward bending (negative curvature) to outward bending (positive curvature) regions, amplifying perturbations [white arrow in Fig.\@~\ref{fig:intuition}(a)].
Using that ${|\nabla\psi| \approx  2|\delta\psi|/\Lambda}$, we estimate the magnitude of the lateral chemotactic fluxes on either side of the interface as
\begin{align}
    J^\pm_\mathrm{ch} 
    = 
    M \,Pe \, \phi_\pm\,\frac{2|\delta\psi|}{\Lambda}
    \, .
\end{align}

Building on our previous arguments, we conclude that the stability of the interface is determined by the competition between active and passive fluxes, both arising from curvature-induced shifts in concentrations. 
Since these fluxes act in opposite directions, an interface instability occurs when ${J^+_\mathrm{ch}+J^-_\mathrm{ch}>2J_\mathrm{eq}}$, which yields the instability criterion (App.\@~\ref{app:InstabilityCriterion})
\begin{align}
     l \, Pe \,  \phi_c> \frac{2}{3}.
     \label{eq:HandwavingInstability}
\end{align}
This relation leads to two important conclusions that we will verify later: (i) the instability condition is independent of the relative mobility $M$, and (ii) there is an inverse relationship between the diffusive length scale $l$ and the Péclet number $Pe$.
Specifically, at larger $l$ a flat interface is destabilized already for moderate P\'eclet numbers. 
In dimensionful form, the above condition translates to
\begin{align}
    \frac{3\beta}{\lambda}\sqrt{\frac{D_p}{\lambda w^2}} > \frac{ D_\rho}{K \rho_c}.
\end{align}
When satisfied, small perturbations grow as mass is transported from positively curved to negatively curved regions.
This further increases local curvature, initiating a positive feedback loop that ultimately destabilizes a flat interface.

The arguments presented so far suggest why chemotactic interactions can destabilize a flat interface. 
Moreover, the criterion in Eq.\@~\eqref{eq:HandwavingInstability}, reasonably predicts the onset of stationary instability, as we demonstrate later on.
However, it does not provide insight into the emergence of capillary waves or oscillatory behavior, which arise from a qualitatively different type of instability mechanism.
In particular, understanding the oscillatory regime requires accounting for memory effects.

Nonetheless, the formation of coherent capillary wave patterns can be understood heuristically:
In our numerical simulations of capillary waves traveling along a flat interface, the density and the chemical field exhibit a phase shift, with density peaks offset from chemical peaks [bottom panel of Fig.\@~\ref{fig:GearsAndBelts}(a)].
Due to the chemo-repulsive interaction, this offset drives the migration of the density field $\phi$ away from regions of high chemical concentration $\psi$, while relaxation towards chemical equilibrium (${\psi=\phi}$) aligns peaks in the chemical and density fields.
This nonreciprocal predator-prey-like interaction drives the formation of traveling waves.
Sustaining a phase shift between the two fields naturally requires slow signal propagation (${M\geq\mathcal{O}(10)}$); otherwise, rapid equilibration of the chemical field suppresses the capillary waves.
Thus, their emergence is closely related to the self-propulsion instability of enzymatically active droplets reported in Refs.\@~\cite{Demarchi.2023,Goychuk.2024}.

In summary, our analysis shows that interface perturbations in the model defined by Eq.\@~\eqref{eq:DynamicsRescaled} can emerge via two distinct mechanisms: 
(i) \textit{Stationary}: Repulsive chemotactic interactions destabilize a flat interface by redistributing mass from regions of negative to regions of positive curvature, thereby amplifying small perturbations.
(ii) \textit{Oscillatory}: When additionally signal propagation is slow, interfacial perturbations may exhibit oscillatory behavior, including the formation of finite-amplitude capillary waves driven by a sustained phase offset between the density field $\phi$ and the chemical field\@$\psi$.
While the stationary instability is independent of the relative mobilities $M$, the oscillatory instability critically depends on a large $M$ to sustain a phase shift between the two fields.
From our heuristic consideration, the interplay between these two mechanisms remains unclear, as does the precise onset of capillary wave formation.
In particular, one anticipates that the destabilizing forces present in the stationary regime are equally present in the oscillatory regime.
However, it is not clear how the additional memory effects due to slow signal propagation affect the lateral mass redistribution and, thus, the overall interface stability.
To resolve these open questions, we now turn to a more rigorous analytical approach based on a systematic expansion in the amplitude of interface undulations.
\section{Stability Analysis}
\label{sec:stabilityAnalysis}
\begin{figure}[tb]
    \centering
    \includegraphics[]{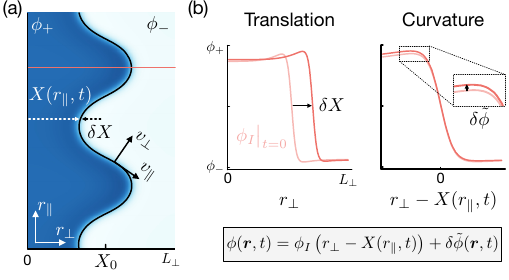}
    \caption{\textbf{Weakly perturbed interfaces}. (a) A weakly perturbed interface (black line) separating a high-density phase $\phi_+$ (dark blue/gray) from the low-density phase $\phi_-$ (light blue/gray) is parametrized in terms of its interface position ${X(r_\parallel,t)=X_0+\delta X(r_\parallel,t)}$. At each point, the interface velocity can be expressed in local coordinates $v_\parallel$ and $v_\perp$, parallel and perpendicular to the interface, respectively. (b) Along the vertical direction $r_\perp$ [red/gray cut-line in (a)], the density profile $\phi(\boldsymbol{r},t)$ (red/gray) consists a translation $\delta X(r_\parallel,t)$ of the unperturbed stationary interface solution $\phi_I(r_\perp)$ (light red/gray) and a curvature induced concentration shift $\delta\tilde\phi(\boldsymbol{r},t)$.}
    \label{fig:calculationFigure}
\end{figure}
In this section, we perform a linear stability analysis of the stationary interface solutions obtained in Sec.\@~\ref{sec:interfaces}, focusing on a flat interface. 
First, we express the interface dynamics in terms of the local currents, using an explicit parametrization.
A more general coordinate-free formulation can be found in Ref.\@~\cite{Würthner.2023}.
Next, we employ a separation ansatz to determine the growth rate of small interface perturbations.
The analysis for a spherical domain proceeds along the same lines, with details provided in App.\@~\ref{app:StabilityAnalysis}.

To parametrize the interface position, we use Cartesian coordinates, ${\boldsymbol{r} = r_\parallel \boldsymbol{\hat r}_\parallel + r_\perp \boldsymbol{\hat r}_\perp}$, where $r_\parallel$ and $r_\perp$ denote the components parallel and perpendicular to the flat unperturbed interface.
This notation emphasizes the generality of our approach, requiring only minimal modifications for curved interfaces.
To analyze interface stability, we consider small transverse perturbations $\delta X(r_\parallel,t)$ of the interface profile ${\boldsymbol{\Phi}_I(r_\perp)=(\phi_I,\psi_I)^T}$. 
As illustrated in Fig.\@~\ref{fig:calculationFigure}(a), the interface position can be parametrized as
\begin{align}
    \boldsymbol{X}(r_\parallel,t)
    =
    r_\parallel \, \boldsymbol{\hat{r}}_\parallel + 
    X(r_\parallel,t) \, \boldsymbol{\hat r}_\perp
    \,,
\label{eq:interfacePositionFlat}
\end{align}
where ${X(r_\parallel,t)=X_0+\delta X(r_\parallel,t)}$, with $X_0$ as the initial position of the flat interface. 
The unit vector $\boldsymbol{\hat r}_\perp$ defines the outward normal of the flat interface, pointing from high- to low-density regions. 
Then, for small deviations, the perturbed interface profile can be expressed as
\begin{align}
    \boldsymbol{\Phi}(\boldsymbol{r},t) 
    =
    \begin{pmatrix}
        \phi_I(r_\perp-\delta X(r_\parallel,t))  \\
        \psi_I(r_\perp-\delta X(r_\parallel,t))
    \end{pmatrix}
    + 
    \begin{pmatrix}
        \delta\tilde\phi(\boldsymbol{r},t) \\
        \delta\tilde \psi(\boldsymbol{r},t)
    \end{pmatrix}
    \,, 
\end{align}
with ${\boldsymbol{\delta\tilde\Phi}=(\delta\tilde\phi,\delta\tilde\psi)^T}$ accounting for shifts in concentrations due to local interface curvature [Fig.\@~\ref{fig:calculationFigure}(b)].
As we have outlined in the discussion of Eqs.\@~\eqref{eq:GibbsThomson} and\@~\eqref{eq:chemicalShift}, to leading order, these shifts are linear in the excess curvature, i.e., of order $\mathcal{O}(\delta X)$.
Therefore, we approximate the general solution as ${\boldsymbol{\Phi}=\boldsymbol{\Phi}_I +\boldsymbol{\delta \Phi}}$, where 
\begin{align}
    \boldsymbol{\delta \Phi}(\boldsymbol{r},t)  = -\partial_\perp \boldsymbol{\Phi}_I(r_\perp) \,\delta X(r_\parallel,t) + \boldsymbol{\delta\tilde\Phi}(\boldsymbol{r},t) \, ,
    \label{eq:PerturbationAnsatz}
\end{align}
denoting differentiation along $\boldsymbol{\hat r}_\perp$ as $\partial_\perp$.
Here the first term corresponds to the translational mode of the stationary interface, and the second term accounts for local curvature as illustrated in Fig.\@~\ref{fig:calculationFigure}(b).
\subsection{Linearized equations of motion}
\label{sec:linearInterfaceDynamics}
As shown in Fig.\@~\ref{fig:calculationFigure}(a), the time evolution of the interface position can always be decomposed as
\begin{align}
    \partial_t \boldsymbol{X}(r_\parallel,t) 
	= v_\parallel \boldsymbol{\hat{e}}_\parallel + v_\perp \boldsymbol{\hat{e}}_\perp \,,
    \label{eq:InterfaceDecomposition}
\end{align}
where ${\boldsymbol{\hat{e}}_\parallel}(r_\parallel,t) $ and ${\boldsymbol{\hat{e}}_\perp}(r_\parallel,t) $ denote the \textit{local} time- and space-dependent unit vectors parallel and perpendicular to the interface.
These are distinct from, and should not be confused with, the time-independent unit vectors $\boldsymbol{\hat r}_\parallel$ and $\boldsymbol{\hat r}_\perp$, which are defined relative to the initial flat interface.
Independently of the coordinate system, the tangential unit vector is given by 
\begin{align}
    \boldsymbol{\hat e}_\parallel = \frac{\partial_\parallel \boldsymbol{X}}{\left|\partial_\parallel \boldsymbol{X}\right|}\, ,
\end{align}
where $\partial_\parallel$ denotes {differentiation along $\boldsymbol{\hat r}_\parallel$.
The explicit expressions for $\boldsymbol{\hat e}_\parallel$ and $\boldsymbol{\hat e}_\perp$ are provided in App.\@~\ref{app:StabilityAnalysis}.
Taking a time derivative of Eq.\@~\eqref{eq:interfacePositionFlat} implies 
\begin{align}
    \partial_t \boldsymbol{X}(r_\parallel,t) = \partial_t\delta X(r_\parallel,t) \, \boldsymbol{\hat r}_\perp \, ,
    \label{eq:secondNormalEvolution}
\end{align}
which is complementary to Eq.\@~\eqref{eq:InterfaceDecomposition}.
Notably, the former relation indicates that the interface position only changes along the normal of the flat interface.
Thus, projecting Eq.\@~\eqref{eq:InterfaceDecomposition} onto $\boldsymbol{\hat r}_\parallel$ one finds
\begin{align}
    v_\parallel= v_\perp \partial_\parallel\delta X \, .
\end{align}
Similarly, projecting Eq.\@~\eqref{eq:InterfaceDecomposition} onto $\boldsymbol{\hat r}_\perp$ and comparing with Eq.\@~\eqref{eq:secondNormalEvolution} yields
\begin{align}
    \partial_t \delta X(r_\parallel,t) =
    v_\perp \sqrt{1+\left(\partial_\parallel\delta X\right)^2} \, .
    \label{eq:timeEvolNormalFrame}
\end{align}
While this relation holds as long as the interface is uniquely parametrized by Eq.\@~\eqref{eq:interfacePositionFlat}, determining an exact expression for $v_\perp$ is challenging, as it requires detailed knowledge of the full interface profile.
To simplify the analysis, we assume a separation of length scales ${l \gg 1}$ so that the interface can be considered infinitely sharp. 
Then, mass conservation of the phase-separating material implies that, for an interface moving at velocity $v_\perp$, the local mass flux across the interface is given by
\begin{align}    
     \Delta_\phi \, v_\perp = \boldsymbol{\hat e}_\perp \cdot \big(\boldsymbol{j}_\mathrm{in}-\boldsymbol{j}_\mathrm{out}\big) \, .
    \label{eq:normalVeloctiy} 
\end{align}
where $\boldsymbol{j}_\text{in/out}$ denote the currents right at the inner and outer side of the interface, respectively.
These currents are determined by evaluating the density current in Eq.\@~\eqref{eq:DenistyRescaled} at ${r_\perp=X^{\pm}}$. 
The quantity ${\Delta_\phi := \phi_\text{in} - \phi_\text{out}}$ on the left-hand side of Eq.\@~\eqref{eq:normalVeloctiy} denotes the density jump between the two phases.
In an infinite system and for an infinitesimal shift $v_\perp \mathrm{d}t$ of the interface, the shape of the density profile is essentially maintained, and mass conservation implies 
\begin{align}
    \Delta_\phi 
    = \phi_I \big\rvert_{r_\perp=-\infty} - \phi_I \big\rvert_{r_\perp= \infty} \, .
    \label{eq:DensityJump}
\end{align}
That is, the density jump $\Delta_\phi$ is given by the non-equilibrium plateau densities on either side of the interface (Fig.\@~\ref{fig:interfaces}), which are obtained from the stationary solution $\phi_I$ (App.\@~\ref{app:1dInterface}).

This contrasts with equilibrium systems, where the Gibbs-Thomson relation symmetrically shifts the densities and the interface contrast reduces to ${\Delta\phi = \phi_+-\phi_-}$\@~\cite{Bray.2002}.
For finite systems, Eq.\@~\eqref{eq:DensityJump} remains a good approximation as long as the diffusive length scale $l$ is significantly smaller than the system's extension $L_\perp$ perpendicular to the interface.
However, if ${l/L_\perp=\mathcal{O}(1)}$, the density profile $\phi_I$ away from the interface cannot fully relax to the stationary plateau values.
Fortunately, as detailed in Sec.\@~\ref{sec:interfaces}, this also implies that the deviation from the equilibrium plateau values within each phase remains relatively small, allowing us to use the equilibrium value ${\Delta_\phi = \phi_+ - \phi_-}$, which becomes exact when ${l\rightarrow\infty}$ or ${Pe\rightarrow 0}$.
Therefore, when considering finite-size systems, and if not stated otherwise, we will make use of the latter choice.

Combining the equation of motion for the interface undulation $\delta X$, Eq.\@~\eqref{eq:timeEvolNormalFrame}, and mass conservation, Eq.\@~\eqref{eq:normalVeloctiy}, the time evolution of the interface position reads
\begin{align}
    \partial_t \delta X(r_\parallel,t) 
    = 
    \sqrt{1+\left(\partial_{\parallel}\delta X\right)^2} \, \, \dfrac{\boldsymbol{\hat e}_\perp \cdot \left (\boldsymbol{j}_\mathrm{in} - \boldsymbol{j}_\mathrm{out}\right)}{\Delta_\phi}
    \, .
    \label{eq:InterfaceEvolution}
\end{align}

In the absence of any perturbations $\delta X$, the interface must be stationary. 
Hence, the difference in the currents is at least of order ${\mathcal{O}(\delta X)}$. 
Thus, Eq.\@~\ref{eq:InterfaceEvolution} simplifies to leading order to
\begin{align}
    \partial_t \delta X(r_\parallel,t) 
    = 
    \dfrac{\boldsymbol{\hat r}_\perp \cdot \left (\boldsymbol{j}_\mathrm{in} - \boldsymbol{j}_\mathrm{out}\right)}{\Delta_\phi} \, ,
    \label{eq:LinearInterfaceEvolution}
\end{align}
where we used that at zeroth order in $\delta X$ the normal directions of the perturbed and the stationary interface coincide, i.e., ${\boldsymbol{\hat e}_\perp=\boldsymbol{\hat r}_\perp}$.
Equation\@~\eqref{eq:LinearInterfaceEvolution} corresponds to the naive time evolution one would write down based solely on mass conservation of the phase-separating species.
The above careful analysis demonstrates that, to leading order in $\delta X$, it is indeed the correct expression. 

To determine the currents $\boldsymbol{j}_\mathrm{in/out}$, we linearize the dynamical equations, Eqs.\@~\eqref{eq:DynamicsRescaled}, inside the high- and low-density domains:
\begin{subequations}
\label{eq:linearizedAll}
\begin{align}
    \partial_t \delta\phi(\boldsymbol{r},t) 
    &= M_\pm \boldsymbol{\nabla}^2 \delta\phi + K_\pm \boldsymbol{\nabla}^2 \delta\psi
    \label{eq:linearizedDensity} \, , \\
    \partial_t \delta\psi(\boldsymbol{r},t) 
    &= \boldsymbol{\nabla}^2 \delta\psi
    + \frac{1}{l^2} \big(\delta\phi
    - \delta\psi\big) \, ,
    \label{eq:linearizedChemical}
\end{align}
\end{subequations}
where ${M_\pm = 2M}$, ${K_\pm = MPe\,\phi_\pm}$, and we have neglected terms of order ${\mathcal{O}\left(\boldsymbol{\nabla}^4\right)}$, which are due to gradient terms in the free energy functional, Eq.\@~\eqref{eq:FreeEnergy}. 
This results in
\begin{align}
    \boldsymbol{j}_\mathrm{in/out} (\boldsymbol{r},t)
    = 
    - M_\pm \boldsymbol{\nabla}\delta\phi(\boldsymbol{r},t) 
    - K_\pm \boldsymbol{\nabla}\delta\psi(\boldsymbol{r},t) \, .
    \label{eq:linearCurrents}
\end{align}   
By substituting these relations into Eq.\@~\eqref{eq:LinearInterfaceEvolution}, one obtains the linearized equation for the interface displacement $\delta X(r_\parallel,t)$.
Along with the linearized dynamics for the fields, Eq.\@~\eqref{eq:linearizedAll}, these form the complete set of equations describing the linearized time evolution of the system.
\subsection{Spectral analysis}
\label{sec:spectralAnalysis}
To perform a spectral analysis of Eqs.\@~\eqref{eq:LinearInterfaceEvolution}--\eqref{eq:linearizedAll} we make the separation ansatz
\begin{align}
    \begin{pmatrix}
        \delta \phi(\boldsymbol{r},t) \\
        \delta \psi(\boldsymbol{r},t) \\
        \delta X(r_\parallel,t)
    \end{pmatrix}
    = 
    \varepsilon_k \exp(\sigma_k t) \, p_k(r_\parallel)
    \begin{pmatrix}
        \delta \phi(r_\perp) \\
        \delta \psi(r_\perp) \\
        1
    \end{pmatrix}\, ,
    \label{eq:LinearAnsatz}
\end{align}
decomposing the interface profile into a set of basis functions ${p_k (r_\parallel)=\cos(k r_\parallel)}$,  where ${k = 2 \pi n / L_\parallel}$, ${n\in \mathbb{Z}\setminus\{0\}}$ and $\varepsilon_k$ denotes the (small) amplitude of the perturbation.
The time evolution of each mode is characterized by the growth rate $\sigma_k$.
Inserting this ansatz, Eq.\@~\eqref{eq:LinearAnsatz}, into the linearized Eqs.\@~\eqref{eq:linearizedAll}, one is left with a set of coupled differential equations
\begin{align}
    0 =
    \begin{pmatrix}
          \sigma_k - M_\pm  \hat L_k & -K_\pm  \hat L_k \\[2mm]
         -l^{-2} &  \sigma_k - \hat L_k + l^{-2}
    \end{pmatrix} 
    \begin{pmatrix}
        \delta \phi(r_\perp) \\[2mm]
        \delta \psi(r_\perp) 
    \end{pmatrix},
    \label{eq:LinearEqPerpDirection}
\end{align}
where ${\hat L_k (\partial_\perp)= -k^2 + \partial_\perp^2}$ denotes a linear differential operator.
A solution to Eq.\@~\eqref{eq:LinearEqPerpDirection} can be readily found in terms of a superposition of eigenfunctions of $\hat L_k$ (see App.\@~\ref{app:StabilityAnalysis}).
The coefficients of this superposition have to be chosen such that the solution ${(\delta\phi,\delta\psi)}$ remains an eigenvector of the matrix in Eq.\@~\eqref{eq:LinearEqPerpDirection} and, at the same time, fulfills a set of proper boundary conditions.
Details of this calculation are provided in App.\@~\ref{app:StabilityAnalysis}, and the choice of proper boundary conditions is discussed in the following subsection.
\subsubsection{Boundary conditions} 
Similar to Sec.\@~\ref{sec:interfaces}, we impose that the field $\psi$ and its perpendicular derivative are continuous at the interface:
\begin{subequations}
\label{eq:BCChemical}
    \begin{align}
    \psi \big\rvert_{r_\perp=X^-} 
    &= \psi \big\rvert_{r_\perp=X^+} 
    \, , \\
    \partial_\perp \psi \big\rvert_{r_\perp=X^-} 
    &= \partial_\perp \psi \big\rvert_{r_\perp=X^+}
    \, . 
\end{align} 
\end{subequations}
As outlined in Sec.\@~\ref{sec:interfaces},  the chemical field $\psi$ is effectively constant on the scale of the interface width $w$, implying that the chemotactic current, which is proportional to $\boldsymbol{\nabla}\psi$ (cf.\@~Eq.\@~\eqref{eq:ChemotaxisCurrent}), is negligible.
Therefore, the interface profile of the density $\phi$ is governed solely by the chemical potential $\mu$ [cf.\@~Eq.\@~\eqref{eq:chemicalPotential}].
As a result, the densities just inside and outside of the interface are the same as for an equilibrium system, which we account for by imposing a local Gibbs-Thomson relation (App.\@~\ref{app:GibbsThomsonSection})
\begin{subequations}
\label{eq:BCDensity}
\begin{align}
    \phi \big\rvert_{r_\perp=R^-} 
    &= \phi_+ + \frac{\gamma}{2\Delta\phi} H[X]
    \, , \\
    \phi \big\rvert_{r_\perp=R^+} 
    &= \phi_- + \frac{\gamma}{2\Delta\phi} H[X]
    \, .
\end{align}
\end{subequations}
The above relation is equivalent to Eq.\@~\eqref{eq:GibbsThomson}, where $H$ denotes the local interface curvature, with its reciprocal corresponding to the radius of curvature $R$ used in Sec.\@~\ref{sec:HeuristicAnalysis}.
The explicit representation of $H$ in terms of the interface position $X$ is provided in App.\@~\ref{app:GibbsThomsonSection}.

Since we consider periodic boundary conditions in the direction parallel to the interface, it remains to specify the boundary conditions in the transverse direction.
Depending on the system size, we distinguish two cases.
For a large system, where the diffusive length scale $l$ is much smaller than the system size $L_\perp$ transverse to the interface (${l\ll L_\perp}$), the density profiles relax exponentially to their non-equilibrium plateau values (Fig.\@~\ref{fig:interfaces}), allowing us to consider a quasi-infinite system.
In this limit, the solution ${(\delta\phi,\delta\psi)}$ to the linearized problem, Eq.\@~\eqref{eq:LinearEqPerpDirection}, needs to remain bounded as ${r_\perp \rightarrow\pm\infty}$.
For finite systems where ${l/L_\perp = \mathcal{O}(1)}$, the density profiles away from the interface cannot fully relax and one has to account for system size effects induced by no-flux boundary conditions
\begin{align}
    \partial_\perp \phi \big\rvert_{r_\perp=(0,L_\perp)} = \partial_\perp \psi \big\rvert_{r_\perp=(0,L_\perp)} = 0
    \, .
    \label{eq:BCSystem}
\end{align}
Together, Eqs.\@~\eqref{eq:BCChemical}--\eqref{eq:BCSystem} form a complete set of boundary conditions, ensuring unique solutions $(\delta\phi,\delta\psi)$ to the linear eigenvalue problem, Eq.\@~\eqref{eq:LinearEqPerpDirection}, which are constructed in App.\@~\ref{app:StabilityAnalysis}.
\subsubsection{Eigenvalue equation}
In the final step, we insert the separation ansatz, Eq.\@~\eqref{eq:LinearAnsatz}, into the evolution equation of the interface position, Eq.\@~\eqref{eq:LinearInterfaceEvolution}:
\begin{align}
    \Delta_\phi \, \sigma_k = 
    &-\left[M_+ \partial_\perp \delta\phi(r_\perp) 
    +K_+ \partial_\perp \delta\psi(r_\perp)\right]_{r_\perp=X_0^-} \notag \\
    &+\left[M_- \partial_\perp \delta\phi(r_\perp) 
    +K_- \partial_\perp \delta\psi(r_\perp)\right]_{r_\perp=X_0^+}.
    \label{eq:ConsistencyCondition}
\end{align}
Since the right-hand side of this equation was determined by solving the eigenvalue problem given by Eq.\@~\eqref{eq:LinearEqPerpDirection}, it implicitly depends on $\sigma_k$. 
Thus, Eq.\@~\eqref{eq:ConsistencyCondition} represents a self-consistency condition for the growth parameter $\sigma_k$.
Solving this condition yields the dispersion relation that characterizes the initial time evolution of a perturbation with mode number $k$.

In the equilibrium limit (${Pe=0}$) and for a flat interface, the self-consistency relation for the growth rates $\sigma_k$, Eq.\@~\eqref{eq:ConsistencyCondition}, can be solved analytically.
For large wavelengths, one recovers the classical capillary wave relation of model B\@~\cite{Bray.2001} (App.\@~\ref{app:modelBDispersion}).
\begin{align}
    \partial_t \delta X(r_\parallel,t ) = \sigma_k \, \delta X (r_\parallel,t ) ,
    \quad \sigma_k = - \frac{2\gamma M}{\Delta\phi^2} \left|k\right|^3,
    \label{eq:modelBDispersion}
\end{align}
i.e., small amplitude perturbations always decay with a characteristic time scale that is proportional to $\left|k\right|^{-3}$.
The fact that all eigenmodes show negative growth rates is in line with our heuristic analysis in Sec.\@~\ref{sec:HeuristicAnalysis}, where we argued that for an equilibrium system, a flat interface should always be stable.

In the presence of chemotactic interactions (${Pe>0}$), an analytic solution to the consistency condition, Eq.\@~\eqref{eq:ConsistencyCondition}, is no longer feasible and we resort to solving it numerically for each value of ${k = 2 \pi n / L_\parallel}$, ${n\in \mathbb{Z}\setminus\{0\}}$\@~\cite{ZN}.
The stability of the interface against small perturbations is inferred from the sign of the real part of $\sigma_k$. 
When ${\mathrm{Re}(\sigma_k) > 0}$ for any $k$, small perturbations tend to amplify and thereby destabilize a flat interface.
The analysis of the different dispersion relations obtained for a flat interface is presented in the next section, Sec.\@~\ref{sec:DispersionFlat}.
The details of the calculations, as well as the corresponding results for curved interfaces, can be found in  Apps.\@~\ref{app:StabilityAnalysis}--\ref{app:CurvedInterfaces}.

In summary, we have considered a linear perturbation theory around the non-homogeneous stationary interface solutions obtained in Sec.\@~\ref{sec:interfaces}.
Our ansatz for the deviations from the stationary profiles, Eq.\@~\eqref{eq:PerturbationAnsatz}, incorporates local translations of the interface as well as curvature-induced perturbations. 
By combining the time evolution of the interface, Eq.\@~\eqref{eq:LinearInterfaceEvolution}, with the linearized mass currents, Eq.\@~\eqref{eq:linearCurrents}, we derived a self-consistency relation, Eq.\@~\eqref{eq:ConsistencyCondition}, which allows us to determine the growth rate of small perturbations, Eq.\@~\eqref{eq:LinearAnsatz}.
While for an equilibrium interface this growth rate is strictly negative [Eq.\@~\eqref{eq:modelBDispersion}], the next section investigates the mechanisms by which chemotactic interactions lead to more complex dispersion relations, resulting in various types of instabilities.
\section{Linear mode analysis and bifurcation diagram}
\label{sec:DispersionFlat}
\begin{figure*}[tb]
    \centering
    \includegraphics[]{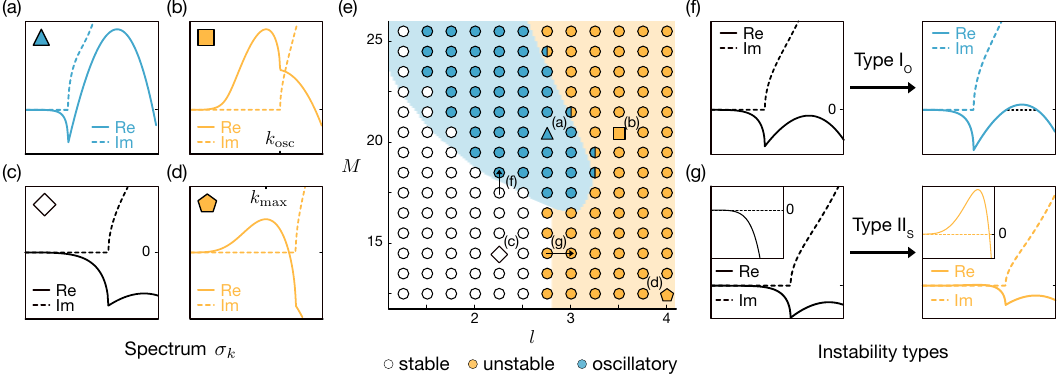}
    \caption{
    \textbf{Dispersion relations and phase diagram}.
    (a--d) Dispersion relations $\sigma_k$ obtained by numerically solving the consistency condition Eq.\@~\eqref{eq:ConsistencyCondition} for parameter combinations $(M,l)$ indicated by the corresponding markers (triangle, square, diamond, pentagon) in the phase diagram (e), for solutions with a density jump across the interface as specified in Eq.\@~\eqref{eq:DensityJump}.
    Each panel shows the maximum real (solid line) and imaginary part (dashed line) of the spectrum, starting from ${\sigma_k=0}$ at ${k=0}$. Panels (b) and (d) further indicate the landmark points $k_\mathrm{osc}$ and $k_\mathrm{max}$, corresponding to the onset of oscillatory behavior and the fastest growing mode, respectively.
    (e) Phase diagram obtained from numerical simulations (bullets) and linear stability analysis (background color/shading) in a two-dimensional system of size ${L_\parallel=150}$ and ${L_\perp=50}$ for a range of values of $M$ and $l$. The other parameters are ${Pe=0.42}$ and  ${\phi_c=\bar \phi=0.7}$. 
    All simulations were initialized with the corresponding stationary profiles obtained from solving Eq.\@~\eqref{eq:stationaryProblem} and evolved using a finite element scheme with periodic boundary conditions in the direction parallel to the interface and no-flux boundary conditions in the perpendicular direction (App.\@~\ref{app:Numerics}). 
    Each interface is categorized as stable (white bullets) or unstable (yellow/gray bullets). Cases where traveling waves [Video\@~S2(c)] or temporally modulated amplitudes [Video\@~S2(b)] dominate the initial dynamics are classified as oscillatory (blue/gray bullets). 
    Mixed-color bullets represent inconclusive classifications.
    The theoretical phase diagram (background color/shading) is constructed by evaluating $\sigma_k$, classifying the interface as stable when Re${(\sigma_k)<0}$ (white), unstable when Re${(\sigma_k)>0}$ and ${k_\mathrm{osc}>k_\mathrm{max}}$ (yellow/light gray) or oscillatory when Re${(\sigma_k)>0}$ and ${k_\mathrm{osc}<k_\mathrm{max}}$ (blue/gray).
    (f)--(g) Illustration of the evolution of the spectrum $\sigma_k$ when transitioning from a stable state to the unstable (g) or oscillatory (f) regime, following the black arrows in the phase diagram (e). The insets of panel (g) zoom in on the spectrum near ${k = 0}$, highlighting detailed behavior in this region.}
    \label{fig:phaseDiagram}
\end{figure*}
In this section, we examine the dispersion relation $\sigma_k$ obtained for a flat interface in the presence of non-vanishing chemotactic interactions (${Pe > 0}$). 
To achieve this, we numerically solve the self-consistency relation, Eq.\@~\eqref{eq:ConsistencyCondition}, across various parameter regimes using a Newton root-finding algorithm\@~\cite{Inc..2024,ZN} (App.\@~\ref{app:StabilityAnalysis}).
Before proceeding with a detailed analysis of the dispersion relation, we begin with a qualitative overview of the different types of behavior that emerge, focusing particularly on how the diffusive length scale $l$, the P\'eclet number $Pe$, and the relative mobility $M$ affect the stability of the interface.

Depending on these parameters, we identify four distinct classes of dispersion relations $\sigma_k$, as illustrated in Fig.\@~\ref{fig:phaseDiagram}(a–d).
All of them share the feature of a vanishing growth rate at ${k=0}$, indicating the presence of a conservation law.
This zero mode corresponds to a homogeneous translation of the interface, which is marginally stable but dynamically inaccessible as it violates mass conservation of the phase-separating density $\phi$.

For small signaling length scales $l$ and weak chemotactic interactions $Pe$, the real part of the spectrum Re${(\sigma_k)}$ is strictly negative [Fig.\@~\ref{fig:phaseDiagram}(c)], indicating that small perturbations are exponentially suppressed and flat interfaces remain stable.
However, for large enough wave numbers\@~$k$, a finite imaginary part Im$(\sigma_k)$ emerges.
This implies that, unlike in the equilibrium case [Eq.\@~\eqref{eq:modelBDispersion}], short wavelength perturbations undergo damped oscillations while the interface approaches a flat state.

As either $l$ or $Pe$ increase, the system transitions into a regime where a finite band of unstable modes with Re${(\sigma_k)>0}$ appears, indicating the onset of an interface instability [Fig.\@~\ref{fig:phaseDiagram}(a--b),(d)].
Beyond a critical wave number $k_\mathrm{osc}$, the spectrum develops a non-vanishing imaginary part, signaling the presence of oscillatory dynamics.
If there is no overlap between the oscillatory and the unstable regime [Fig.\@~\ref{fig:phaseDiagram}(d)], the instability can be classified as non-oscillatory or \textit{stationary} (S) in the nomenclature of Cross and Hohenberg\@~\cite{Cross.1993}.
However, for small diffusive length scales $l$ and sufficiently large relative mobilities $M$, the oscillatory regime (${k>k_\mathrm{osc}}$) may overlap with the band of unstable modes [Figs.\@~\ref{fig:phaseDiagram}(a--b)].
In this case, our further classification of the instability depends on the relative position of the critical wavenumber to the fastest-growing mode 
\begin{align}
    {k_{\mathrm{max}}=\text{argmax}_k\left[\text{Re}(\sigma_k)\right]}\,.
    \label{eq:kmax}
\end{align}

When there is an overlap between the oscillatory regime and the unstable range of wavevectors, but ${k_\mathrm{osc}>k_\mathrm{max}}$ the dominant mode grows in a non-oscillatory fashion [Fig.\@~\ref{fig:phaseDiagram}(b)].
In this case, we also classify the instability as non-oscillatory or stationary. 
In contrast, when ${k_\mathrm{osc}<k_\mathrm{max}}$, the fastest growing mode $k_\mathrm{max}$ exhibits a non-vanishing imaginary part [Fig.\@~\ref{fig:phaseDiagram}(a)].
Hence, we refer to this as an \textit{oscillatory} (O) instability\@~\cite{Cross.1993}. 
Physically, this means that perturbations at the interface not only grow in amplitude but also undergo sustained periodic motion, which can drive propagating capillary waves as well as other dynamic interfacial phenomena.

To summarize, we have identified three distinct classes of dispersion relations, corresponding to stable or unstable interfaces, where the instability can be classified as either stationary or oscillatory.
Next, we provide a more quantitative analysis of the associated bifurcation diagram, highlighting the nature of the bifurcations as well as the typical length scales and velocities of the emerging interfacial perturbations. 
Moreover, to validate our approximations, i.e., the sharp interface limit and the linearization around the equilibrium plateau values, we perform extensive numerical simulations and compare them with our analytical predictions.
Lastly, we compare the insights from the perturbative approach outlined in the previous section~\ref{sec:stabilityAnalysis}, with the heuristic instability criterion, Eq.\@~\eqref{eq:HandwavingInstability}, obtained in Sec.\@~\ref{sec:HeuristicAnalysis}.
\subsection{Bifurcation diagram}
\label{sec:PhaseDiagram}
The transition lines between differently shaded areas in the diagram, Fig.\@~\ref{fig:phaseDiagram}(e), define boundaries between the associated instabilities, with lines separating the stable from the stationary or oscillatory unstable regimes corresponding to bifurcation points.
For the given parameters and ${M \lesssim  16.5}$, one observes that the onset of the instability is independent of $M$ and devoid of any oscillatory behavior, consistent with the heuristic instability condition, Eq.\@~\eqref{eq:HandwavingInstability}.
The transition from the stable to the non-oscillatory/stationary unstable regime proceeds via the emergence of a small band of unstable modes [Fig.\@~\ref{fig:phaseDiagram}(g)] that extends towards ${k=0}$; see the insets of Fig.\@~\ref{fig:phaseDiagram}(g).
That is, in the terminology of Cross and Hohenberg\@~\cite{Cross.1993}, the system undergoes a large scale type-II$_\text{s}$ instability, which is of the same type as the one observed in the Cahn-Hilliard model describing spinodal decomposition\@~\cite{Cahn.1958,Bray.2002}.
At larger mobilities and small diffusive length scales $l$, an oscillatory instability sets in, extending the unstable regime beyond the critical value of ${l\approx 2.8}$ found for ${M\lesssim 16.5}$ [Fig.\@~\ref{fig:phaseDiagram}(e)]. 
Thus, we conclude that slower signal propagation further destabilizes a flat interface.
Other than the transition to the stationary unstable regime, the transition from the stable towards the oscillatory unstable regime happens at a finite wavevector and the band of unstable modes does not extend towards ${k=0}$ [Fig.\@~\ref{fig:phaseDiagram}(f)].
However, as pointed out before, due to mass conservation of the phase-separating density, the zero mode remains marginal---that is, the system undergoes a conserved type-I$_\text{o}$\@~\cite{Cross.1993}, or \textit{conserved wave}\@~\cite{Frohoff-Hülsmann.2023}, bifurcation.

To probe the validity of our approximations, especially the sharp interface approximation, we next compare the bifurcation diagram obtained from our perturbative analysis [Fig.\@~\ref{fig:phaseDiagram}(e)] with numerical simulations.
Following the procedure outlined in Sec.\@~\ref{sec:Gears&Belts}, we initialize the system with the exact flat interface solution to Eq.\@~\eqref{eq:stationaryProblem} and evolve it numerically using a finite element scheme to integrate the dynamical equations, Eqs.\@~\eqref{eq:DynamicsRescaled}.
Details on the simulation times and further specifics are provided in App.\@~\ref{app:Numerics}. 
At the end of each simulation, the interface position—defined as the ${\phi = \phi_c}$ level-set is analyzed for deviations from planarity.
If the deviation exceeds one interface width, the interface is considered unstable.
Moreover, when the onset of the instability exhibits either traveling waves [Fig.\@~\ref{fig:categorization}(c), Video\@~S2(c)] or oscillating undulations [Fig.\@~\ref{fig:categorization}(b), Video\@~S2(b)] of the interface, we classify it as oscillatory.
Importantly, in this approach, we do not introduce any perturbations beyond those naturally arising from the finite element mesh used in the numerical solutions.
To ensure the robustness of our results and eliminate any potential bias from the initial conditions, we also performed simulations in which we perturbed the stationary solution $\phi_I$ by uniform white noise ${\zeta \in [-0.01, 0.01]}$.
Both approaches produced equivalent results.

As illustrated in Fig.\@~\ref{fig:phaseDiagram}(e), the numerical results show good agreement with the theoretical predictions.
Our perturbative approach, though, tends to slightly overestimate the regime of the oscillatory instability.
This discrepancy is likely due to the sharp interface approximation becoming less reliable in the parameter regime where ${l \approx 1}$, consistent with the trends observed for the stationary interface solutions in Fig.\@~\ref{fig:interfaces}(b).

\subsection{Capillary wavelength}
\label{sec:wavelength}
\begin{figure}
    \centering
    \includegraphics[]{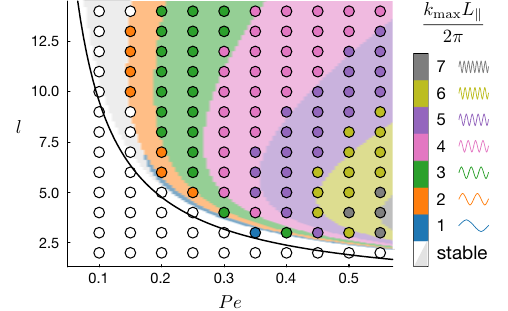}
\caption{
\textbf{Fastest growing linear modes}. 
Mode numbers ${n=k_\mathrm{max}L_\parallel/(2\pi)}$ of the fastest growing mode $k_\mathrm{max}$ as inferred from numerical simulations (bullets) and the analytical spectrum $\sigma_k$ (background colors) in a system of size ${L_\parallel=200}$ and ${L_\perp=50}$ for ${M = 10}$, ${\bar\phi=\phi_c=0.7}$ and various values of $Pe$ and $l$. 
All simulations were initialized with the corresponding stationary profiles obtained from solving Eq.\@~\eqref{eq:stationaryProblem} and perturbed by random white noise ${\zeta\in [-0.01,0.01]}$. The initial state is evolved using a finite element scheme with periodic boundary conditions in the direction parallel to the interface and no-flux boundary conditions in the perpendicular direction. 
The fastest growing mode was inferred by fitting an exponential growth law to the time-dependent power spectrum Eq.\@~\eqref{eq:powerSpectrum} of the ${\phi=\phi_c}$ level-set (App.\@~\ref{app:fastestMode}). 
The color bar on the right indicates the wave number $n$ of the fastest growing mode together with a qualitative representation of its shape (colored lines).
For grayscale rendering, alternating light and dark regions in the mode diagram represent increasing wavenumbers (from left to right).
The heuristic instability criterion, Eq.\@~\eqref{eq:HandwavingInstability}, (solid black line) approximates the unstable regime (light gray area) for a system with infinite extension in the direction perpendicular to the interface.}
\label{fig:modeDiagram}
\end{figure}
Beyond identifying the boundaries of the unstable regimes, the dispersion relation $\sigma_k$ also determines the dominant length scale $\Lambda$ of the emerging initial patterns.
Due to the exponential time evolution of small interface perturbations in Eq.\@~\eqref{eq:LinearAnsatz}, the fastest-growing mode, $k_\mathrm{max}$, dominates over other unstable modes when their initial amplitudes $\varepsilon_k$ are comparable. 
Therefore, for an initial perturbation $\delta X(r_\parallel)$ with a sufficiently uniform spectrum, the typical wavelength of the deformation is given by ${\Lambda = 2\pi/k_\mathrm{max}}$.
As seen in Fig.\@~\ref{fig:modeDiagram}, $k_\mathrm{max}$ increases with the Péclet number $Pe$ but tends to decrease with the diffusive length scale $l$, implying that stronger chemotactic interactions reduce the typical length scale $\Lambda$, while larger interaction ranges tend to increase $\Lambda$.

To validate these analytical predictions, we compare them with numerical simulations\@~\cite{ZN} of the time evolution of flat interface solutions to the profile equations, Eqs.\@~\eqref{eq:stationaryProblem}, using the full dynamics described by Eq.\@~\eqref{eq:DynamicsRescaled}. 
To ensure a sufficiently homogeneous initial spectrum of perturbations, we perturb the initial density profile by uniform random noise ${\zeta\in[-0.01,0.01]}$. 
Throughout the simulation, we track the interface position $X(r_\parallel,t)$, defined as the ${\phi=\phi_c}$ level set, and calculate the power spectrum of the interface undulations
\begin{align}
    S_k(t) 
    = 
    \big| \hat X(k,t) \big|^2 
    \, .
\label{eq:powerSpectrum}
\end{align}
which is defined via the Fourier-transformed interface profile $\hat X(k,t)$.
For each combination of parameters, we average the results over three independent runs, which further minimizes the influence of the initial perturbation. 
Based on our separation ansatz, Eq.\@~\eqref{eq:LinearAnsatz}, the power spectrum of each unstable mode is expected to exhibit exponential growth. 
Therefore, we fit an exponential growth law of the form ${S_k(t) = S_k(0) \exp(\omega_k t)}$, allowing us to infer the empirically fastest-growing mode as ${k_{\mathrm{max}}=\text{argmax}_k\left(\omega_k\right)}$; 
for a detailed explanation see App.\@~\ref{app:fastestMode}.

A comparison between the theoretical prediction based on our perturbative approach and the numerically obtained values is presented in Fig.\@~\ref{fig:modeDiagram}. 
For better illustration, we choose to display the associated mode numbers ${n_\mathrm{max}=k_\mathrm{max}L_\parallel/(2\pi)}$ which only take integer values.
For small wavenumbers ${n\leq4}$ we find almost perfect agreement between the numerically obtained fastest growing modes and our theoretical predictions.
For larger wave numbers ${n\geq5}$, in some instances, the (integer) mode number of the empirically fastest growing mode deviates by plus one from its theoretically predicted value.
We attribute these discrepancies to two factors: 
The colored regions in Fig.\@~\ref{fig:modeDiagram} mark the parameter regime where a given mode is dominant.
At the boundary of two such domains, the growth rates of the respective modes take very similar values.
In this case, due to the limited number of statistical realizations and the finite resolution of the numerical simulation, it can be difficult to accurately pinpoint the fastest-growing mode within the finite observation window.
Additionally, as discussed previously in Sec.\@~\ref{sec:interfaces}, shorter diffusive length scales $l$ and higher P\'eclet numbers $Pe$, reduce the quality of the sharp interface approximation (${l\gg1}$) and lead to larger deviations from the equilibrium plateau values $\phi_\pm$, respectively. 
As a consequence, in this parameter regime, our perturbative approach is expected to yield less accurate results and seems to slightly underestimate the wavenumber of the fastest-growing mode.
Nevertheless, our numerical study consistently supports the overall trend that $n_\mathrm{max}$ increases with $Pe$ and decreases with $l$; cf.\@~Fig.\@~\ref{fig:modeDiagram}.

Beyond this, Fig.\@~\ref{fig:modeDiagram} allows us to evaluate the accuracy of our heuristic instability criterion, Eq.\@~\eqref{eq:HandwavingInstability}. 
As can be seen in the figure, the perturbative approach accurately predicts the onset of the instability, whereas the heuristic criterion overestimates the unstable regime for large diffusive length scales $l$. 
These discrepancies are due to finite system size effects, which were not accounted for in the heuristic analysis. 
To illustrate this, we also compared the heuristic instability criterion, Eq.\@~\eqref{eq:HandwavingInstability}, with the instability threshold obtained from our perturbative approach in the limit of an infinitely extended system perpendicular to the interface (${L_\perp \rightarrow \infty}$). 
In this limit, finite system size effects should not play a role, and, indeed, Fig.\@~\ref{fig:modeDiagram} shows a good agreement between the instability thresholds obtained from our heuristic as well as the perturbative approach (gray area).
We conclude that, for large enough systems (${l\ll L_\perp}$), the heuristic instability criterion, Eq.\@~\eqref{eq:HandwavingInstability}, predicts the onset of stationary interface instabilities reasonably well.
\subsection{Capillary wave velocity}
\label{sec:WaveVelocity}
\begin{figure}
    \centering
    \begin{overpic}[]{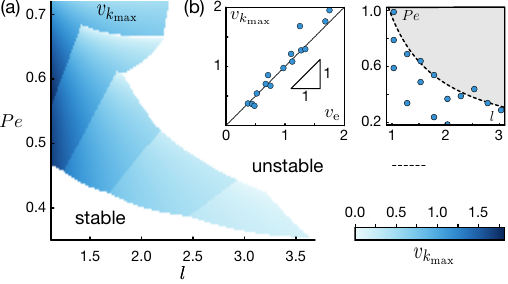}
        \put(86.5,22){Eq.\@~\eqref{eq:HandwavingInstability}} 
    \end{overpic}
    \caption{\textbf{Capillary wave velocities.} (a) Capillary wave velocities $v_{k_\mathrm{max}}$ at ${M=20}$ and ${\phi_c=\bar\phi=0.7}$, as determined from the dispersion relation $\sigma_k$, using Eq.\@~\eqref{eq:ConsistencyCondition}, for various values of $Pe$ and $l$ in a system of size ${L_\parallel=100}$. The system's extension in the direction perpendicular to the interface is considered infinite. A vanishing velocity (white areas) indicates that either the interface remains flat (stable) or is unstable but the fastest-growing mode has a vanishing phase velocity (Im${(\sigma_{k_\mathrm{max}})=0}$). The numerical value of the theoretical phase velocity is encoded by the color bar on the lower right. (b) Comparison of the theoretically predicted ($v_{k_\mathrm{max}}$) and numerically inferred velocities ($v_e$) of (transiently) stable capillary waves in a system of size ${L_\perp=40}$ and ${L_\parallel=100}$ for 15 different combinations of $M$, $Pe$, and $l$ (left panel). A detailed overview of the corresponding parameters is provided in Tab.\@~\ref{tab:VelocityParams}, and their location in the $l$-$Pe$-plane is shown in the right panel. The parameters cluster in a region (white) which is bounded by the heuristic instability criterion, Eq.\@~\eqref{eq:HandwavingInstability}, (dashed black line).}
    \label{fig:velocityDiagramm}
\end{figure}
Within the oscillatory regime, the spectrum $\sigma_k$ also provides insight into the lateral phase velocity of unstable modes, given by\@~\cite{Saarloos.2003}
\begin{align}
    v_k=\frac{\left|\text{Im}(\sigma_k)\right|}{k}.
    \label{eq:Velocity}
\end{align}
If the interface develops a coherently moving capillary wave pattern, as shown in Figs.\@~\ref{fig:categorization}(c) and\@~\ref{fig:GearsAndBelts}(a), its wavelength and velocity are expected to closely match those associated with the fastest-growing mode $k_\mathrm{max}$.
Consequently, the velocity of persistent capillary waves can be approximated by $v_{k_\mathrm{max}}$, whose theoretical values are shown in Fig.\@~\ref{fig:velocityDiagramm} for various values of $Pe$ and $l$ and a semi-infinite system with discrete modes ${k=2\pi n/L_\parallel, n \in \mathbb{Z}\setminus\{0\}}$.
From this analysis, one observes that the theoretical phase velocity of capillary waves is only weakly affected by the P\'eclet number $Pe$ but decreases significantly with the diffusive length scale\@~$l$.

To test our theoretical predictions for capillary wave velocities, we selected fifteen different parameter combinations (cf.\@~Tab.\@~\ref{tab:VelocityParams} in App.\@~\ref{app:phaseVelocities}), each exhibiting a (transiently) stable monochromatic capillary wave pattern.
These combinations were chosen to sample a broad region of the relevant parameter space, with representative values of $Pe$ and $l$ spanning ${Pe \in [0.2, 1]}$ and ${l \in [1, 3]}$ [see Fig.\@~\ref{fig:velocityDiagramm}(b)].
The parameter $M$ was then selected based on the requirement that a stable wave pattern emerge---a necessary condition to reliably infer a phase velocity (App.\@~\ref{app:phaseVelocities}). 
If no stable wave pattern emerged for a given pair of $Pe$ and $l$, we slightly adjusted these parameters and repeated the procedure.
As shown in the right panel of Fig.\@~\ref{fig:velocityDiagramm}(b), the resulting parameter combinations cluster within a region approximately bounded by the heuristic instability criterion, Eq.\@~\eqref{eq:HandwavingInstability}.
This clustering is consistent with expectations: Eq.\@~\eqref{eq:HandwavingInstability} marks the stationary instability threshold (gray area), beyond which one no longer expects the formation of capillary waves.

A comparison of the numerically observed (``empirical'') velocities, $v_\mathrm{e}$, obtained from solving the full dynamics Eq.\@~\eqref{eq:DynamicsRescaled}, with the corresponding theoretical prediction, $v_{k_\mathrm{max}}$, is shown on the left of Fig.\@~\ref{fig:velocityDiagramm}(b). 
The data points cluster around the diagonal ${v_\mathrm{e}=v_{k_\mathrm{max}}}$, with a mean relative deviation of only 12\%. 
Given that we are applying our analytical results to a parameter regime where the sharp interface approximation loses accuracy, this level of agreement is quite satisfactory.
Moreover, the empirical velocities were inferred from (transiently) stable, finite-amplitude wave patterns (Fig.\@~\ref{fig:GearsAndBelts}), whereas our perturbative approach is limited to small-amplitude deformations.
That is, nonlinear effects could influence the wave dynamics and potentially contribute to the observed deviations.
We conclude that if the interface develops a transiently stable and coherent wave pattern, its velocity is well approximated by the theoretical phase velocity of the fastest-growing mode.

Additionally, Fig.\@~\ref{fig:velocityDiagramm}(a) illustrates that, for the respective parameters, the oscillatory regime—characterized by a non-zero phase velocity—is confined to short diffusive length scales of ${l \leq 3.6}$.
This observation holds in general and is likewise seen in Fig.\@~\ref{fig:phaseDiagram}(e).
Intuitively, the coherent motion of capillary waves requires lateral mass redistribution over a length scale that is set by the capillary wavelength $\Lambda$.
In the previous section, Sec.\@~\ref{sec:wavelength}, we showed that this wavelength is directly related to the diffusive length scale\@~$l$.
Consequently, when increasing $l$ and, thus, also $\Lambda$, maintaining a coherently moving wave pattern requires ever faster diffusive transport.
This explains why beyond a critical length scale $l_c$ capillary waves cease to exist [Fig.\@~\ref{fig:phaseDiagram}(e), Fig.\@~\ref{fig:velocityDiagramm}(a)].
Moreover, this dependence rationalizes the observed trend that the phase velocity $v_{k_\mathrm{max}}$ is inversely related to the diffusive length scale $l$ [Fig.\@~\ref{fig:velocityDiagramm}(a)]:
The larger the wavelength of the interface deformation, the longer it takes to redistribute mass from the back toward the front of a perturbation pattern, which decreases the emergent wave velocity.
A similar argument applies when the interface does not develop a coherent capillary wave pattern but shows an oscillatory instability characterized by non-monotonous growth of individual modes [Fig.\@~\ref{fig:categorization}(b)].
In this case, the emergent dynamics still require lateral mass redistribution over a range set by the characteristic length scale of the interface perturbation, which—as before—is directly related to the diffusive length scale $l$ (Sec.\@~\ref{sec:interfaces}).

In summary, we investigated the stability of a flat interface in the presence of chemotactic interactions by numerically solving Eq.\@~\eqref{eq:ConsistencyCondition} to derive the dispersion relation $\sigma_k$. 
Through an analysis of the real and imaginary components of $\sigma_k$, we identified stable as well as oscillatory and non-oscillatory (stationary) unstable regimes. 
This classification enabled us to construct a bifurcation diagram (Fig.\@~\ref{fig:phaseDiagram}), which was validated through extensive numerical simulations. 
A key result is that the formation of capillary waves---or, more generally, the onset of oscillatory instability---requires large relative mobilities $M$ and short diffusive length scales $l$, consistent with the heuristic analysis in Sec.\@~\ref{sec:HeuristicAnalysis}. 
Additionally, our heuristic instability criterion, Eq.\@~\eqref{eq:HandwavingInstability}, was shown to predict the onset of the stationary instability reasonably well. 
Finally, we showed that, within the unstable regimes, the fastest-growing mode $k_\mathrm{max}$ determines both the characteristic wavelength (Fig.\@~\ref{fig:modeDiagram}) and the lateral phase velocity (Fig.\@~\ref{fig:velocityDiagramm}) of the evolving interfacial perturbations.
\section{Long-term stability of capillary waves}\label{sec:StabilityWavelengthSelection}
\label{sec:LongTimeStability}
\begin{figure*}
    \centering
    \includegraphics{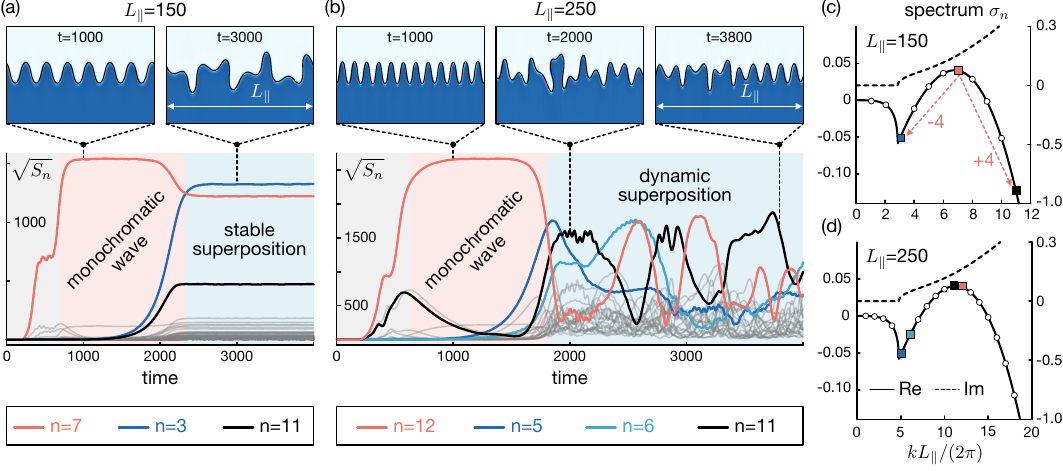}
    \caption{\textbf{Stability of capillary waves.}
    (a--b) Time-resolved power spectrum $S_n(t)$ [cf.\@~Eq.\@~\eqref{eq:powerSpectrum}] of the interface profile ${\phi=\phi_c}$ (black lines in the insets) for different mode numbers ${n=kL_\parallel/(2\pi)}$ obtained from numerical simulations in systems of size ${L_\perp=50}$, ${L_\parallel=150}$ (a) and ${L_\parallel=250}$ (b), respectively. 
    Simulations were initialized with the corresponding stationary profiles obtained from solving Eq.\@~\eqref{eq:stationaryProblem} and evolved using a finite element scheme with periodic boundary conditions in the direction parallel to the interface and no-flux boundary conditions in the perpendicular direction. 
    The insets show snapshots from the simulated interface dynamics at various time points, the full dynamics are shown in Video\@~S4 of the supplementary material\@~\cite{SM}.
    Subpanel (a) displays the time-dependent power spectrum (thin gray lines) for each mode number $n$, whereas subpanel (b) only shows trajectories if $\sqrt{S_n(t)}>200$ at any point in time.
    Trajectories are highlighted by color (red/gray, dark blue/gray, light blue/gray, black) when, at any point, $\sqrt{S_n(t)}>300$ (a) or $\sqrt{S_n(t)}>1500$ (b).
    For better illustration, curves were smoothed using a Gaussian-weighted moving average with a window size of ${\Delta t=60}$ and a standard deviation of ${\sigma=12}$.
    (c--d) Real- (solid line, left axis) and imaginary part (dashed line, right axis) of the spectrum $\sigma_n$ obtained by numerically solving the consistency condition Eq.\@~\eqref{eq:ConsistencyCondition}. The results are shown for an infinite system (black lines), as well as for a system of size ${L_\parallel=150}$ (c) and ${L_\parallel=250}$ (d) with infinite extension in the direction perpendicular to the interface (bullets).
    The growth rates of modes highlighted in subpanels (a,b) are indicated as squares, with the color code corresponding to the one used in subpanels (a,b), respectively. 
    The parameters used for all subpanels are ${M=22.5}$, ${l=1.75}$, ${Pe=0.42}$, and ${\bar\phi=\phi_c=0.7}$.
    }
    \label{fig:superposition}
\end{figure*}
Having analyzed the onset of interface instabilities, we now turn to their long-term behavior.
As outlined in Sec.\@~\ref{sec:Gears&Belts}, interface perturbations typically grow until they reach the system boundary or eventually destroy the interface.
Our simulations show that this behavior consistently occurs for the stationary instability [Fig.\@~\ref{fig:categorization}(a), Video\@~S4\@~\cite{SM}].
However, near the onset of the oscillatory instability [i.e., at the left boundary of the blue region in Fig.\@~\ref{fig:phaseDiagram}(e)], we frequently observe the formation of finite amplitude capillary waves [Fig.\@~\ref{fig:categorization}(c), Video\@~S4\@~\cite{SM}].
The long-term evolution of these capillary waves is governed by a secondary instability:
An initially monochromatic wave with a fixed wavelength either transitions into a stationary superposition of multiple propagating modes [Fig.\@~\ref{fig:superposition}(a)] or evolves into a more complex, quasi-chaotic state [Fig.\@~\ref{fig:superposition}(b)].
In the latter case, multiple modes with different wavelengths continuously appear and disappear, dynamically reshaping the interface profile over time.

What mechanisms stabilize finite-amplitude capillary waves, and what determines the selection of dominant modes in the long-term dynamics?
A comprehensive quantitative treatment of these questions would require a detailed analysis that accounts for nonlinear effects, which are beyond the scope of this manuscript.
In the following, we present a phenomenological description of the behaviors observed in our numerical simulations and offer an intuitive explanation for the underlying mechanism.
\subsection{Long-term wavelength selection}
To assess the long-term behavior of finite amplitude capillary waves, we numerically studied their time evolution over extended time scales, employing a finite element scheme to integrate the dynamical equations, Eq.\@~\eqref{eq:DynamicsRescaled}.  
The simulations were conducted for two different system sizes ${L_\parallel\in\left\{150,250\right\}}$ while keeping all other parameters fixed.
In both cases, the system is initialized with the corresponding stationary interface profiles given by Eq.\@~\eqref{eq:stationaryProblem}.
As before, the position of the interface profile, i.e.\@~the ${\phi=\phi_c}$ level-set, is tracked over time, and its power spectrum $S_n(t)$, Eq.\@~\eqref{eq:powerSpectrum}, is computed for mode numbers ${n=kL_\parallel/(2\pi)}$.
The results, shown in Figs.\@~\ref{fig:superposition}(a--b), indicate that the initially flat interface is unstable, with multiple modes gaining spectral weight [gray shaded areas in Figs.\@~\ref{fig:superposition}(a--b)].
However, these modes are quickly out-competed by a single, fastest-growing mode, which has a comparable wavelength in both systems.
The fastest-growing mode aligns well with the linear spectrum of the flat interface [Figs.\@~\ref{fig:superposition}(c--d)], obtained by numerically solving the consistency condition Eq.\@~\eqref{eq:ConsistencyCondition}:
While for the smaller system (${L_\parallel=150}$), our perturbative theory correctly predicts the fastest growing mode ${n_\mathrm{max}=7}$, for the larger system (${L_\parallel=250}$) we predict ${n_\mathrm{max}=11}$ though simulations show ${n_\mathrm{max}=12}$.
As discussed in Sec.\@~\ref{sec:wavelength}, this small deviation likely arises from either biased initial conditions or a breakdown of the sharp interface approximation used in our analytical approach.

After the initial instability [gray shaded area in Figs.\@~\ref{fig:superposition}(a--b)], perturbations stop growing and the amplitude of the fastest growing mode stabilizes while the spectral weight of the other unstable modes decreases.
The system thus settles into a monochromatic wave with a fixed wavelength [red shaded area in Figs.\@~\ref{fig:superposition}(a--b)].
However, this state is not fully stationary. 
After some time, a secondary instability emerges, leading to the growth of additional modes that reduce the dominance of the initially fastest-growing mode.
For both system sizes, this secondary instability gives rise to modes that lie outside the band of linearly unstable modes predicted for the flat interface [Figs.\@~\ref{fig:superposition}(c--d)], highlighting its non-linear character.

Depending on the system size, one finds two qualitatively different kinds of long-term behavior.
For the smaller system ${L_\parallel = 150}$, the interface dynamics reaches a time-independent configuration in the form of a stable superposition of modes with wavenumbers ${n = (3, 7, 11)}$.
Notably, the newly emerging modes ${n = (3, 11)}$ are not only linearly stable [Fig.\@~\ref{fig:superposition}(c)] but also obey the relation ${n = n_\mathrm{max} \pm \Delta n}$, where ${\Delta n = 4}$.
While this suggests that nonlinear interactions allow the dominant mode (${n_\mathrm{max} = 7}$) to excite otherwise stable modes (${n = 3, 11}$), it remains unclear whether this process explicitly requires the involvement of the ${n = 4}$ mode.

For the larger system, ${L_\parallel = 250}$, the observed dynamics are more complex, and the system does not approach a steady state within the simulation time.
Instead, the interface evolves as a time-dependent superposition of modes, in which spectral weights fluctuate dynamically, while the overall amplitude of the interface profile remains approximately conserved.
Unlike the smaller system (${L_\parallel = 150}$), where only a few modes have significant spectral weight,
Fig.\@~\ref{fig:superposition}(b) shows that in the larger system, multiple linearly stable and unstable modes contribute to the interface dynamics.

Our simulations show that this behavior is generic:
Given sufficient simulation time, fixed-amplitude capillary waves consistently evolve into a superposition of modes.
This complex interface dynamics arises from non-linear interactions of linearly stable and unstable modes. 
Increasing the system size parallel to the interface shifts the system from a stable superposition of modes towards a more dynamic regime, where many modes contribute to the interface profile.
From the perspective of linear mode stability, the system size primarily influences the number of linearly unstable modes.
However, it is not clear if and how the presence of additional unstable modes can fully explain the observed behavior.
In particular, our observations suggest the presence of nontrivial nonlinear effects that warrant further investigation.
In the present work, we limit our analysis to highlighting the existence of the secondary instability and instead focus on identifying the mechanism that stabilizes the finite-amplitude capillary waves. 
\subsection{Stability of capillary waves}
\label{sec:StabilityOfWaves}
In our model, Eqs.\@~\eqref{eq:DynamicsRescaled}--\eqref{eq:FreeEnergy}, only two mechanisms can potentially inhibit the growth of interface perturbations and thereby stabilize finite-amplitude capillary waves: 
(i) Finite system size effects due to interactions with the system boundaries and 
(ii) surface tension.
As we will argue in the following, the no-flux boundary conditions in our simulations effectively impose a weakly repulsive force on the phase-separating density at the system boundaries, which could suppress further vertical growth of perturbations.
Additionally, surface tension always tends to minimize the length (and curvature) of the interface, thereby counteracting the growth of ever-larger perturbations.

To assess, which of these options is the more reasonable one, we recall two important observations:
(i) Finite-amplitude interface perturbations occur only in the oscillatory instability regime, suggesting their formation is tied to features unique to this instability.
(ii) Finite amplitude capillary waves are only found close to the onset of the oscillatory instability, i.e., to the left of the blue regime in Fig.\@~\ref{fig:phaseDiagram}(e).

We first consider whether finite system size effects, arising from boundary interactions, could account for the emergence of finite-amplitude capillary waves. 
In all simulations shown in Figs.\@~\ref{fig:overview}, \ref{fig:GearsAndBelts}, and \ref{fig:phaseDiagram}--\ref{fig:superposition}, periodic boundary conditions are imposed in the parallel direction, and no-flux boundary conditions are applied in the transverse direction.

Analogous to the mirror charge principle\@~\cite{Griffiths.2019}, vanishing normal gradients of $\phi$ and $\psi$ at the ``top'' and ``bottom'' boundaries can be interpreted as arising from a mirrored (virtual) distribution of $\phi$ on the opposite side of the boundary.
This interpretation follows from the fundamental idea that a vanishing derivative locally enforces symmetry, leading to a mirrored field configuration across the boundary.
In electrostatics, a similar argument applies to a grounded conductor, where the mirror charge ensures the correct boundary conditions\@~\cite{Griffiths.2019}.
Due to the repulsive nature of chemotactic interactions, this virtual distribution exerts a repulsive force on the density field\@~$\phi$.
That i,s the horizontal boundaries---those parallel to the interface---effectively repel growing ``fingers'' of the high-density phase, potentially inhibiting further growth in the vertical direction.
However, this interaction has a characteristic screening length $l$, which, in our simulations, is typically an order of magnitude smaller than the system’s transverse extension.
As a result, boundary effects are exponentially suppressed in the regime of capillary wave formation.
Additionally, the proposed mechanism does not discriminate between the stationary and oscillatory instabilities and therefore fails to explain why finite-amplitude perturbations occur only in the oscillatory regime.
Consequently, we can rule out system-size effects as the primary cause of finite-amplitude capillary waves.

This leaves surface tension as the most likely mechanism responsible for stabilizing finite-amplitude growth.
In Sec.\@~\ref{sec:HeuristicAnalysis}, we argued that surface tension always tends to minimize the length and curvature of an interface, thereby opposing the growth of interface perturbations.
This consideration helps rationalize why finite-amplitude perturbations occur only in the oscillatory regime.
Specifically, we demonstrated that the stationary instability arises from a competition between repulsive chemotactic interactions and surface tension.
Both effects exhibit the same leading-order dependence on interface curvature:
Curvature-induced shifts in the density, Eq.\@~\eqref{eq:GibbsThomson}, and chemical concentration, Eq.\@~\eqref{eq:chemicalShift}, scale linearly with the inverse radius of curvature.
Consequently, when repulsive interactions are sufficiently strong to destabilize an initially flat interface, surface tension alone is insufficient to restore stability.

In contrast, Fig.\@~\ref{fig:phaseDiagram}(e) shows that the oscillatory regime extends beyond the regime where we expect an interface instability based on our heuristic criterion, Eq.\@~\eqref{eq:HandwavingInstability}.
This indicates that the oscillatory instability is not solely governed by the competition between chemotactic interactions and surface tension.
In particular, as shown in Sec.\ref{sec:PhaseDiagram} and App.\@~\ref{app:CurvedInterfaces}, slow signal propagation ($M \gg 1$) further destabilizes the interface, potentially facilitating capillary wave formation as follows:}
Initially, the latter effect may be strong enough to destabilize an otherwise stable interface. 
If its influence diminishes as the amplitude (and thus the curvature) of the developing pattern increases, the system could reach a stable configuration where surface tension precisely balances the destabilizing active currents.
This hypothesis is supported by the observation that finite amplitude waves emerge only near the onset of the oscillatory instability, where the interface is only weakly unstable.

In summary, we have demonstrated that, close to the onset of the oscillatory instability, an initially flat interface can settle into a state of finite-amplitude capillary waves.
While these waves initially exhibit a characteristic wavelength determined by the fastest-growing linear mode, a secondary instability gradually introduces additional modes.
For smaller systems, this leads to a stable superposition of modes, whereas larger systems exhibit more complex, time-dependent superpositions involving contributions from several linearly stable and unstable modes.
These findings highlight the importance of nonlinear interactions between different modes, suggesting a rich dynamic behavior that warrants further investigation.
Motivated by this, we addressed what stabilizes finite-amplitude capillary waves against further growth in the vertical direction.
While we ruled out finite system size effects, we identified a plausible scenario in which the competition between slow signal propagation and surface tension gives rise to the formation of finite amplitude capillary waves.
Notably, this requires that the destabilizing effect of slow signal propagation weakens with increasing amplitude of the capillary wave pattern.
However, fully validating this hypothesis, again, requires taking into account (weakly) nonlinear effects, which is beyond the scope of this manuscript.
Therefore, we now conclude our analysis of the formation and long-term stability of capillary waves and shift focus to potential use cases and applications.
\section{Liquid Mechanics}
\label{sec:liquidMechanics}
In the remainder of this study, we explore whether the inherently non-equilibrium nature of our model, Eqs.\@~\eqref{eq:DynamicsRescaled}--\eqref{eq:FreeEnergy}, could enable the conversion of chemical energy consumed at the micro-scale into mechanical work extracted at a meso- or macroscopic scale.
Specifically, we investigate whether and under what conditions actively rotating gears and conveyor belts (Sec.\@~\ref{sec:Gears&Belts}) can facilitate localized and directional transport of the phase-separating species.
\subsection{Conveyor belts}
As discussed in Sec.\@~\ref{sec:Gears&Belts}, capillary waves traveling along the interface of a phase-separated domain are not merely traveling excitations; they also induce a non-vanishing net particle current.
Figure\@~\ref{fig:GearsAndBelts} shows that this current is sharply localized near the interface and rapidly decays to zero within the high- and low-density phases.
As a result, these liquid conveyor belts exhibit highly localized and directional mass transport.

However, as detailed in App.\@~\ref{app:conveyorTransport}, the associated transport velocities remain relatively slow.
This is because the capillary waves described in this work do not behave as rigid, coherently moving objects.
Rather than moving collectively at a uniform velocity, the capillary waves arise from continuous diffusive mass redistribution between different segments of the wave pattern.
This process significantly reduces overall transport efficiency, likely ruling out liquid conveyor belts as a viable mechanism for efficient active mass transport.
\subsection{Boundary induced transport}
\begin{figure}
    \centering
    \includegraphics[]{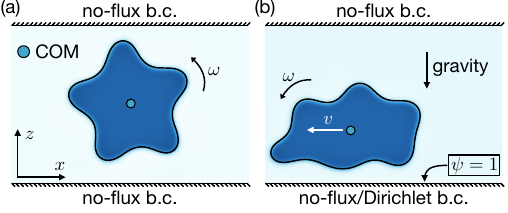}
    \caption{\textbf{Active gear spinning between two parallel walls} with (a) symmetric and (b) asymmetric boundary conditions (b.c.). (a) For the symmetric case, at both walls, we employ no-flux boundary conditions for the phase-separating density $\phi$ and the chemical field $\psi$. (b) In the asymmetric case, a gravity-like interaction, Eq.\@~\eqref{eq:gravity}, breaks the up-down symmetry. To stabilize the system, we additionally employ a Dirichlet boundary condition (${\psi=1}$) for the chemical field at the bottom wall, while all other boundary conditions remain the same as for\@~(a). Due to the broken symmetry, a gear rotating with an angular velocity $\omega$ acquires a finite center of mass (COM, blue/gray dot) velocity $v$.}
    \label{fig:gearsBC}
\end{figure}
Alternatively, one can ask whether the chirality of liquid rotating gears can be exploited to generate forward motion, similar to a driven solid gear rolling on a surface.
This type of motion differs from the self-propulsion of active droplets discussed in Sec.\@~\ref{sec:overview}.
It offers the advantage that the gear can be easily guided to move along a solid obstacle, whereas guiding the motion of active droplets requires carefully engineered chemical gradients.

To explore this possibility, it is essential to account for gear–boundary interactions, since an isolated gear in an unbounded system always resides at the same position; cf.\@~Sec.\@~\ref{sec:overview}.
The presence of a system boundary induces a mismatch between the far-field configuration of an isolated gear and the field values imposed by the boundary conditions.
In the previous section (Sec.\@~\ref{sec:StabilityOfWaves}), we outlined how, in the case of no-flux boundary conditions, this can be interpreted as a repulsive interaction exerted by the wall onto the phase-separating density.
Here, we investigate how this interaction, when combined with the intrinsic chirality of a gear, may generate a torque that results in a propulsive force along the system boundary.

Specifically, we consider a single rotating gear positioned between two identical walls with no-flux boundary conditions, as shown in Fig.\@~\ref{fig:gearsBC}(a).
For clarity, we will refer to this type of boundary condition as ``weakly repulsive,'' to distinguish it from a different set of boundary conditions that will be discussed later.
In this symmetric configuration, the weakly repulsive gear–wall interactions naturally center the rotating gear equidistant between the two walls; see Fig.\@~\ref{fig:gearsBC}(a) and Video\@~S5\@~\cite{SM}.
Because the forces exerted by the walls are equal in magnitude but opposite in direction, the gear is not expected to exhibit net motion.

\begin{figure*}
    \centering
    \includegraphics[]{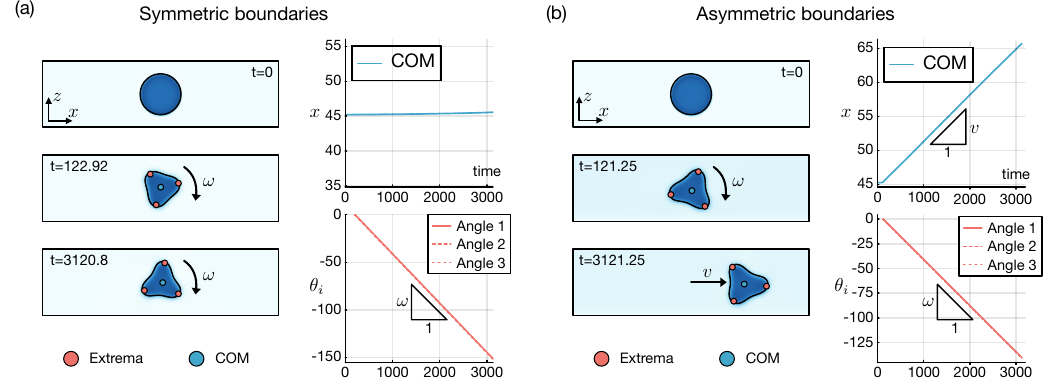}
    \caption{\textbf{Center of mass ($v$) and angular ($\omega$) velocity of an active gear} spinning in between two parallel walls. For symmetric boundary conditions, the gear remains stationary (a), whereas it propels along the $x$-direction when the symmetry of the system is broken by different boundary conditions at the two walls (b); cf.\@~Video\@~S5\@~\cite{SM}. The translational velocity is obtained by tracking the $x$-component of the gear's center of mass (COM, blue/gray dot in the center of the gear) and fitting a linear function. The rotational velocity $\omega$ is determined by tracking the extrema (red/gray dots at the gear's boundary) of the interface profile (black line) and calculating their relative azimuthal angle $\theta_i$, ${i\in\{1,2,3\}}$ (in units of $2\pi$) to the center-of-mass, termed angle 1--3 in the bottom right of subpanels (a) and (b). Averaging over the different angles and fitting a linear function, one obtains the angular velocity $\omega$. All results are obtained from finite element simulations at parameters corresponding to parameter set 7 of Tab.\@~\ref{tab:GearParameters} and initialized with a single spherical droplet.}
    \label{fig:gearVelocity}
\end{figure*}

To break the symmetry between the two walls, we augment our free energy functional, Eq.\@~\eqref{eq:FreeEnergy}, with a weak gravity-like interaction of the form
\begin{align}
    {\cal F}_G[\phi] 
    = g_G
    \int \mathrm{d}^2 r \,
    z \,\phi \, ,
    \label{eq:gravity}
\end{align}
where $z$ denotes the coordinate perpendicular to the walls.
This additional interaction causes the gear’s center of mass (COM) to deviate from the mid-plane; see Fig.\@~\ref{fig:gearsBC}(b) and Video\@~S6\@~\cite{SM}.
Although this setup can, in principle, produce crawling motion under certain conditions, the gravity-like interaction often destabilizes the system, causing the gear to collide with the bottom wall.
Therefore, to enhance the gear's stability, we impose a strongly repulsive interaction at the bottom boundary by setting ${\psi\big\rvert_{z=0} = 1}$ [cf.\@~Fig.\@~\ref{fig:gearsBC}(b)].
This boundary condition fixes the chemical concentration at the bottom wall, effectively making it a source of the chemical species.

While this choice may appear somewhat ad hoc, its sole purpose is to demonstrate that breaking the symmetry of gear-wall interactions can lead to emergent propulsion---a behavior absent in systems with symmetric gear-wall interactions [Fig.\@~\ref{fig:gearsBC}(a)] or in isolated gears [Fig.\@~\ref{fig:overview}(d)].
In fact, removing the second wall already breaks this symmetry.
Therefore, in systems with a large bulk, we expect crawling motion to arise whenever there is an interaction that weakly aligns the active gears with system boundaries.
A promising candidate for such an aligning force is depletion effects, which are commonly observed in active matter systems.

To investigate the emergent behavior under both types of boundary conditions, we performed numerical simulations, initializing the system with a single spherical droplet and evolving it using a finite element scheme with periodic boundary conditions in the direction parallel to the solid walls\@~\cite{ZN}.
Results for a representative set of parameters (parameter set 7 in Tab.\@~\ref{tab:GearParameters}) are shown in Fig.\@~\ref{fig:gearVelocity}, with the corresponding dynamics displayed in Video\@~S5 of the supplementary material\@~\cite{SM}.

In all simulations, the spherical solution is unstable, and the system evolves into a rotating gear state.
For each simulation, we determine the gear's angular velocity by tracking the extrema of its interface profile and calculating their azimuthal angle relative to the gear’s center of mass as a function of time.
The resulting trajectories are shown in the bottom right of Figs.\@~\ref{fig:gearVelocity}(a--b).
In both cases, the gear spins with a constant angular velocity, which depends only weakly on the particular boundary conditions.

To measure the gear's translational velocity, we track the $x$-component of its center of mass.
As expected, for symmetric boundary conditions, the gear remains stationary [Fig.\@~\ref{fig:gearVelocity}(a)], whereas in the asymmetric case, its COM propagates with a constant velocity [Fig.\@~\ref{fig:gearVelocity}(b)].
We conclude that gear–wall interactions can indeed convert the chiral motion of a rotating gear into a net translation of the whole phase-separated domain, while only marginally affecting its angular velocity.

To verify our previous results and gain further insights, we repeated the analysis for ten different parameter combinations [cf.\@~Tab.\@~\ref{tab:GearParameters}], each time extracting the gear's angular and COM velocity under both symmetric and asymmetric boundary conditions.
The parameters were selected similarly to those in Fig.\@~\ref{fig:velocityDiagramm}, ensuring that (i) the gears remain stable throughout the simulation and (ii) they do not crash into the walls.
Moreover, the parameters were chosen to cover a broad range of gear morphologies, i.e., varying numbers of arms.

\begin{figure}
    \centering
    \includegraphics[]{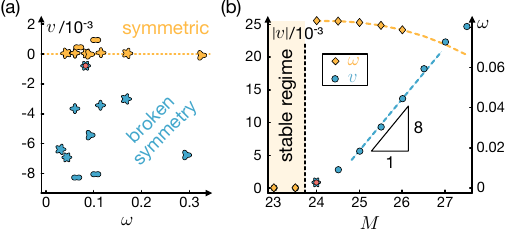}
    \caption{\textbf{Symmetry breaking leads to spontaneously rotating gears crawling along a wall}. (a) Velocity $v$ and angular velocity $\omega$ of active gears with symmetric (yellow/light gray) and non-symmetric (blue/gray) wall interactions for ten different parameter combinations which result in different shapes of the gear, with the number of arms indicated by the marker shapes. The parameters were chosen such that the gear is stable over the whole simulation period and does not collide with the bottom wall. A complete list of parameters is provided in Tab.\@~\ref{tab:GearParameters}. 
    (b) For a single parameter combination, marked by a red dot in Figure\@~(a), the same quantities ($v,\omega$) are shown at varying mobility $M$. In the stable regime (yellow/light gray area), there is no oscillatory instability and, accordingly, no capillary waves and no center of mass propagation. The angular velocity for the three rightmost data points was obtained by extrapolating the previous ones. For all simulations, we keep the product ${M\, g_G=0.002}$ fixed and choose the direction of the $x$-axis such that the gear spins clockwise. A representative time evolution is shown in Video\@~S6 for ${M=26.5}$\@~\cite{SM}.}
    \label{fig:gearData}
\end{figure}

As shown in Fig.\@~\ref{fig:gearData}(a), the results confirm our earlier findings:
Under asymmetric boundary conditions, the gears acquire a non-zero translational velocity $v$, whereas under symmetric boundary conditions they remain stationary (i.e., ${v = 0}$).
The only exceptions are the two dumbbell-shaped gears, which show a small COM velocity even in the symmetric case.
However, as we show in App.\@~\ref{app:translationGears}, this is due to a long-time instability in which the gears eventually deviate from the mid-plane and collide with one of the walls.
This instability does not correspond to a persistent translational motion of the COM and therefore does not contradict our previous conclusions.

Additionally, Fig.\@~\ref{fig:gearData}(a) shows a perfect anti-correlation between the signs of the gear’s translational and angular velocities:
Gears rotating counterclockwise move to the left, while those rotating clockwise move to the right.
Beyond this anti-correlation, however, no clear relationship between angular and translational velocities emerges.
In particular, a larger angular velocity does not imply a higher center of mass velocity.
Likewise, for the parameters considered in Fig.\@~\ref{fig:gearData}(a), no convincing connection between the gear’s velocity and its shape can be drawn.

As one might have expected, the highly nonlinear dynamics and the multitude of parameters make it challenging to disentangle and independently control the gear’s angular and translational velocities.
Nevertheless, we identify the relative mobility $M$ as a key determinant of the emergent center of mass velocity.
As shown in Fig.\@~\ref{fig:gearData}, increasing the mobility beyond the onset of capillary wave formation leads to the emergence of a rotating gear with angular velocity $\omega$.
Close to this onset, the translational velocity remains small.

Further increasing $M$, we observe two effects:
The translational velocity $v$ increases almost linearly, while the angular velocity $\omega$ shows little variation.
The weak dependence of $\omega$ on $M$ can be understood from the fact that the velocity of capillary waves is only marginally influenced by the relative mobility [Sec.\@~\ref{sec:WaveVelocity}].
At the same time, $M$ determines how strongly the gear responds to the torque generated by the gear–wall interactions.
We conclude that varying $M$ allows for tuning the crawling velocity $v$ over approximately one order of magnitude without significantly affecting the rotation velocity $\omega$.
However, when $M$ is increased further, the gear becomes unstable and tends to fragment into smaller domains.

It is important to note that, similar to the mass transport observed at the interface of a conveyor belt, the propagation of gears along a wall is relatively inefficient.
Typically, many full rotations are required for the gear to advance significantly.
However, for the rightmost data point in Fig.\@~\ref{fig:gearData}(b), we find ${2\pi v / \omega \approx 2.33}$—meaning that a single rotation advances the gear by more than two unit lengths, corresponding to approximately $5\%$ of its linear extension.
This result raises the possibility that transport efficiency could be further enhanced through parameter optimization, although such an investigation lies beyond the scope of this manuscript.

In summary, we demonstrated how the emergence of capillary waves can be exploited to extract work from a chemically active system.
As a specific example, we analyzed a single phase-separated gear rotating between two solid walls.
In the absence of additional interactions that break the symmetry between the walls, the gear’s center of mass remains stationary.
By introducing a weak gravity-like interaction and asymmetric boundary conditions, we showed that the gear’s chiral motion can be translated into a constant center of mass velocity, with gears rotating counterclockwise (clockwise) always propagating to the left (right).
Moreover, we found that while it only weakly affects the gear’s angular velocity $\omega$, tuning the relative mobility $M$ can increase the gear’s translational velocity $v$ by at least one order of magnitude.
\section{Discussion}
\label{sec:Discussion}
In this manuscript, we have studied the interplay between short-ranged phase separation and long-ranged repulsive chemical signaling.
As a specific example, we considered chemically active droplets that catalyze a conversion reaction, resulting in chemical gradients that induce chemophoretic flows\@~\cite{Demarchi.2023,Goychuk.2024}.
This process establishes a continuous feedback loop, in which the phase-separated droplets alter the local chemical composition, which in turn influences their migration behavior.
We demonstrated that this feedback mechanism gives rise to an emergent chemotactic interaction, closely related to the classical Keller–Segel model of chemotaxis\@~\cite{Keller.1970, Keller.1971}.
Accordingly our minimal model class, defined by Eqs.\@~\eqref{eq:DynamicsRescaled}--\eqref{eq:FreeEnergy}, should be understood as a generalized Keller–Segel dynamics that incorporates long-ranged chemical interactions alongside short-range forces driving phase separation.

From a broader perspective, our model features a conserved, phase-separating density field that is minimally coupled to self-generated gradients in a second, non-conserved field.
This type of dynamics arises in a variety of systems, including transcriptional condensates\@~\cite{Henninger.2021}, chemotactic motility-induced phase separation\@~\cite{Zhao.2023}, and micro-phase separation of living cells\@~\cite{Carrère.2023}, underscoring the generic nature of our model.
\subsection{Summary}
We have demonstrated that the interplay between phase separation and chemotaxis results in a wide spectrum of non-equilibrium behaviors.
These range from stationary configurations of equally sized droplets to dynamical patterns, including self-propelled stripes and droplets, as well as capillary waves propagating along the interface of phase-separated domains.

When the propagation of chemical signals is significantly faster than the diffusive redistribution of the phase-separating species, the chemical field can be considered quasi-static.
In this limit, chemotactic interactions effectively act as an instantaneous long-range potential\@~\cite{Jäger.1992,Kolk.2023}.
This explains the emergence of stationary microphase-separated states, similar to those observed in (non-)equilibrium phase-separating systems with (effective) long-range repulsive interactions\@~\cite{Liu.1989,Glotzer.1994,Kumar.2023,Yu.2025,Winter.2025,Muratov.2002,Zwicker.2015}.
In contrast, when the chemical field cannot be assumed quasi-static, it introduces a memory effect, giving rise to the dynamical phenomena listed above.

In this study, we focused on the dynamic regime, with particular emphasis on the emergence of interfacial instabilities and capillary waves.
To analyze these phenomena, we employed a heuristic scaling argument along with a more rigorous perturbative approach to assess the stability of both flat and curved interfaces.
Through this analysis, we identified two distinct mechanisms that can destabilize an otherwise stationary interface: stationary and oscillatory instabilities.

The stationary instability, reminiscent of the classical Mullins–Sekerka instability, is characterized by finger-like perturbations that grow predominantly perpendicular to the interface.
These perturbations increase in amplitude until they either reach the system boundary or completely disrupt the interface, often promoting the formation of traveling stripe patterns.
The onset of this instability is reasonably well predicted by a heuristic instability criterion, derived by comparing the magnitudes of surface-tension- and chemotaxis-induced mass fluxes near the interface.
A similar instability has been reported in the context of chemotactic agents propagating along chemoattractant gradients\@~\cite{Alert.2022}, although prior studies have primarily focused on different sensing mechanisms rather than on physical interactions.

In contrast, the \textit{oscillatory} instability leads to more intricate dynamics, including non-monotonic amplitude growth of individual modes and the formation of capillary waves propagating along the interface.
Unlike the stationary instability, the oscillatory one critically depends on the relative mobility of the two fields, requiring the chemical field to propagate more slowly than the density field.

To further characterize both instabilities, we developed a perturbative approach based on small undulations of flat and curved interfaces.
By applying the sharp interface limit, we derived a consistency condition for the growth rate of small interface perturbations.
Solving this equation enabled us to distinguish between stationary and oscillatory instabilities: the real part of the growth rate determines the stability, while a nonzero imaginary part signals the presence of oscillatory modes.
Using this framework, we constructed a bifurcation diagram, which we validated through extensive numerical simulations.
Additionally, we showed that the fastest-growing perturbation mode provides a reliable estimate for the characteristic wavelength and phase velocity of the resulting interface modulations.

Having analyzed the onset of interfacial instabilities, we next examined their long-term behavior, with special emphasis on the oscillatory regime.
Notably, close to the onset of the oscillatory instability, we frequently observed the formation of capillary wave patterns with a well-defined wavelength and finite amplitude.
For initially flat or constantly curved interfaces, we refer to the resulting structures as conveyor belts'' and actively rotating gears'', respectively.
This behavior contrasts sharply with that observed in the stationary regime, where perturbations grow without bounds.

Surprisingly we found that these monochromatic finite-amplitude waves are only transiently stable and eventually evolve into more complex patterns characterized by the superposition of different modes.
Through numerical simulations, we identified two generic long-term scenarios:
Either the system settles into a superposition of capillary waves with constant spectral weights, or it exhibits persistent dynamics marked by continual mode growth and decay.
In both cases, linearly stable modes contribute significantly to the final pattern, underscoring the nonlinear nature of the secondary instability.
Understanding the mechanisms that stabilize finite-amplitude capillary waves---as well as the dynamics underlying the secondary instability and associated mode selection---remains an open challenge and poses interesting avenues for future research.
Addressing these questions will require the incorporation of nonlinear effects, for example through weakly nonlinear analysis\@~\cite{Cross.1993,Hoyle.2006,Greve.2024}.

A set of static and dynamic patterns similar to those reported here has also been observed in the context of chemotactic motility-induced phase separation (MIPS)\@~\cite{Zhao.2023}.
In that study, the authors considered active Brownian particles (ABPs) undergoing MIPS while interacting chemotactically.
Specifically, they explored a system in which ABPs consume a chemical species that is continuously replenished.
Their results included arrested labyrinth and droplet patterns, as well as self-propelled stripes and droplets—features also observed in our model—along with dynamical spirals, which appear to be absent in our case.

Although the model in Ref.\@~\cite{Zhao.2023} includes a nonlinear reaction term and represents MIPS via an asymmetric non-equilibrium potential, it shares key structural similarities with our approach.
In particular, both models involve the same number of conservation laws and a comparable coupling mechanism between the chemical and phase-separating components.
Naturally, this raises the question of whether chemotactic ABPs also exhibit capillary wave formation and related interfacial phenomena—features that have not yet been reported.
Addressing this question, potentially building on Ref.\@~\cite{Zhao.2023}, could open an intriguing direction for future research.
\subsection{Liquid nanomachines}
The liquid nature of active phase-separated assemblies, such as the gears and conveyor belts studied here, offers intriguing opportunities for regulation and control.
Their self-assembly proceeds via spontaneous nucleation and growth, which can be modulated by external cues such as catalytic sites or chemical gradients\@~\cite{Zwicker.2014,Weber.2019}, allowing for the positioning of liquid domains at predefined locations.

In addition, when non-equilibrium currents counteract the equilibrium coarsening process, arrested coarsening can serve as a self-regulated size-selection mechanism.
Moreover, the formation of phase-separated domains relies only on short-range interactions, whereas the emergence of capillary waves or self-propelled domains requires out-of-equilibrium processes—such as the chemotactic interactions considered in this work.
These chemotactic processes, in turn, require the consumption of chemical energy, typically provided in the form of a chemical fuel such as ATP\@~\cite{Zwicker.2022}.
This opens the possibility of designing protocols in which the system behavior is regulated by the availability of fuel.
For example, in chemically fueled droplets, tuning the concentration of the fuel reservoir has been shown to directly control droplet growth and shrinkage\@~\cite{Donau.2023}.

The ability to actively control the positioning, self-assembly, and behavior of active phase-separated assemblies raises the intriguing question of whether they could serve as a platform for designing nanomachines that extract work by consuming chemical energy at the microscale. 
As a proof of principle, protein patterns have recently been shown to facilitate the directional transport of unrelated cargo on membranes through a diffusiophoretic mechanism\@~\cite{Ramm.2021}.

As a specific example, we explored whether capillary waves could facilitate molecular transport, either along the interface (as in a conveyor belt) or through phase-separated domains crawling along a solid object (like a rotating gear).
While we demonstrated the principle possibility of highly localized transport, the associated currents and transport velocities are relatively small, making the process inefficient in practice.

From a broader perspective, the absence of a shear modulus in these liquid-like systems acts as a limiting factor for the efficiency and stability of mechanical transport processes, since the fluid-like nature of the assemblies makes it difficult to resist deformations and sustain directed forces over time.
\subsection{Coacervates and droplet division}
\begin{figure}[tb]
    \centering
    \includegraphics{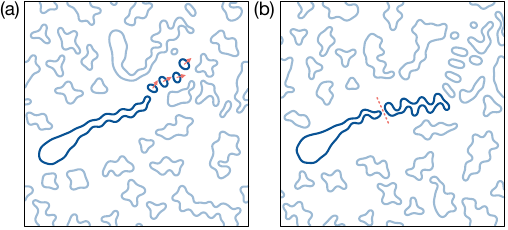}
    \caption{\textbf{Droplet division.} Snapshots from numerical simulations of a two-dimensional system showing ejection events (a) and the scission (b) of stationary droplets caused by capillary waves. Interfaces are determined by the ${\phi=\bar\phi}$ level-sets (blue/gray lines). Both snapshots are from the same finite element simulation with parameters chosen as ${\phi_c=0.7}$, ${Pe=0.81}$, ${l=1}$, ${M=12}$ and ${\bar \phi=0.59}$, in a quadratic domain with linear extension ${L=200}$ and periodic boundary conditions.}
    \label{fig:dropletBreakage}
\end{figure}
The previous discussion was centered around the nature and emergence of capillary waves, and explored their potential role in facilitating localized molecular transport.
Another important concept discussed in the context of active condensates is their potential role in the origin of life.
Whereas the idea of coacervates as precursors of modern cells was popularized long ago\@~\cite{Haldane.1929,Oparin.1952,Slootbeek.2022}, recent studies have provided new insights.
It has been demonstrated that both the active turnover of phase-separating densities\@~\cite{Zwicker.2017} and enzymatic activity\@~\cite{Demarchi.2023} can drive droplet division in three dimensions.

In particular, the instability driving droplet division in Ref.\@~\cite{Zwicker.2017} was associated with a Mullins–Sekerka instability\@~\cite{Mullins.1963}, while Ref.\@~\cite{Demarchi.2023} emphasized its resemblance to the classical pearling instability\@~\cite{Rayleigh.1878}.
In both cases, the observed instability is stationary, meaning it lacks oscillatory behavior.
Since our model represents a limiting case of the one discussed in Ref.\@~\cite{Demarchi.2023}, the droplet division observed there can be considered the three-dimensional analog of the stationary instability of two-dimensional interfaces reported here.

For the stationary droplet instabilities reported in Refs.\@~\cite{Demarchi.2023,Zwicker.2017}, no instances of droplet division have been observed in two spatial dimensions.
It was argued that, in three dimensions, surface tension strongly disfavors the formation of thin necks connecting two spherical droplets and thereby promotes division.
In contrast, in two dimensions, where the curvature of a narrow ``bridge'' vanishes, a droplet typically deforms into a dumbbell-like shape without fully dividing\@~\cite{Bauermann.2022r1c}.

In light of this, a key question arising in the context of our work is whether droplet division can occur in two dimensions if the initial instability is oscillatory rather than stationary.
Indeed, we frequently observe (stationary) large phase-separated domains ejecting smaller droplets [Fig.\@~\ref{fig:dropletBreakage}(a)] or breaking apart [Fig.\@~\ref{fig:dropletBreakage}(b)], even in two dimensions; see also Video\@~S7\@~\cite{SM}.
This phenomenon appears to result from the ``interference'' of capillary waves emerging from different regions of the interface. 
When local oscillations on opposite sides of a domain lead to the formation of a thin neck [Fig.\@~\ref{fig:dropletBreakage}(a)], a self-propulsion instability of the resulting small domain can trigger actual breakage (cf.\@~Video\@~S7\@~\cite{SM}). 
This process results in the repeated ejection of smaller domains, representing a novel mechanism of droplet division in which smaller, self-propelled droplets emerge from a larger ``mother droplet''.

We speculate that such a mechanism could have interesting implications for the exchange of material and information between disjoint coacervates or protocell populations\@~\cite{Cremer.2012,Laurent.2019}, and suggest that this process deserves further investigation.
\subsection{Conservation laws in non-equilibrium field theories}
A particular class of active systems that have recently gained significant attention are systems with so-called ``non-reciprocal'' interactions\@~\cite{You.2020,Fruchart.2021}.
This line of research is informed by and builds upon, the broad body of knowledge developed in the field of population dynamics in spatially extended systems\@~\cite{Frey.2010}, characterized by interactions such as predation\@~\cite{Täuber.2012}, mutualism\@~\cite{Lavrentovich.2014}, parasitism, cyclic competition\@~\cite{Reichenbach.2007,Reichenbach.2008}, and competition for resources.
With the exception of fully symmetric mutualism, all of these interactions are non-reciprocal and play a crucial role in shaping the spatial distribution of species and community structure.
Similarly, non-reciprocal interactions drive phenomena such as the collective motion of chemophoretic colloids\@~\cite{Agudo-Canalejo.2019,Saha.2019}, or give rise to exotic material properties such as those observed in rotating starfish embryo crystals\@~\cite{Tan.2022} and robotic metamaterials\@~\cite{Brandenbourger.2019}.

In our model, non-reciprocity emerges because the phase-separating density actively migrates away from regions of high chemical concentration.
Meanwhile, the system’s tendency toward chemical equilibrium causes higher chemical concentrations to accumulate in areas of increased density, creating a dynamic feedback loop similar to the classical predator-prey paradigm.
While in many instances non-reciprocity gives rise to dynamical patterns\@~\cite{You.2020,Saha.2020,Frohoff-Hülsmann.2021,Schueler.2014}, it does not necessarily lead to complex or emergent behavior.
For example, in the classical Keller–Segel model with repulsive interactions\@~\cite{Keller.1970,Keller.1971}, the above feedback mechanism leads to a stable homogeneous state unless additional nonlinearities are introduced\@~\cite{Liebchen.2018}.

For systems involving two scalar fields---for example, a population of agents interacting through a chemical field---the requirement of non-reciprocity and nonlinear time evolution can be fulfilled by coupling a field $\phi$ to a second field $\psi$ whose time evolution is completely linear.
The resulting minimal model can be organized into four main categories, depending on the number of conservation laws present:
\begin{subequations}
    \label{eq:GeneralModel}
    \begin{align}
        \partial_t \phi(\boldsymbol{r},t) &= \boldsymbol{\nabla}^{a}\left(M\boldsymbol{\nabla}^{a}\mu[\phi]+G\boldsymbol{\nabla}^{a}\psi\right) \label{eq:NRCH1}, \\
    \partial_t \psi(\boldsymbol{r},t) &= 
    \boldsymbol{\nabla}^2\psi + H\boldsymbol{\nabla}^{2b}\phi +\frac{c}{l^2}(\phi-\psi). \label{eq:NRCH2}
    \end{align}
\end{subequations}
Here, $G$ comprises a minimal (linear) coupling, and $H$ corresponds to a cross-diffusion term, which we have thus far not considered.
Just as in our model, Eq.\@~\eqref{eq:DynamicsRescaled}, nonlinear effects are encoded in an effective chemical potential ${\mu = \delta \mathcal{F} / \delta \phi}$, which is constructed to ensure a finite pattern amplitude.
The parameters ${{a,b,c} \in {0,1}}$ allow us to interpolate between theories with different numbers of conservation laws.

When ${a = b = 1}$ and ${c = 0}$, both fields are conserved, and one recovers the model discussed in Ref.\@~\cite{Brauns.2024}, which is a simplified version of the non-reciprocal Cahn–Hilliard model\@~\cite{Saha.2020,Frohoff-Hülsmann.2021}.
Non-reciprocity in this case manifests via mass-conserving predator–prey-like currents when ${G \neq H}$.
Conversely, when ${a = 0}$ and ${c = 1}$, there is no conservation law.
If both kinetic coefficients $M$ and $G$ are negative and there is no cross-diffusion ${(H = 0)}$, the equations reduce to the spatially extended \textit{FitzHugh–Nagumo} (FHN) model\@~\cite{FitzHugh.1961,Nagumo.1962}.
Originally developed as a model for neuronal excitability\@~\cite{Izhikevich.2006}, the FHN model has since become a widely studied paradigm in various fields, including chemical pattern formation\@~\cite{Zheng.2015,Shoji.2007} and wave propagation\@~\cite{Rocşoreanu.2000,Bär.2003,Mikhailov.2006,Ziepke.2016,Gelens.2014}.
For the FHN model, non-reciprocity follows the classical activator-inhibitor scheme of Turing patterns\@~\cite{Turing.1952}:
The field $\phi$ is actively degraded in regions of high $\psi$ concentrations, whereas $\psi$ is produced where there is a lot of $\phi$.
Our model Eq.\@~\eqref{eq:DynamicsRescaled}, is recovered for ${a=c=1}$ and ${H=0}$.
It exhibits only a single conservation law and thus bridges the gap between the case of a fully conserved system and a system without any conservation law.
Additionally, there is a second class of models with a single conservation law when ${a=c=0}$ and ${b=1}$.
It corresponds to a non-conserved field $\phi$ with model A\@~\cite{Hohenberg.1977} like phase ordering kinetics that is degraded by a second conserved field $\psi$.

Despite differences in the number of conservation laws, different realizations of the model class defined by Eqs.\@~\eqref{eq:GeneralModel} show a very similar phenomenology, including the formation of stationary stripe and droplet patterns, self-propelled domains\@~\cite{Brauns.2024,Cebrián-Lacasa.2024} and interface instabilities\@~\cite{Ohta.1989,Petrich.1994,Goldstein.1996,Brauns.2024}.
Notably, capillary wave formation has been reported for the fully conserved case\@~\cite{Brauns.2024} whereas for the non-conserved FHN model the interface instability seems to be stationary, except for a ``breathing mode'', which does not break the interface's symmetry\@~\cite{Ohta.1989}.
Exploring in detail the relevance of the coupling scheme compared to the amount of conservation laws could be a promising avenue for future research.

\begin{acknowledgments} 
We thank Henrik Weyer and Fridtjof Brauns for stimulating discussions, and Richard Swiderski and Tobias Roth for their insightful comments on our manuscript.  
This research was supported in part by the National Science Foundation under Grant No.\@~PHY-2309135, through the Kavli Institute for Theoretical Physics (KITP).  
We further acknowledge financial support from the Deutsche Forschungsgemeinschaft (DFG, German Research Foundation) through the Excellence Cluster ORIGINS under Germany’s Excellence Strategy (EXC-2094-390783311).  
EF acknowledges additional support from the John Templeton Foundation and the Chan--Zuckerberg Initiative\@~(CZI).
\end{acknowledgments}


\cleardoublepage
\appendix
\section{Surface tension}
\label{app:SurfaceTension}
At an interface between two coexisting phases, molecular interactions are less favorable than in the bulk.
Specifically, the reduced number of alike interaction partners induces an energy penalty that generally depends on the local curvature of the interface. 
This effect gives rise to the notion of surface tension, which describes the energetic cost of an interface per surface area element.

From a continuum perspective, when the radius of the interface curvature exceeds the \textit{Tolman} length \cite{Tolman.1949,Blokhuis.2006}, the effective surface tension of a binary mixture is well approximated by that of a flat interface.
For completeness and convenience of the reader, we briefly outline the standard derivation \cite{Bray.2002,Weber.2019} of the surface tension in dimensionless form.
For a Ginzburg-Landau free energy functional of the form 
\begin{align}
    \mathcal{F}[\rho] 
    = \int \mathrm{d}^d r \, 
    \bigg[ 
    \frac{\kappa}{2}\left(\boldsymbol{\nabla}\rho\right)^2 + f(\rho)
    \bigg] \, ,
    \label{app:FreeEnergy}
\end{align}
with the (symmetric) local free energy 
\begin{equation}
    f(\rho) = - \frac{r}{2}\left(\rho-\rho_c\right)^2 + \frac{u}{4}\left(\rho-\rho_c \right)^4 \, ,
    \label{app:FreeEnergyLocal}
\end{equation}
the stationary profile of a flat interface at thermal equilibrium, positioned at ${z=0}$, is given by
\begin{align}
    \rho_I(z) 
    = 
    \rho_c + \sqrt{\frac{r}{u}} \,  \tanh\left( \frac{z}{w}\right)
    \, ,
    \label{app:1dInterface}
\end{align}
where $z$ denotes the direction perpendicular to the flat interface, and ${w=\sqrt{2 \kappa/r}}$ is the width of the interface.
Far from the interface at ${z=0}$, the profile becomes flat, approaching $\rho_\pm = \rho_c \pm \sqrt{r/u}$ as ${z \to \pm \infty}$. 
In these plateau regions, the free energy of the field configuration is dominated by the contributions from the local free energy\@~\eqref{app:FreeEnergyLocal}.
In contrast, directly at the interface the major contribution is due to the gradient term in Eq.\@~\eqref{app:FreeEnergy}.
To calculate the overall energy cost for the interface, we consider the total free energy for the interface profile $\rho_I (z)$ and subtract from it the plateau contributions.
Defining surface tension as the cost per surface area element, one has 
\begin{align}
    \gamma &=
    \int_{-\infty}^\infty \mathrm{d} z \, \bigg[ 
    \frac{\kappa}{2}
    \big( \partial_z \rho_I (z) \big)^2 + f(\rho_I(z))  \bigg] \notag \\
    &- \int_{-\infty}^0\mathrm{d} z \, 
    f(\rho_-)
    - \int_{0}^\infty\mathrm{d} z\, 
    f(\rho_+)
    = \frac{2}{3} \frac{r^2w}{u} \, .
    \label{app:definitionSurfaceTension}
\end{align}
In our non-dimensionalized theory, cf.\@~Sec.\@~\ref{sec:rescaling}, the basic energy scale is set by the typical interaction strength in the high-density phase ${r\rho_+^2w^d}$, whereas distances are measured in units of the equilibrium interface width $w$.
Further using that $r/u=(\rho_+-\rho_c)^2$, the dimensionless surface tension reads
\begin{equation}
    \gamma = \frac{2}{3} \frac{(\rho_+-\rho_c)^2}{\rho_+^2}=\frac16 \, \Delta \phi^2 \, ,
\label{app:SurfaceTensionRescaled}
\end{equation}
where we reused the symbol $\gamma$ and, in the second identity, employed the relation ${\Delta\phi=2(\rho_+-\rho_c)/\rho_+}$ [see Eq.\@~\eqref{eq:DeltaPhi} and the comment below it].
This shows that the dimensionless form of the surface tension is entirely determined by $\Delta \phi$, that is, the density contrast in the phase-separated state.

\section{Gibbs-Thomson relation}
\label{app:GibbsThomsonSection}
Throughout this manuscript, we frequently use a linearized form of the Gibbs-Thomson relation \cite{Eyre.1993,Garcke.2000} to specify boundary conditions for the density field $\phi$ at sharp interfaces. 
The Gibbs-Thomson relation quantifies how the equilibrium chemical potential at a curved interface, such as a droplet, differs from that at a flat interface, with higher curvature leading to increased chemical potential.
Given the central importance of this concept to our theoretical framework, we include a concise summary of its key principles and derivation. To this end, we closely follow the approach outlined in Ref.\@~\cite{Bray.2002}.

Since it is the main focus of our study, we consider a two-dimensional system (${d=2}$) and a setup as described in Sec.\@~\ref{sec:stabilityAnalysis}, where a single interface separates a high-density phase from a low-density phase. 
Our starting point is the non-dimensionalized chemical potential, given by
\begin{align}
    \mu = - \dfrac{1}{2}\boldsymbol{\nabla}^2\phi + f^\prime(\phi) \, ,
\label{app:rescaledChemicalPotential}
\end{align}
where the local free energy density $f(\phi)$ is defined in Eq.\@~\eqref{eq:FreeEnergy}.
At each point along the interface, we define a local coordinate system consisting of two directions: one tangential and one normal to the interface.
The corresponding unit vectors are denoted as $\boldsymbol{\hat e}_\parallel$ and $\boldsymbol{\hat e}_\perp$, with $\boldsymbol{\hat e}_\perp$ pointing from the high- to the low-density phase.
Then, close to the interface, the Laplacian of the density field can be expressed as \cite{Bray.2002}
\begin{align}
    \boldsymbol{\nabla}^2 \phi = \partial_\perp^2 \phi - (\boldsymbol{\nabla}\boldsymbol{\hat e}_\perp) \partial_\perp \phi \, ,
\end{align}
where ${\boldsymbol{\nabla}\boldsymbol{\hat e}_\perp= H}$ denotes the local mean curvature $H$ of the interface.
This expression introduces the interplay between the variation of the density $\phi$ and the interface curvature.
To relate it to the chemical potential, we multiply Eq.\@~\eqref{app:rescaledChemicalPotential} by $\partial_\perp \phi$ and integrate along the normal direction of the interface, which yields
\begin{align}
    \int_{-\varepsilon}^\varepsilon \mathrm{d}r_\perp \,  \mu \,  \partial_\perp \phi
    = &- \frac{1}{4}\int_{-\varepsilon}^\varepsilon \mathrm{d}r_\perp \, \partial_\perp\left( \partial_\perp\phi\right)^2 \notag \\
    &+ \frac{1}{2} \int_{-\varepsilon}^\varepsilon \mathrm{d}r_\perp \,  \boldsymbol{\nabla}\boldsymbol{\hat e_\perp} \left(\partial_\perp \phi\right)^2 \notag \\
    &+  \int_{-\varepsilon}^\varepsilon \mathrm{d}r_\perp \, f^\prime(\phi) \partial_\perp \phi \, ,
\label{app:projection}
\end{align}
where $\varepsilon$ is a small distance of the order of the interface width $w$ and the interface is positioned at ${r_\perp=0}$.

To proceed with the derivation, we make the following three key assumptions:
(i) \textit{Local equilibrium}: The chemical potential is assumed to be constant across the interface, ensuring that the system is in local thermodynamic equilibrium.
(ii) \textit{Weak variations in the bulk}: Within each phase (high- and low-density), the density field $\phi$ is assumed to vary only weakly, which we will confirm self-consistently.
(iii) \textit{Fast relaxation}: The density field $\phi$ locally relaxes much faster than the interface profile. 
Since the latter is driven by mass redistribution between different segments of the interface, it has a characteristic timescale ${T_D\sim 1/\Lambda^2}$, where $\Lambda$ denotes the characteristic length scale of the interface undulations, cf.\@~Fig.\@~\ref{fig:intuition}.
That is, for late times and weakly curved interfaces, the third assumption always holds.
As a result, within the bulk phases, $\phi$ obeys a stationary diffusion equation
\begin{align}
    \boldsymbol{\nabla}^2\phi=0
    \label{app:stationaryDiffusion}\, .
\end{align}
With these assumptions, Eq.\@~\eqref{app:projection} simplifies to 
\begin{align}
    \mu \, \phi\Big\rvert_{-\varepsilon}^{\varepsilon}
    \approx f(\phi) \Big\rvert_{-\varepsilon}^{\varepsilon} + \left( \int_{-\varepsilon}^\varepsilon \mathrm{d}r_\perp \, \frac{1}{2} \left(\partial_\perp \phi\right)^2\right) H \, ,
\end{align}
where we further used that $\partial_\perp\phi$ is sharply peaked around the interface.
While this relation is general, further progress can be made by exploiting the symmetry of the free energy,  $f(\phi)$, as defined in Eq.~\eqref{eq:freeEnergyDensity}.
Using equations Eq.\@~\eqref{app:rescaledChemicalPotential} and Eq.\@~\eqref{app:stationaryDiffusion} together with the assumption (i), we can relate the densities $\phi_\mathrm{in}$ and $\phi_\mathrm{out}$ on the two sides of the interface via the equation
\begin{align}
    f^\prime(\phi_\mathrm{in}) = f^\prime(\phi_\mathrm{out}).
\end{align}
Expanding the two densities around the minima of the free energy, $\phi_\mathrm{in/out}=\phi_\pm + \delta\phi_\mathrm{in/out}$, one finds
\begin{align}
    \delta\phi_\mathrm{in}=\delta\phi_\mathrm{out}=\delta\phi \, ,
\end{align}
where we have used that ${f^{\prime\prime}(\phi_+)=f^{\prime\prime}(\phi_-)}$.
That is, the shift in concentrations is symmetric on both sides of the interface.
Further exploiting the fact that, within the bulk phases, the chemical potential and the shift in concentration are linearly related, we write ${\mu=f^{\prime\prime}(\phi_\pm)\delta\phi=2\delta\phi}$.
Using this relationship, we finally arrive at the desired expression
\begin{align}
    \delta \phi =  \frac{1}{\Delta\phi}\left( \int_{-\varepsilon}^\varepsilon \mathrm{d}r_\perp \, \frac{1}{2} \left(\partial_\perp \phi\right)^2\right) H\, ,
\end{align}
with $\Delta\phi$ the difference between the minima of the free energy, see Eq.\@~\eqref{eq:DeltaPhi}.
For weakly curved interfaces and within the sharp interface limit ${\varepsilon\rightarrow0}$, we can further approximate the term in the brackets by the non-dimensional surface tension $\gamma$ as defined in App.\@~\ref{app:SurfaceTension}.
To that end, we note that for a flat interface 
\begin{align}
    \frac{1}{4} \left(\partial_\perp \phi\right)^2 = f(\phi)-f(\phi_\pm)\, ,
    \label{app:flatRelation}
\end{align}
which follows from integrating the stationarity condition ${\mu=0}$ once.
Substituting this into the expression for the integral yields
\begin{align}
    \int_{-\varepsilon}^\varepsilon \mathrm{d}r_\perp \, \frac{1}{2} \left(\partial_\perp \phi\right)^2 \approx 
    \int_{-\infty}^\infty \mathrm{d}r_\perp \, \frac{1}{2} \left(\partial_\perp \phi\right)^2 &\approx  \notag \\
    \int_{-\infty}^\infty \mathrm{d}r_\perp \, \frac{1}{4} \left(\partial_\perp \phi\right)^2 + f(\phi)-f(\phi_\pm) &= \gamma \, ,
\end{align}
where in the first step we used the fact that $\partial_\perp\phi$ is sharply peaked around ${r_\perp=0}$ to extend the integral boundaries to infinity.
In the second step, we assume the interface is only weakly curved, allowing us to use the flat-interface relation, Eq.\@~\eqref{app:flatRelation}.
The final integral corresponds to the definition of the surface tension $\gamma$  for a flat interface, Eq.\@~\eqref{app:definitionSurfaceTension}.
Combining these results, we obtain the linearized Gibbs-Thomson relation \cite{Eyre.1993,Garcke.2000}, frequently used throughout this manuscript:
\begin{align}
    \delta \phi =  \frac{\gamma}{2\Delta\phi} H\, ,
    \label{app:GibbsThomson}
\end{align}
which relates the shift in concentration $\delta \phi$ near the interface to the local interface curvature $H$. 
Physically, it implies that regions of higher curvature experience a more pronounced shift in concentration, reflecting the balance between surface tension and bulk phase equilibrium.
Moreover, it confirms that within the regime of weakly curved interfaces, the density field $\phi$ undergoes only small modulations, consistent with our second assumption.

We now provide explicit expressions for the two interface configurations considered in this manuscript.
To do so, we adopt the coordinate system introduced in Sec.\@~\ref{sec:stabilityAnalysis}, where $r_\parallel$ and $r_\perp$ denote the global coordinates parallel and perpendicular to the unperturbed flat or spherical interface, respectively.
For the case of a flat interface with weak undulations, the interface position is parameterized as ${\boldsymbol{X}=r_\parallel \boldsymbol{\hat r}_\parallel + (X_0+\delta X(r_\parallel) )\boldsymbol{\hat r}_\perp}$; see Sec.\@~\ref{sec:stabilityAnalysis}.
The local curvature in this case is given by
\begin{align}
    H(r_\parallel)
    &= - \frac{\partial_\parallel^2\delta X}{\Big(1+\left(\partial_\parallel\delta X\right)^2\Big)^{3/2} }\notag \\
    &\approx -\partial_\parallel^2\delta X(r_\parallel) + \mathcal{O}(\delta X^2) \, .
    \label{app:LocalGTFlat}
\end{align}
Consequently, to linear order, the density shift close to the interface reads
\begin{align}
    \delta \phi(r_\parallel) = -\frac{\gamma}{2\Delta\phi}\partial_\parallel^2 \delta X (r_\parallel) \, .
    \label{app:DensityShiftFlat}
\end{align}
For a spherical interface parametrized as ${\boldsymbol{X}=X(r_\parallel)\boldsymbol{\hat r}_\perp}$ with ${X(r_\parallel)=X_0+\delta X(r_\parallel)}$ and where $r_\perp$ and $r_\parallel$ denote the radial and azimuthal direction, respectively, one finds
\begin{align}
    H(r_\parallel)
    &=\frac{X^2+2 (\partial_\parallel X)^2 - X\partial_\parallel^2 X}{\left(X^2 +\left(\partial_\parallel X \right)^2 \right)^{3/2}} \notag \\
    &\approx \frac{1}{X_0} - \frac{ \delta X(r_\parallel)+\partial_\perp^2 \delta X(r_\parallel)}{X_0^2} + \mathcal{O}(\delta X ^2) \, 
    \label{app:LocalGTSpherical}
\end{align}
Here, the first term represents the curvature of the unperturbed spherical interface, while the second term accounts for the contribution of local undulations.
Thus, to leading order, the shift in concentration due to perturbations of the spherical interface is given by
\begin{align}
    \delta \phi(r_\parallel) 
    = -\frac{\gamma}{2\Delta\phi} \frac{ \delta X(r_\parallel)+\partial_\perp^2 \delta X(r_\parallel)}{X_0^2} \, .
    \label{app:DensityShiftCurved}
\end{align}
Throughout this manuscript, we use the above relations, Eq.\@~\eqref{app:DensityShiftFlat} and Eq.\@~\eqref{app:DensityShiftFlat}, to provide boundary conditions for the linearized density field $\phi$ at the sharp interface.
Although originally derived for an equilibrium system (${Pe=0}$), these relations can be extended to our non-equilibrium model under the assumption of local equilibrium. Specifically, they remain valid when the chemical field varies on a length scale much larger than the equilibrium interface width (${l\gg 1}$), ensuring that chemotactic effects are negligible at the interface. This separation of length scales justifies the application of equilibrium boundary conditions even in the presence of non-equilibrium dynamics.
\section{Linear stability of the homogeneous steady state}
\label{app:phaseDiagramLSA}
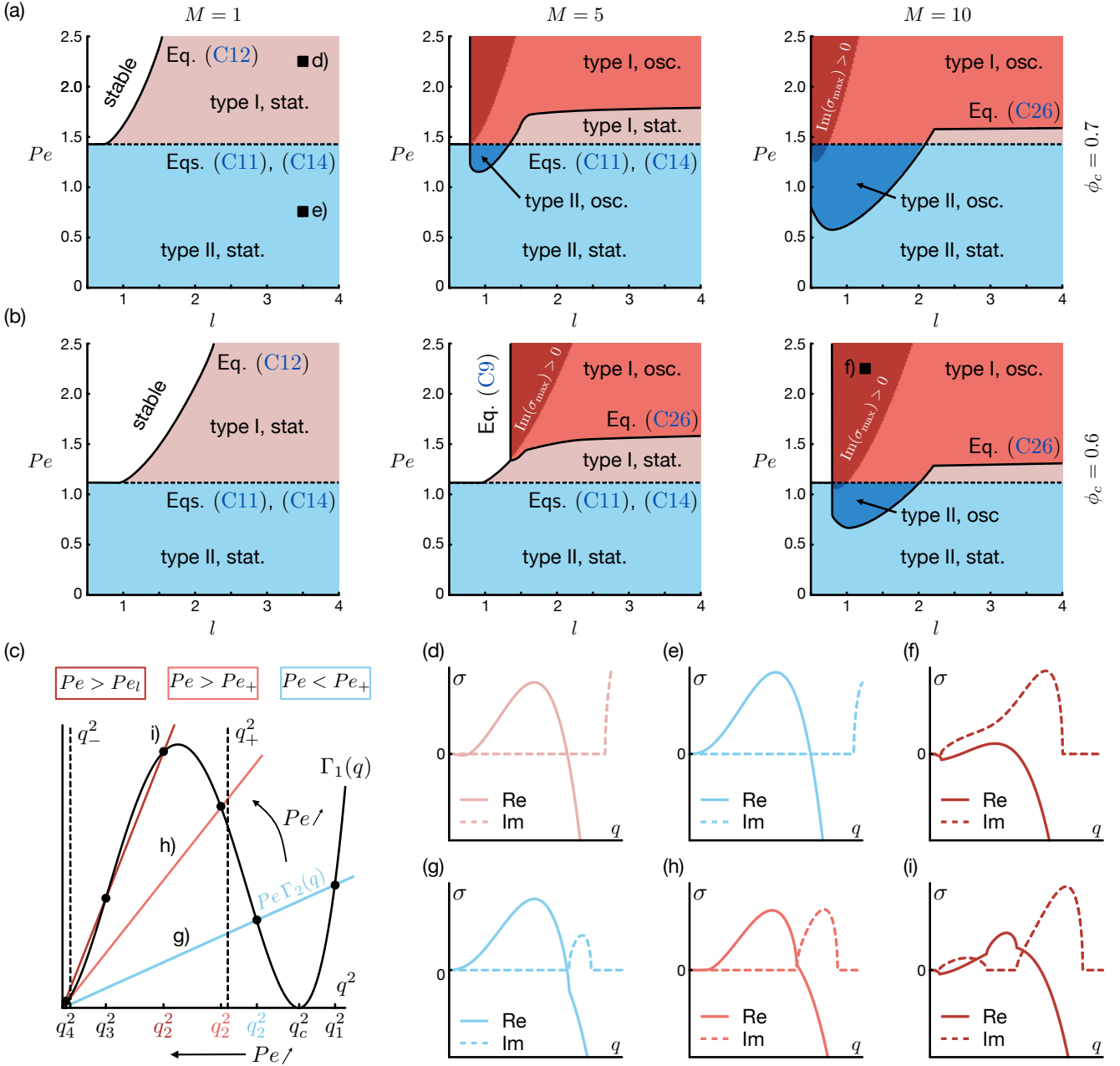
\begin{figure*}
    \centering
    \begin{overpic}[]{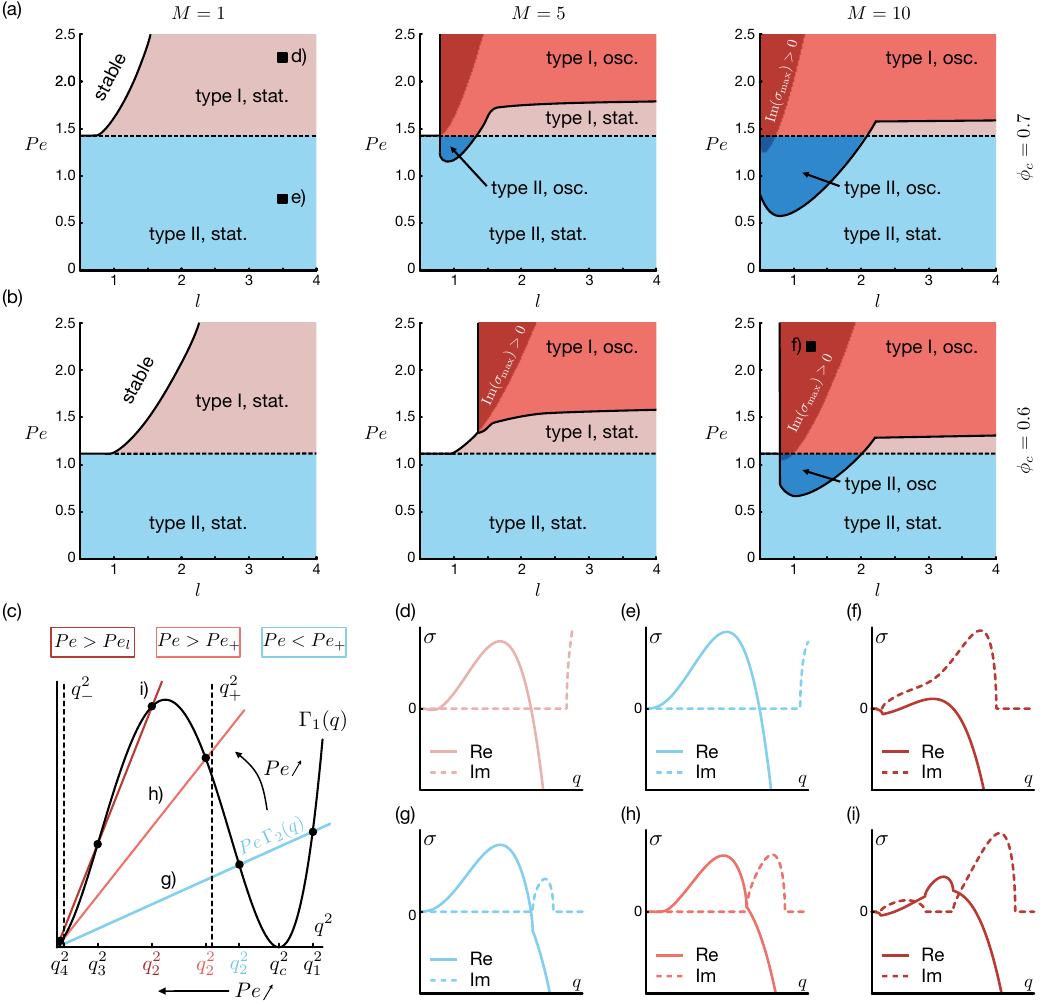}
        \put(14.5,89){\normalsize \sffamily Eq.\@~\eqref{app:InstabilityCon3}} 
        \put(14.5,79.5){\normalsize \sffamily Eqs.\@~\eqref{app:InstabilityCon2},\@~\eqref{app:typeIvsII}} 
        \put(19,61.75){\normalsize \sffamily Eq.\@~\eqref{app:InstabilityCon3}} 
        \put(14.5,49.6){\normalsize \sffamily Eqs.\@~\eqref{app:InstabilityCon2},\@~\eqref{app:typeIvsII}} 
        \put(42,56){\rotatebox{90}{\normalsize \sffamily Eq.\@~\eqref{app:InstabilityCon1}}}
        \put(46.5,49.6){\normalsize \sffamily Eqs.\@~\eqref{app:InstabilityCon2},\@~\eqref{app:typeIvsII}} 
        \put(46.5,79.5){\normalsize \sffamily Eqs.\@~\eqref{app:InstabilityCon2},\@~\eqref{app:typeIvsII}} 
        \put(53.4,56.75){\normalsize \sffamily Eq.\@~\eqref{app:OsciThreshold}} 
        \put(85.4,84){\normalsize \sffamily Eq.\@~\eqref{app:OsciThreshold}} 
        \put(85.4,54.4){\normalsize \sffamily Eq.\@~\eqref{app:OsciThreshold}} 
    \end{overpic}
    \caption{\textbf{Linear stability analysis} of the homogeneous steady state ${\phi=\psi=\bar\phi}$. 
    (a--b) Bifurcation diagram in the $l$-$Pe$ plane for ${\phi_c=0.7}$, ${M=\{1,5,10\}}$ at (a) ${\bar\phi=0.7}$ and (b) ${\bar\phi=0.6}$. Each diagram illustrates the phase boundaries [black lines, Eqs.\@~\eqref{app:InstabilityCon1},\@~\eqref{app:InstabilityCon2},\@~\eqref{app:InstabilityCon3}] between stable (white) and unstable domains (colored/shaded areas), as well as the boundaries [dashed lines, Eq.\@~\eqref{app:typeIvsII}] between type-I (red) and type-II (blue) instabilities. The shading of each domain indicates whether the steady state exhibits a stationary instability (light blue/red/gray) or unstable oscillatory modes (darker blue/red/gray). The darkest domains (dark blue/red/gray) indicate the parameter regime where also the fastest growing mode has a finite imaginary part Im$(\sigma_\mathrm{max})>0$.
    When displayed in grayscale, the same distinctions hold, with light gray areas corresponding to stationary instabilities and darker gray shades indicating oscillatory modes.
    For large diffusive length scales, the boundary between the stationary and oscillatory domains approaches a horizontal asymptote given by Eq.\@~\eqref{app:OsciThreshold}.
    (c) Graphical construction of the oscillatory modes for ${M=7}$, ${l=1.5}$, and ${\bar\phi=\phi_c=0.7}$. 
    The plot shows the intersection points $q^2_{1-4}$ between the curves $\Gamma_1(q)$ (black) and ${Pe\,\Gamma_2(q)}$, see Eqs.\@~\eqref{app:Gamma1}–\eqref{app:Gamma2}, for different values of $Pe$. The intersections are shown for ${Pe=0.5}$ (light blue/gray), ${Pe=1.5}$ (red/gray), and ${Pe=2.8}$ (dark red/gray). 
    (d--i) Largest eigenvalue $\sigma(q)$ of perturbations around the homogeneous steady state [cf.\@~Eq.\@~\eqref{app:LinearStabilityMatrix}] for parameters corresponding to the square markers in panels (a--b) and the colored lines in panel (c).}
    \label{figAPP:homLSA}
\end{figure*}
In Sec.\@~\ref{sec:overview}, we introduced the various dynamical and stationary states exhibited by our model, highlighting the influence of key parameters without constructing a complete phase diagram.
Here, we take a more systematic approach by performing a linear stability analysis of the homogeneous steady state ${\phi=\psi}$.
By analyzing the dispersion relation of small perturbations around this state, we identify different types of instabilities that mark the onset of pattern formation.
Following the nomenclature of Cross and Hohenberg \cite{Cross.1993}, we classify these as long-wavelength (type-II) and short-wavelength (type-I) instabilities, further distinguishing between stationary and oscillatory modes.
To assess the validity of the linear theory, in the next section, we compare its predictions with numerical simulations of the full nonlinear dynamics, Eq.\@~\eqref{eq:DynamicsRescaled}.
This allows us to verify the linear approximations and to identify regimes where nonlinear effects dominate.
By combining both approaches, we construct a comprehensive phase diagram that captures both the onset and fully developed dynamics of pattern formation.
\subsection{Linearized dynamics}
For small perturbations around the homogeneous steady state ${\phi=\psi=\bar\phi}$, with $\bar\phi$ denoting the average density, we consider perturbations of the form
\begin{align}
    \begin{pmatrix}
        \delta \phi(\boldsymbol{x},t) \\[2mm]
        \delta \phi(\boldsymbol{x},t)
    \end{pmatrix}
    = 
     \exp(\sigma t) \exp(i \boldsymbol{q}\cdot\boldsymbol{x})
    \begin{pmatrix}
        \delta \phi \\[2mm]
        \delta \phi
    \end{pmatrix} \, ,
\end{align}
with $\sigma$ indicating the growth rate of individual spatial Fourier modes $\boldsymbol{q}$.
Substituting this ansatz into the dynamical equations, Eq.\@~\eqref{eq:DynamicsRescaled}, and retaining terms to linear order in the perturbations, we obtain a linear set of equations
\begin{align}
    \sigma \begin{pmatrix}
        \delta \phi \\[2mm]
        \delta \psi
    \end{pmatrix} 
    = 
    \begin{pmatrix}
        -\frac{q^2M}{2}\left(q^2+2f^{\prime\prime}\right) & -q^2M\,Pe \, \bar\phi \\[2mm]
        l^{-2} & -q^2 - l^{-2}
    \end{pmatrix}
    \begin{pmatrix}
        \delta \phi \\[2mm]
        \delta \psi
    \end{pmatrix} \, .
    \label{app:LinearStabilityMatrix}
\end{align}
We emphasize that since the chemical field $\psi$ enters the dynamics of $\phi$ linearly, see Eq.~\eqref{eq:DenistyRescaled}, this result remains valid for more general reaction kinetics than the linear mass-action law employed in this work.
Specifically, after rescaling, one obtains the same form of linearized equations, even when choosing non-linear reaction terms.
Thus, our minimal model correctly describes the pattern-forming onset of a much broader class of systems.

By solving the above eigenvalue problem for each mode $q$, we derive a dispersion relation $\sigma(q)$, which informs about the stability of the homogeneous steady state:
If Re${(\sigma)>0}$ for any mode $q$, the homogeneous state is unstable as small perturbations tend to grow.
A non-vanishing imaginary part further indicates that the corresponding mode is subject to oscillatory behavior.
For a general ${2\times2}$ matrix, the eigenvalues can be calculated as 
\begin{align}
    \sigma_{1/2}(q)= \dfrac{\tau(q) \pm \sqrt{\tau(q)^2-4\delta(q)}}{2} \, ,
    \label{app:eigenvalue2x2}
\end{align}
where $\tau(q)$ and $\delta(q)$ are the matrix' trace and determinant, respectively:
\begin{align}
    \tau(q) &= -\frac{M}{2} q^4 - \left(Mf^{\prime\prime} +1 \right)q^2 - \frac{1}{l^2} \label{app:trace} \, ,\\
    \delta(q) &= Mq^2\left[\frac{q^4}{2} +\left(\frac{1}{2l^2}+f^{\prime\prime}\right)q^2 + 
    \frac{f^{\prime\prime} +Pe \, \bar\phi}{ l^2}\right] \label{app:determinant} \, .
\end{align}

We use the above representation, Eqs.\@~\eqref{app:trace}--\eqref{app:determinant}, to determine:
(i) whether the system is linearly unstable, 
(ii) if the instability affects short- or long-wavelength modes, i.e., whether the band of unstable modes extends to ${q=0}$, and
(iii) whether it is oscillatory or stationary.
This classification allows us to construct a bifurcation diagram as shown in Fig.\@~\ref{figAPP:homLSA}(a,b) for different values of $M$ and $\phi_c$.
Proceeding along the above classification steps, we will provide a detailed analysis of these diagrams, highlighting the different effects shaping their topology.
\subsection{Stability}
The homogeneous state is linearly unstable if, for any $q$, either ${\tau(q) >0}$ or ${\delta(q)<0}$.
From the above equations, one infers that these conditions are only satisfied when ${f^{\prime\prime}(\bar\phi) < 0}$, that is within the spinodal regime of the free energy, Eq.\@~\eqref{eq:freeEnergyDensity}---the emergence of patterns is always driven by the phase-separation dynamics in the system.
Moreover, since ${\tau(0)<0}$ and ${\delta(0)=0}$, it follows that the larger eigenvalue obeys ${\sigma_1(0)=0}$, implying that the homogeneous mode ${q=0}$ is marginally stable, as expected from the conservation of $\phi$.

To proceed, we note that both the trace $\tau(q)$ and the determinant $\delta(q)$ are polynomials in $q^2$.
This allows us to apply a quadratic closure for each of them, resulting in parabolic expressions of the form 
\begin{align}
    \tau(q) &= a_\tau (q^2-q_{c,\tau} ^2)^2 + b_\tau  \, ,\\
    \dfrac{\delta(q)}{Mq^2} &= a_\delta (q^2-q_{c,\delta}^2)^2  + b_\delta\, ,
\end{align}
where the coefficients $a_{\tau/\delta}$ and $b_{\tau/\delta}$ will be determined explicitly in the following analysis.
From this, we infer that the homogeneous steady state is unstable if ${q_{c,\tau/\delta}^2 > 0}$ and at least one of the two parabolas intersects the $q^2$-axis, as this implies that either ${\tau(q) >0}$ or ${\delta(q)<0}$ for some $q$.
For the trace $\tau(q)$, we find
\begin{align}
    \tau_q =& -\frac{M}{2}\left(q^2+\frac{Mf^{\prime\prime}+1}{M}\right)^2 \notag \\
    &+ \frac{(Mf^{\prime\prime}+1)^2}{2M}-\frac{1}{l^2} \, .
    \label{app:TauNormalForm}
\end{align}
Since this describes a downwards opening parabola, $\tau_q$ intersects the $q^2$-axis whenever
\begin{align}
    Mf^{\prime\prime}+1 < - \sqrt{\frac{2M}{l^2}} \,,
    \label{app:InstabilityCon1}
\end{align}
which yields our first instability condition.
Notably, this criterion is independent of the P\'eclet number $Pe$ and, thus, independent of the strength of chemotactic interactions.
Furthermore, since $f^{\prime\prime}(\bar\phi)$ is bounded from below by -1 (see Eq.\@~\eqref{eq:freeEnergyDensity}) the condition, Eq.\@~\eqref{app:InstabilityCon1}, requires ${M>1}$, corresponding to slow propagation of the chemical field.
In Fig.\@~\ref{figAPP:homLSA}(a--b), the instability condition, Eq.\@~\eqref{app:InstabilityCon1}, corresponds to the vertical boundary separating stable and unstable regions in the subpanels with ${M>1}$.
Applying a quadratic closure to the determinant $\delta(q)$ yields
\begin{align}
    \dfrac{\delta(q)}{Mq^2} =& \frac{1}{2}\left(q^2+ \frac{1}{2l^2}+f^{\prime\prime}\right)^2 \notag\\
    &+ \frac{Pe \,\bar\phi+f^{\prime\prime}}{l^2}-\frac{1}{2}\left(\frac{1}{2l^2}+f^{\prime\prime}\right)^2 \,.
\end{align}
This implies that $\delta(q)$ is guaranteed to be negative when
\begin{align}
    Pe \, \bar \phi + f^{\prime\prime} < 0 \, 
    \label{app:InstabilityCon2}
\end{align}
corresponding to an instability around ${q=0}$; cf.\@~Eq.\@~\eqref{app:trace}.
If this condition is not met, the homogeneous steady state may still be unstable if
\begin{align} 
    \frac{1}{2l^2}+f^{\prime\prime} < -\sqrt{2\frac{Pe \,\bar\phi+f^{\prime\prime}}{l^2}} \, .
    \label{app:InstabilityCon3}
\end{align}
The first condition, Eq.\@~\eqref{app:InstabilityCon2}, compares the strength of phase separation $f^{\prime\prime}$ to the strength of the chemotactic interactions $Pe$:
For sufficiently weak chemotactic repulsion, the homogeneous steady state is unstable, indicated by the horizontal solid lines in Fig.\@~\ref{figAPP:homLSA}(a--b).
The latter condition, Eq.\@~\eqref{app:InstabilityCon3}, is the only one that depends on both, the interaction strength $Pe$ and the diffusive length scale $l$.
Specifically, it implies that even if Eq.\@~\eqref{app:InstabilityCon2} is not fulfilled, instability can still occur for sufficiently large diffusive length scales $l$; see\@~Fig.\@~\ref{figAPP:homLSA}.

In summary, we have derived three conditions, Eq.\@~\eqref{app:InstabilityCon1}, Eq.\@~\eqref{app:InstabilityCon2} and Eq.\@~\eqref{app:InstabilityCon3}, which determine the onset of pattern formation and mark the boundaries of the colored (unstable) regimes in Fig.\@~\ref{figAPP:homLSA}.
\subsection{Long- vs.\@~short-wavelength instabilities}
Next, we analyze whether the instabilities exclusively drive the growth of short-wavelength modes (type-I) or whether the band of unstable modes extends to ${q=0}$, thus including long-wavelength modes (type-II).
Exploiting that the eigenvalues are also functions of $q^2$, we classify an instability as type-I, when 
\begin{align}
    \partial_{q^2}\mathrm{Re}(\sigma) \big\rvert_{q=0} < 0 \, , 
\end{align}
implying that modes close to the origin ${q=0}$ are stable.
Here $\partial_{q^2}$ denotes the partial derivative with respect to\@~$q^2$. 
Using Eqs.\@\eqref{app:eigenvalue2x2}--\eqref{app:determinant}, this condition is equivalent to ${\partial_{q^2}\delta\big\rvert_{q=0}>0}$, which simplifies to
\begin{align}
    f^{\prime\prime}+Pe\,\bar\phi > 0 \,.
    \label{app:typeIvsII}
\end{align}
Notably, this is the opposite of the second instability condition, Eq.\@~\eqref{app:InstabilityCon2}, meaning that for sufficiently strong repulsive interactions, the band of unstable modes is always bounded away from ${q=0}$.
Such a behavior is often considered indicative for arrested coarsening, i.e., the formation of patterns with a characteristic length scale, which is in contrast to a purely phase separating system which only exhibits type-II instabilities \cite{Cross.1993}.
In the $l$-$Pe$ plane, the boundary between type-I and type-II instabilities appears as a horizontal line, which can be confirmed by inspecting Fig.\@~\ref{figAPP:homLSA}(a--b).
\subsection{Stationary vs.\@~oscillatory instabilities}
Lastly, we investigate whether and under which conditions the system can exhibit an oscillatory instability. 
Specifically, we explore for which parameter range the band of unstable modes overlaps with the range of modes where Im${(\sigma)>0}$.
The latter requires the discriminant in Eq.\@~\eqref{app:eigenvalue2x2} to be negative, i.e.\@ ${\Delta(q)=\tau(q)^2-4\delta(q)<0}$, which in turn implies ${\delta(q)>0}$.
Recalling that an instability requires either ${\tau(q)>0}$ or ${\delta(q)<0}$, the homogeneous state can only be unstable if ${\tau(q)>0}$, corresponding to the condition in Eq.\@~\eqref{app:InstabilityCon1}.

Solving for both of these conditions simultaneously turns out cumbersome, as the discriminant is a fourth-order polynomial in $q^2$.
However, we can use a graphical construction, shown in Fig.\@~\ref{figAPP:homLSA}(c), to identify when the system may exhibit an oscillatory instability.
To that end, we emphasize that any potential mode has to lie within the unstable interval ${[q_-,q_+]}$, where $q_\pm$ are given by the roots of $\tau(q)$ [see Eq.\@~\eqref{app:TauNormalForm}]
\begin{align}
    q_\pm^2 =  \frac{-(1+Mf^{\prime\prime}) \pm \sqrt{(1+Mf^{\prime\prime})^2 -2\frac{M}{l^2}}}{M} \, ,
    \label{app:UnstableInterval}
\end{align}
where Eq.\@~\eqref{app:InstabilityCon1} ensures that the right-hand side is strictly positive.
Importantly, the interval boundaries $q_\pm$ do not depend on the strength $Pe$ of the chemotactic interaction. This will be a key point in the further discussion.
We proceed by re-writing the discriminant as
\begin{align}
    \Delta(q) &=\Gamma_1(q) - Pe\, \Gamma_2(q) \, \label{app:Discriminant} ,\\
    \Gamma_1(q) &= \left( Mq^2\left(f^{\prime\prime}+\frac{q^2}{2}\right)-\left(q^2+\frac{1}{l^2}\right)\right)^2  \, , \label{app:Gamma1}\\
    \Gamma_2(q) &=\frac{4M\bar\phi}{l^2} q^2 \, , \label{app:Gamma2}
\end{align}
where the auxiliary functions $\Gamma_1$ and $\Gamma_2$ are strictly positive and also do not depend on the P\'eclet number $Pe$.
That is, when plotted against $q^2$, the chemotactic interaction strength $Pe$ controls the slope of the linear function $Pe\, \Gamma_2(q)$, as shown in Fig.\@~\ref{figAPP:homLSA}(c).
The form of Eq.\@~\eqref{app:Discriminant} further implies that in the absence of chemotactic interactions (${Pe=0}$), the discriminant $\Delta(q)$ is always positive, ruling out the emergence of oscillatory modes in an equilibrium system.
For finite interaction strength, unstable oscillatory modes (${\Delta(q)<0}$) only emerge if, within the unstable interval ${[q_-,q_+]}$, the curve $\Gamma_1(q)$ lies below the curve defined by ${Pe\,\Gamma_2(q)}$; see Fig.\@~\ref{figAPP:homLSA}(c). 
Since ${\Gamma_1(0)>\Gamma_2(0)=0}$, this condition can only be fulfilled for finite ${q>0}$, whereas asymptotically one always has ${\Delta(q)>0}$, since $\Gamma_1$ grows as ${\Gamma_1\sim q^8}$.

We now explore how the number of intersection points between the curves $\Gamma_1(q)$ and ${Pe\,\Gamma_2(q)}$ changes as we increase $Pe$ starting from zero.
If for any $Pe$ one of these intersection points resides within the unstable interval ${[q_-^2,q_+^2]}$, we know that $\Delta(q)$ changes its sign, implying the existence of a range of oscillatory modes with ${\Delta(q)<0}$.
From Fig.\@~\ref{figAPP:homLSA}, it is clear that $\Gamma_1(q)$ has a single root at 
\begin{align}
    q_c^2 = \frac{(1-Mf^{\prime\prime})+\sqrt{(1-Mf^{\prime\prime})^2+2M/l^2}}{M} \, ,
\end{align}
which also comprises the only intersection point at vanishing $Pe$.
One finds that at ${q^2 =q_c^2}$ ${\Delta(q_c)<0}$ holds trivially whenever ${Pe>0}$.
However, comparing $q_c^2$ to the expression for $q^2_+$ in Eq.\@~\eqref{app:UnstableInterval}, we see that it always lies outside of the unstable interval.

For small $Pe$, Fig.\@~\ref{figAPP:homLSA}(c) illustrates that $\Gamma_1(q)$ and ${Pe \, \Gamma_2(q)}$ intersect only twice (blue line), implying that $\Delta(q)$ has two roots $q^2_{1/2}$, emerging from $q_c^2$ when ${Pe=0}$.
While the larger ($q_1^2$) of these roots always lies beyond $q_c^2$ and, thus, outside of the unstable interval ${[q_-,q_+]}$, the smaller one can eventually fall into the unstable interval as $Pe$ increases.
However, as $Pe$ increases, a competing effect occurs:
By increasing the slope of the linear curve, at some point, two new intersection points $q_{3/4}^2$, with ${q_3^2>q_4^2}$, emerge, creating two regions where ${\Delta(q)<0}$ [see the red lines in Fig.\@~\ref{figAPP:homLSA}(c)].
This regime with four intersection points is bounded by the cases where $Pe \, \Gamma_2(q)$ is tangential to $\Gamma_1(q)$, implying that
\begin{align}
    \left(\partial_{q^2} \Gamma_1 - Pe\, \partial_{q^2}\Gamma_2\right)\big\rvert_{q_\mathrm{l/r}^2} &=  0 \\
    \left(\Gamma_1-Pe\, \Gamma_2\right) \big\rvert_{q_\mathrm{l/r}^2}&=  0 \, .
\end{align}
The conditions are simultaneously fulfilled at
\begin{align}
    q_\mathrm{r/l}^2= \frac{(1-Mf^{\prime\prime}) \pm \sqrt{(1-Mf^{\prime\prime})^2-6M/l^2}}{3M}\, ,
\end{align}
representing the left and right tangential points, which only exist for ${(1-Mf^{\prime\prime})^2-6M/l^2>0}$.
Using the equality of the slopes, we can further deduce that at the tangential points
\begin{align}
    Pe_\mathrm{r/l} = \frac{l^2}{4M \bar \phi} \, \partial_{q^2} \Gamma_1\big\rvert_{q_{r/l}^2} \, .
    \label{app:boundaryTangent}
\end{align}
Increasing $Pe$ beyond this regime, causes the intersection points $q_2^2$ and $q_3^2$ to disappear.
That is, for large enough values of $Pe$ only the left- and right-most intersection points ($q_{1/4}^2$) persist and all modes in the unstable interval ${[q_-,q_+]}$ have a finite imaginary part [${\Delta(q)<0}$].
In particular, from Eq.\@~\eqref{app:boundaryTangent}, we can deduce an upper bound ${Pe>Pe_l}$, beyond which all unstable modes are oscillatory.
This comprises the main finding of our graphical analysis: 
As long as there exists a band of unstable modes where Eq.\@~\eqref{app:InstabilityCon1} is fulfilled, the homogeneous steady state always exhibits unstable oscillatory modes if the chemotactic interaction strength $Pe$ is large enough.

For the onset of unstable oscillatory modes, several possible scenarios arise.
To analyze these scenarios one by one, we recall that the unstable interval ${[q_-^2,q_+^2]}$ does not depend on $Pe$.
Then, if initially, all intersection points reside outside this interval, one of them will eventually enter it as the interaction strength $Pe$ increases.
Since $q_2^2$ emerges from $q_c^2$, it necessarily enters the unstable interval from above, whereas $q_{3/4}^2$ could have emerged above or below it and, thus, enter the unstable interval from any side.
In any case, each of these scenarios requires that at the boundaries $q_\pm^2$ of the unstable interval one has ${\Gamma_1(q) - Pe\,\Gamma_2(q)<0}$, or equivalently ${Pe>Pe_\pm}$, with
\begin{align}
    Pe_\pm = \frac{l^2}{4M\bar\phi}\frac{\Gamma_1(q_\pm)}{q_\pm^2} \, .
    \label{app:Pepm}
\end{align}
Depending on which of these conditions is satisfied first, the system either exhibits a band of unstable oscillatory modes ${[q_-^2,q_3^2]}$ [see Fig.\@~\ref{figAPP:homLSA}(c,h)], i.e., at larger wavelengths, or ${[q_{2,4}^2,q_+^2]}$, i.e., at shorter wavelengths.
Further increasing the interaction strength $Pe$ can give rise to a situation with two bands of unstable oscillatory modes [see Fig.\@~\ref{figAPP:homLSA}(c,i)], which eventually merge.
Alternatively, there is a scenario, where $q_3^2$ and $q_4^2$ jointly emerge within the range of unstable modes.
As they originate from $q_l^2$, this necessarily requires that ${q_l^2 > q_-^2}$. 
Since the condition for the existence of $q_{3/4}^2$ was given by Eq.\@~\eqref{app:boundaryTangent}, we conclude that the latter scenario is possible if ${q_l^2 > q_-^2}$ and ${Pe>Pe_l}$.

In total, we find that the homogeneous steady state is subject to unstable oscillatory modes if ${Pe > Pe_c}$, where the critical threshold ${Pe_c}$ is given by 
\begin{align}
    Pe_c = 
    \begin{cases}
        \mathrm{min}\{Pe_-,Pe_+,Pe_l\}\, , \quad   &\text{if} 
        \hspace{2mm} \exists \, q_{r/l}^2 \hspace{2mm}  \And  \hspace{2mm}  q_l^2> q_-^2 \, , \\[2mm]
        \mathrm{min}\{Pe_-,Pe_+\}\, , \quad &\text{otherwise.}
    \end{cases}
\end{align}
Which scenario occurs first critically depends on the values of the remaining parameters and has to be checked from case to case.
However, we find a tendency that for small $l$ unstable oscillatory modes dominantly emerge by $q_2^2$ entering the unstable interval from above [see Fig.\@~\ref{figAPP:homLSA}(c)], whereas for larger $l$ the $q_3^2$ intersection point enters from below, driven by the fact that ${q_-^2\rightarrow0}$ as ${l\rightarrow \infty}$.
For the latter case, we can expand the expression for $Pe_-$ in Eq.\@~\eqref{app:Pepm} to leading order in $l^{-2}$, which yields
\begin{align}
    Pe_\infty = - \frac{\left(f^{\prime\prime}\right)^2M}{\left(1+f^{\prime\prime}M \right) \bar\phi} \, .
    \label{app:OsciThreshold}
\end{align}
At the same time, one can check that both $Pe_+$ and $Pe_l$ diverge as ${l\rightarrow\infty}$.
This constitutes the second core finding of our graphical analysis:
While for short diffusive length scales $l$ the transition from stationary to oscillatory behavior follows a complex curve, (see Fig.\@~\ref{figAPP:homLSA}), it becomes particularly simple as $l\rightarrow\infty$.
Specifically, for large $l$ unstable oscillatory modes exist whenever ${Pe>Pe_\infty}$.
That is, the boundary between the stationary and the oscillatory regime approaches a horizontal line as can also be seen in Fig.\@~\ref{figAPP:homLSA}(a--b).

Finally, we remark that the conditions derived in the previous paragraphs only predict the existence of unstable oscillatory modes at some unspecified wavelength. 
In particular, they do not imply that the fastest growing mode, where Re($\sigma$) is maximal, also exhibits a non-vanishing imaginary part.
As illustrated in Fig.\@~\ref{figAPP:homLSA}, the latter is only true close to the left onset of the oscillatory regime, meaning for small diffusive length scales $l$.
Nevertheless, we will speak of an oscillatory instability whenever the band of unstable and oscillatory modes overlap, additionally highlighting whenever the fastest growing mode falls into the overlapping regime.
With this final remark, we conclude our analysis of the onset of pattern formation and turn our attention towards the static and dynamic patterns observed at late times.
\section{Phase diagram}
\label{app:phaseDiagramNumerics}
\begin{figure*}
    \centering
    \includegraphics[]{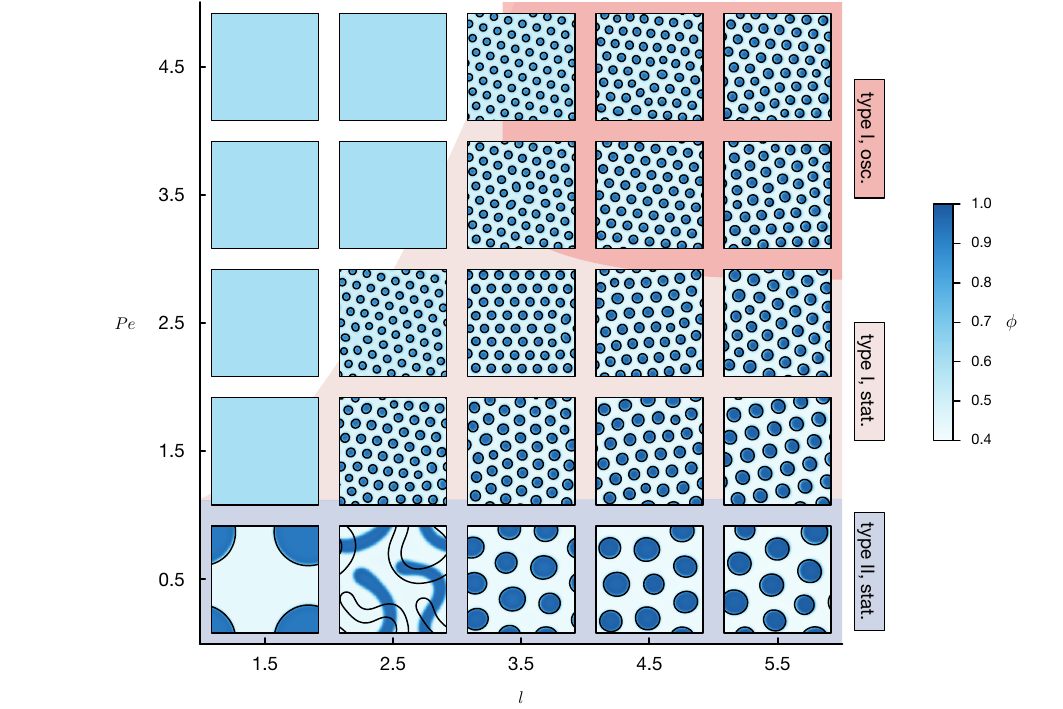}
    \caption{\textbf{Phase diagram} obtained from numerically solving the dynamical equations, Eq.\@~\eqref{eq:DynamicsRescaled} at ${M=2.5}$, using a finite element scheme for a quadratic system of size ${L=75}$ with periodic boundary conditions.
    Each simulation was initialized with an approximately homogeneous state ${\phi=\psi=\bar\phi=0.6}$, subject to weak, uniformly distributed perturbations with a relative amplitude of $10^{-3}$.
    The insets display the system's state at ${t=10^6}$, where light (dark) blue areas correspond to the low- (high-) density phase (see colorbar), as well as the interface profile (${\phi=\phi_c}$ level-set) at an earlier time ${t=10^5}$ (black lines).
    The background colors indicate the type of instability, determined by solving the linearized problem, Eq.\@~\eqref{app:LinearStabilityMatrix}; see Fig.\@~\ref{figAPP:homLSA}.
    Within the explored parameter range, the homogeneous steady state is either stable (white) or exhibits a stationary type-II (purple, light blue in Fig.\@~\ref{figAPP:homLSA}), stationary type I (light red) or oscillatory type-I (red) instability.
    When displayed in grayscale, the same correspondence holds, with light and dark gray shades indicating different dynamical regimes, matching the labels in the small boxes on the right.
    For all simulations, as well as for the linear stability analysis, we kept the parameters ${\phi_c=0.7}$ and ${\bar\phi=0.6}$ fixed, while varying $l$ and $Pe$.}
    \label{figAPP:phaseDiagram1}
\end{figure*}
\begin{figure*}
    \centering
    \includegraphics[]{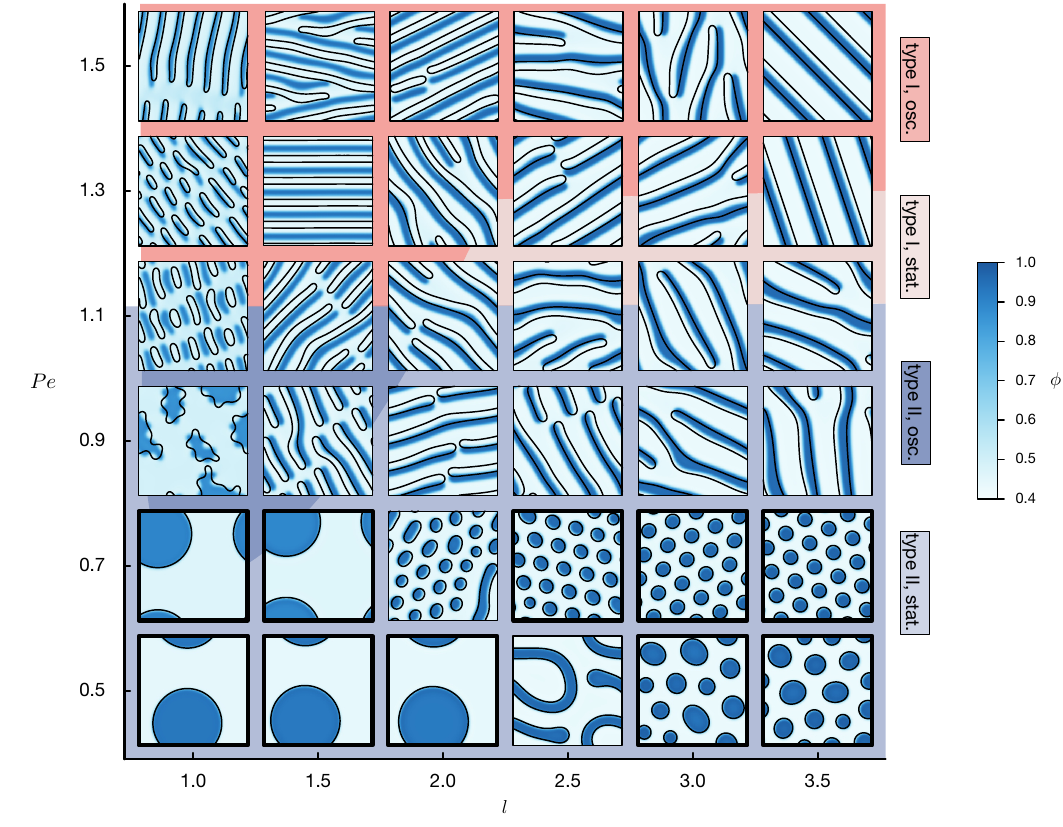}
    \caption{\textbf{Phase diagram} obtained from numerically solving the dynamical equations, Eq.\@~\eqref{eq:DynamicsRescaled}, at ${M=10}$ using a finite element scheme for a quadratic system of size ${L=75}$ with periodic boundary conditions.
    Each simulation was initialized with an approximately homogeneous state ${\phi=\psi=\bar\phi=0.6}$, subject to weak, uniformly distributed perturbations with a relative amplitude of $10^{-3}$.
    The insets display the system's state at ${t=6000}$ for the dynamic patterns (thin frames) and ${t=10^5}$ for the stationary patterns (thick frames).
    The interface profiles ${\phi=\phi_c}$ (black lines) separating the high- (dark blue/gray) from the low-density domains (light blue/gray) is shown at a previous time ${t=5995}$ (${t=99,500})$.
    The background colors indicate the type of instability, determined by solving the linearized problem, Eq.\@~\eqref{app:LinearStabilityMatrix}; see Fig.\@~\ref{figAPP:homLSA}.
    Within the explored parameter range, the homogeneous steady state is either stable (white) or exhibits a stationary type-II (purple, light blue in Fig.\@~\ref{figAPP:homLSA}), oscillatory type II (dark purple, dark blue in Fig.\@~\ref{figAPP:homLSA}), stationary type-I (light red) or oscillatory type-I (red) instability.
    When displayed in grayscale, the same correspondence holds, with light and dark gray shades indicating different dynamical regimes, matching the labels in the small boxes on the right.
    For all simulations, as well as for the linear stability analysis, we kept the parameters ${\phi_c=0.7}$ and ${\bar\phi=0.6}$ fixed, while varying $l$ and $Pe$.}
    \label{figAPP:phaseDiagram2}
\end{figure*}
To explore the terminal behavior, we employ numerical simulations initialized with an almost homogeneous state ${\phi=\psi=\bar\phi=0.6}$ subject to weak, uniformly distributed perturbations.
Specifically, we use a perturbation of the form $\phi/\psi=\bar\phi(1+0.001\zeta)$, where $\zeta$ is a random uniform variable distributed between $-1$ and $1$.
The results for ${M=2.5}$ and ${M=10}$, corresponding to intermediate and fast signal propagation, are shown in Fig.\@~\ref{figAPP:phaseDiagram1} and Fig.\@~\ref{figAPP:phaseDiagram2}, respectively.
For ${M=2.5}$, we observe that linear stability analysis correctly predicts the onset of pattern formation, but is, at best, only loosely connected to the observed terminal behavior.
In particular, for nearly all displayed parameter combinations, the final state consists of a stationary pattern with approximately equally sized droplets, whose size decreases with increasing chemotactic interaction strength, $Pe$.
However, we do not observe any significant differences between regions characterized by different types of instability.
In particular, the coexistence of multiple finite-sized droplets is not restricted to the parameter regime exhibiting a type-I instability (red domains).
Instead, it extends into the type-II unstable regime, typically associated with classical coarsening dynamics.
Moreover, in the regime where linear stability analysis, Eq.\@~\eqref{app:LinearStabilityMatrix}, predicts the existence of unstable oscillatory modes, the terminal state does not exhibit any dynamic behavior.
Instead, the only instance of a dynamical pattern is found within the stationary type-II regime (${Pe=0.5}$, ${l=2.5}$).

For ${M=10}$, Fig.\@~\ref{figAPP:phaseDiagram2} reveals a more diverse range of behaviors, with the most striking difference from the previous case (${M=2.5}$, Fig.\@\ref{figAPP:phaseDiagram1}) being the emergence of various dynamical states, ranging from traveling stripe and droplet patterns to the formation of capillary waves.
Thus, comparing both figures reveals that if signal propagation is slow compared to the diffusive redistribution of mass (${M \gg 1}$), it significantly enhances the formation of dynamical states.
Moreover, we find that for weaker chemotactic interactions ${0.5\leq Pe\leq 0.7}$, dynamical patterns are confined to the region between full and arrested coarsening, whereas for stronger interactions they emerge generically, with traveling stripes being the most prevalent form.
Additionally, one observes that for the given parameters, capillary wave formation is confined to a narrow parameter regime around ${Pe=0.9}$ and ${l=1}$---which seemingly coincides with the oscillatory type-II instability.
Similar to the previous figure, the emergence of dynamical states is not restricted to the oscillatory domain but extends into the stationary regime.

In summary, Fig.\@~\ref{figAPP:phaseDiagram1} and Fig.\@~\ref{figAPP:phaseDiagram2} illustrate the rich variety of phenomena captured by our minimal model, Eqs.\@~\eqref{eq:DynamicsRescaled}–\eqref{eq:FreeEnergy}, and provide a comprehensive phase diagram, encompassing both the onset of pattern formation and the terminal behavior.
Since the phenomena observed in numerical simulations only loosely correspond to the predictions of linear stability analysis, we conclude that fully understanding the terminal dynamics requires incorporating nonlinear effects, potentially by expanding around the fully phase-separated state \cite{Wurtz.2018,Demarchi.2023,Goychuk.2024}.
\section{Stationary Interfaces}
\label{app:stationaryInterfaces}
Analyzing the stability of stationary flat and constantly curved interfaces requires knowledge of the corresponding interface profiles.
In Sec.\@~\ref{sec:stationaryInterfaces}, we outlined how these profiles can be approximated by analytically solving the stationary equations Eq.\@~\eqref{eq:DenistyRescaled} and Eq.\@~\eqref{eq:ChemicalRescaled} in the sharp interface limit.
Here, we extend this analysis by providing additional details of the calculation and presenting the full analytical expressions omitted in the main text.

To determine stationary interface solutions, we follow the approach outlined in Secs.\@~\ref{sec:stationaryInterfaces} and\@~\ref{sec:stabilityAnalysis} and describe the interface using a global coordinate system where one unit vector, ${\boldsymbol{\hat r}_\perp}$, is perpendicular to the interface, pointing from the high- towards the low-density phase.
Thus, for a flat interface, ${\boldsymbol{r} = x \boldsymbol{\hat e}_x + z \boldsymbol{\hat e}_z}$, while for a curved interface, we use ${\boldsymbol{r} = r \boldsymbol{\hat e}_r}$.
Here, ${\boldsymbol{\hat e}_x}$ and ${\boldsymbol{\hat e}_z}$ denote the unit vectors in the horizontal and vertical directions, while ${\boldsymbol{\hat e}_r}$ represents the radial unit vector for the curved case.
As before, we collectively refer to the vertical coordinate as ${r_\perp=(z,r)}$.
We have shown, that in this coordinate frame, stationary interface profiles, are solutions to the one-dimensional problem defined in Eqs.\@~\eqref{eq:stationaryProblem}:
\begin{subequations}   
\label{app:stationaryProblem}
\begin{align}
    0 &= \partial_\perp\mu + Pe \, \phi \, \partial_\perp\psi \, , 
    \label{app:stationaryProblemDensity} 
    \\
    0 &= \boldsymbol{\nabla}^2_\perp\psi + \frac{1}{l^2}
    \big( 
    \phi-\psi
    \big) \, ,
    \label{app:stationaryProblemChemical}
\end{align}
\end{subequations}
where $\partial_\perp$ denotes the derivative along $\boldsymbol{\hat r}_\perp$, and ${\boldsymbol{\nabla}_\perp^2}$ is the corresponding Laplacian, taking into account the effects of curvilinear coordinates.
Without loss of generality, we set the interface position to ${r_\perp=X_0}$, corresponding to a flat interface positioned at $X_0$ or a spherical domain of radius $X_0$.

Assuming a separation of length scales, ${w\ll l}$ and ${w\ll R}$, the density field exhibits an almost discontinuous jump across the interface but is only weakly modulated within each of the high- and low-density bulk domains; cf.\@~Fig.\@~\ref{fig:interfaces}(a).
In contrast, the chemical field varies on a much larger scale~$l$, smoothly interpolating between the chemical equilibria ${\psi=\phi}$ in the bulk phases.
Thus, in each domain, we can approximate the interface profile by solving a piecewise linear version of equations Eq.\@~\eqref{app:stationaryProblemDensity} and Eq.\@~\eqref{app:stationaryProblemChemical}:
\begin{subequations}
    \label{app:stationaryProblemLinearized}
\begin{align}
    0 &=  2\partial_\perp \phi + Pe \, \phi_\pm \partial_\perp\psi \, , \\
    0 &= \boldsymbol{\nabla}_\perp^2 \psi + \dfrac{1}{l^2}\left(\phi-\psi\right) \, ,
\end{align}
\end{subequations}
where we linearized around the minima $\phi_\pm$ of the free energy, Eq.\@~\eqref{eq:freeEnergyDensity}, and neglected terms of the order ${\mathcal{O}(\boldsymbol{\nabla}^4)}$.
The solutions to the above equations in each of the domains need to be connected by proper boundary conditions at ${r_\perp=X_0}$.
Since ${l\gg 1}$, the chemical field $\psi$ can be assumed continuous across the interface, which yields
\begin{align}
    \psi\big\rvert_{r_\perp=X_0^\pm}=\psi_0 , \quad
    \partial_\perp\psi\big\rvert_{r_\perp=X_0^-} 
    = \partial_\perp\psi\big\rvert_{r_\perp=X_0^+} \,.
    \label{app:Continuity}
\end{align}
As outlined in Sec.\@~\ref{sec:stationaryInterfaces}, this implies that on the scale of the interface width $w$, chemotactic effects are negligible and the interface profile is determined exclusively by the equilibrium chemical potential, Eq.\@~\eqref{eq:chemicalPotential}.
Thus, the densities close to the interface have to obey the local Gibbs-Thompson relation, Eq.\@~\eqref{app:GibbsThomson}; see App.\@~\ref{app:GibbsThomsonSection}.
Consequently, for a flat interface the density at the interface is given by the minima of the free energy 
\begin{align}
    \phi\big\rvert_{r_\perp=X_0^\pm}=\phi_\mp \, ,
\end{align}
whereas for a spherical interface, local interface curvature introduces an additional shift in the concentrations
\begin{align}
    \phi\big\rvert_{r_\perp=X_0^\pm}=\phi_\mp + \frac{\gamma}{2\Delta\phi}\frac{1}{X_0}\, .
\end{align}
Upon introducing the auxiliary quantities ${k_\pm=Pe\, \phi_\pm/2}$ and the length scales ${l_\pm=l/\sqrt{1+k_\pm}}$, the solution in each domain is, thus, given by
\begin{align}
    \phi_\pm(r_\perp) &= 
    c_\pm- k_\pm \psi_\pm(r_\perp)\, , \\
    \psi_\pm(r_\perp) &= 
    \frac{c_\pm}{1+k_\pm} + a^\pm p_1\left(\frac{r_\perp}{l_\pm}\right) + b^\pm p_2\left(\frac{r_\perp}{l_\pm}\right) \,, 
\end{align}
where we defined the constant 
\begin{align}
    c_\pm &= \left(\phi_\pm + \frac{\gamma}{2\Delta\phi}H\right) + k_\pm \psi_0  \, ,
\end{align}
with ${H\in\{0,X_0^{-1}\}}$ denoting the local interface curvature; see Sec.\@~\ref{app:GibbsThomsonSection}.
The functions $p_{1/2}(r)$ are solutions to the linear problem ${\boldsymbol{\nabla}_\perp^2p=p}$ and depend on the curvature of the interface.
Using that ${\boldsymbol{\nabla}_\perp^2=\{\partial_z^2,\partial_r^2+(1/r)\partial_r\}}$, one finds
\begin{align}
    p_{1/2}(r)=
    \begin{cases}
        \exp(\pm\, r) \, , \quad &\mathrm{flat}\,, \\[2mm] I_0(r),\,  K_0(r) \, , \quad &\mathrm{curved}  \, ,
    \end{cases} 
\end{align}
with $I_n$ and $K_n$ denoting modified Bessel functions of the first and second kind, respectively.
For a curved interface, symmetry and differentiability at the origin require ${\partial_\perp\phi\big\rvert_{r_\perp=0}=\partial_\perp\psi\big\rvert_{r_\perp=0}=0}$ which yields ${b^+=0}$.
However, the prefactors $a^\pm$, and $b^\pm$ are not fully determined by the boundary conditions at the interface alone and additional constraints must be specified at the system boundaries ${r_\perp=0}$ and ${r_\perp=L_\perp}$.
To prevent mass flux across these boundaries, we impose no-flux conditions
\begin{align}
    \partial_\perp\phi\big\rvert_{r_\perp=\{0,L_\perp\}} 
    = \partial_\perp\psi\big\rvert_{r_\perp=\{0,L_\perp\}} 
    = 0 \, .
    \label{app:NoFluxConditions}
\end{align}
With that, the prefactors $a^\pm$, $b^\pm$, as well the chemical density $\psi_0$ at the interface can be determined analytically. 
However, we note that $X_0$ effectively sets the average density in the system through the self-consistency condition
\begin{align}
    \int_V \phi(r_\perp) \,  \mathrm{d}^dr
    = \bar \phi \int_V\mathrm{d}^dr \, ,
\end{align}
where $V$ denotes the system volume.
Since, in the absence of chemotactic interactions, the average density $\bar\phi$ is one of the primary determinants for the spinodal instability, we choose to fix $L_\perp$ and the average density while treating $X_0$ as a free parameter that must be determined self-consistently.
Due to the mathematical complexity of this approach and the length of the resulting expressions, we opt to solve the full problem numerically using Mathematica \cite{Inc..2024}. 
The approximate interface profile obtained in this way is shown in Fig.\@~\ref{fig:interfaces}(a) for a representative set of parameters, alongside the corresponding numerical solutions (bullets).

The above problem can be significantly simplified when ${l\ll L_\perp-X_0}$ and ${l\ll X_0}$.
In this limit, we may consider a (semi-)infinite system, where the boundary conditions, Eq.\@~\eqref{app:NoFluxConditions} are replaced by the requirement, that the solutions in each domain remain finite.
For a flat interface, this leads to ${b^+=a^-=0}$, with ${r_\perp \in [-\infty,\infty]}$ and ${X_0=0}$.
In contrast, for a spherical domain, keeping ${X_0}$ fixed and taking ${L_\perp \to \infty}$ enforces ${a^-=0}$.
Thus, in a (semi-)infinite system, applying the remaining conditions at the interface, one obtains
\begin{widetext}
\begin{align}
    a^+= - \frac{\Delta\phi}{1+k_++(l_-/l_+)(1+k_-)}\,, \quad
    b^-=  \frac{\Delta\phi}{1+k_-+(l_+/l_-)(1+k_+)}\,, \quad
    \psi_0 = \phi_+ - \Delta\phi\left(1+\dfrac{(1+k_-)l_-}{(1+k_+)l_+}\right)^{-1}\,,
    \label{app:interfaceFlatInfinite}
\end{align}
\begin{subequations}
\label{app:interfaceCurvedInfinite}
\begin{align}
    a^+&=  \frac{-\Delta\phi\, K_1(X_0/l_-)}{(1+k_+)I_0(X_0/l_+)K_1(X_0/l_-)+(1+k_-)(l_-/l_+)I_1(X_0/l_+)K_0(X_0/l_-)}\,, \\
    b^-&=   \frac{\Delta\phi\, I_1(X_0/l_-)}{(1+k_-)I_1(X_0/l_+)K_0(X_0/l_-)+(1+k_+)(l_+/l_-)I_0(X_0/l_+)K_1(X_0/l_-)} \,,\\
    \psi_0 &= \left(\phi_++\dfrac{\gamma}{2\Delta\phi}\frac{1}{X_0}\right) - 
    \Delta\phi\left(1+\dfrac{(1+k_-)l_-}{(1+k_+)l_+}\dfrac{I_1(X_0/l_+)K_0(X_0/l_-)}{I_0(X_0/l_+)K_1(X_0/l_-)}\right)^{-1}\,,
\end{align}
\end{subequations}
\end{widetext}
for the flat [Eq.\@~\eqref{app:interfaceFlatInfinite}] and spherical [Eq.\@~\eqref{app:interfaceCurvedInfinite}] interface, respectively.
From these expressions, one finds that in the limit of weak chemotactic interactions (${Pe\rightarrow 0}$), the chemical concentration $\psi_0$ at the interface reduces to ${\psi_0=\phi_c+ \gamma K/(2\Delta\phi)}$, corresponding to the inflection point of a symmetric equilibrium profile $\phi_I$.
This provides a valuable consistency check as it implies that for vanishing chemotactic interactions the field $\psi$ is in chemical equilibrium (${\phi=\psi}$), without perturbing the interface profile.
\section{Instability Criterion}
\label{app:InstabilityCriterion}
Having constructed stationary solutions, we now examine their stability against small perturbations of the interface position.
To this end, we revisit the qualitative arguments outlined in Sec.\@~\ref{sec:HeuristicAnalysis} of the main text and provide the detailed calculations omitted therein.

We consider a flat interface subjected to a weak sinusoidal modulation, as illustrated in Fig.\@~\ref{fig:intuition}(a).
In this setup, local interface curvature induces a shift in the phase-separating density $\phi$, as described by the Gibbs-Thomson relation,\@~\eqref{app:GibbsThomson}.
Similarly, the chemical concentration $\psi$ at the interface also depends on local curvature, since it represents an average of $\phi$ over a spatial region set by the reaction-diffusion length $l$.
Consequently, the chemical concentration at the interface deviates from its value for a perfectly flat interface in a curvature-dependent manner.

These curvature-induced inhomogeneities in $\phi$ and $\psi$ give rise to diffusive fluxes on both sides of the interface.
Depending on the direction of these fluxes, they may either act to restore the flat interface or amplify the perturbations.
The largest deviations in both $\phi$ and $\psi$ occur at points of maximal (positive or negative) interface curvature, which are spaced by half the wavelength of the imposed perturbation; see Fig.\@~\ref{fig:intuition}(a).
By estimating the shifts in density and chemical concentration at these extremal points, we can quantify the resulting fluxes and establish the heuristic instability criterion given in Eq.\@~\eqref{eq:HandwavingInstability}.

At the extremal points of the sine-like perturbation, the shift in density $\delta\phi$ takes values ${\delta\phi=\pm \gamma/(2\Delta\phi R)}$, with $R$ denoting the magnitude of the corresponding radius of curvature indicated in Fig.\@~\ref{fig:intuition}(a).
On each side of the interface, this culminates in a linearized current ${\boldsymbol{J}_\mathrm{eq}=-2M\boldsymbol{\nabla}\phi}$.
To estimate the magnitude of the density gradient, we approximate
\begin{align}
    \left|\boldsymbol{\nabla}\phi\right| \approx \dfrac{\delta \phi - (-\delta \phi)}{\Lambda}
    \approx \frac{\gamma}{\Lambda \Delta\phi R} \, ,
\end{align}
where $\Lambda$ is the lateral distance between the two points of maximal curvature.
Thus, the magnitude of the current on each side of the interface follows as
\begin{align}
    J^\pm_{\text{eq}} \approx \frac{2M\gamma}{\Lambda\Delta\phi R} \, .
    \label{app:EqCurrents}
\end{align}
To also estimate the magnitude of currents resulting from curvature-induced shifts in chemical concentration, we consider a single high-density ($\phi_+$) domain of radius $R$ immersed in an infinite background of low density ($\phi_-$).
As a reference, we compare this scenario to a semi-infinite system where a high-density domain of linear extension $R$ is separated from an infinite low-density phase by a flat interface; see Fig.\@~\ref{fig:intuition}(b).
To isolate curvature-induced effects, we approximate the density profile $\phi$ as a step function transitioning between the equilibrium densities $\phi_\pm$ and solve only the corresponding stationary equation for the chemical field:
\begin{align}
    0 = \boldsymbol{\nabla}^2\psi + \frac{1}{l^2}\left(\phi_\pm-\psi\right) \, .
\end{align}
Requiring continuity of the chemical concentration across the interface and no-flux boundary conditions on the finite half-domain, we obtain the piecewise solution for the flat interface:
\begin{align}
    \psi(r_\perp)&=
    \begin{cases}
        \phi_+ -  \psi_1 \dfrac{\cosh(z/l)}{\cosh(R/l)} \, , \quad &z<R\\[4mm]
        \phi_- + \psi_2 \exp\left((R-z)/l\right) \, , \quad &z>R
    \end{cases}\,,
    \\[3mm]
    \psi_1 &= \dfrac{\Delta\phi}{1+\tanh(R/l)} \, , \ \psi_2 = \dfrac{\Delta\phi \, \tanh(R/l)}{1+\tanh(R/l)}
\end{align}
where $z$ denotes the coordinate perpendicular to the flat interface.
For the spherical domain, the radial symmetry of the problem implies that the chemical concentration $\psi(r)$ depends only on the radial coordinate $r$.
This immediately leads to the boundary condition ${\partial_r\psi\big\rvert_{r=0}=0}$, as any nonzero gradient at the center would introduce a discontinuity in $\psi$.
Solving for $\psi(r)$ in this case, we obtain
\begin{align}
    \psi(r)&=
    \begin{cases}
        \phi_+ -  \psi_1 \dfrac{I_0(r/l)}{I_0(R/l)}\, , \quad &r<R\\[4mm]
        \phi_- +  \psi_2 \dfrac{K_0(r/l)}{K_0(R/l)} \, , \quad &r>R
    \end{cases}\,,
    \\[3mm]
    \psi_1 &= \Delta\phi\left(1+\frac{K_0(R/l)}{K_1(R/l)}\frac{I_1(R/l)}{I_0(R/l)}\right)^{-1} \,, \\[3mm]
    \psi_2 &= \Delta\phi\left(1+\frac{K_1(R/l)}{K_0(R/l)}\frac{I_0(R/l)}{I_1(R/l)}\right)^{-1} \,.
\end{align}
Thus the difference $\delta\psi$ of chemical concentration at the interface between a spherical and flat domain reads
\begin{align}
    \delta\psi
    &= \frac{\Delta\phi}{1+\tanh(R/l)} - \dfrac{\Delta\phi}{1+ \frac{K_0(R/l)}{K_1(R/l)}\frac{I_1(R/l)}{I_0(R/l)}} \notag \\
    &\approx -\frac{\Delta\phi}{4} \frac{l}{R} + \mathcal{O}\left(\frac{l^2}{R^2}\right) \, ,
\end{align}
which provides an estimate for the shift in chemical concentration at the extremal point of the outward-bending section of the interface; see Fig.\@~\ref{fig:intuition}(c).
We can repeat the same analysis for the inward-bending region by considering a low-density domain immersed in a high-density background.
In the preceding calculation, this corresponds to replacing ${\Delta\phi\rightarrow-\Delta\phi}$.
Consequently, the curvature-induced shift in chemical concentration at the extremal point of the inward-bending region has the opposite sign compared to that of the outward-bending region.
As a result, we can estimate magnitude of the gradient in chemical concentration between these two extremal points to be approximately
\begin{align}
    \left|\boldsymbol{\nabla}\psi\right| \approx \frac{\Delta\phi \,l}{2\Lambda R}\,. 
\end{align}
That is, for the the fluxes in both domains, one finds
\begin{align}
    J_\mathrm{ch}^\pm = M Pe\,\phi_\pm  \frac{\Delta\phi\,l}{2\Lambda R} 
    \, .
\label{app:ChemCurrents}
\end{align}
As outlined in Sec.\@~\ref{sec:HeuristicAnalysis}, the equilibrium currents, Eq.\@~\eqref{app:EqCurrents}, act to restore the flat interface, whereas the chemotactic currents, Eq.\@~\eqref{app:ChemCurrents}, promote further growth of the perturbations.
We conclude, that an initially flat interface is unstable against weak perturbations if the strength of the combined chemotactic currents on both sides of the interface exceeds that of the equilibrium currents
\begin{align}
    J_\mathrm{ch}^++J_\mathrm{ch}^- > J_\mathrm{eq}^++J_\mathrm{eq}^-  \, .
\end{align}
Using the expression for the surface tension in rescaled coordinates, Eq.\@~\eqref{app:SurfaceTensionRescaled}, finally yields the heuristic instability criterion given in Eq.\@~\eqref{eq:HandwavingInstability}.
\section{Stability Analysis}
\label{app:StabilityAnalysis}
The previous section provided a qualitative argument for how chemotactic fluxes can destabilize an initially flat interface, leading to the heuristic instability criterion, Eq.\@~\eqref{eq:HandwavingInstability}.
In this section, we revisit the more quantitative framework developed in section \ref{sec:stabilityAnalysis}, providing the derivation for the case of a spherical interface as well as explicit expressions omitted in the main part of this manuscript.

\subsection{Interface Dynamics}
To investigate the interplay of surface tension and non-equilibrium signaling, we need to parametrize the interface position $\boldsymbol{X}$.
As for the construction of stationary interface solutions, we choose a coordinate frame where the local unit vectors are perpendicular ($\boldsymbol{\hat r}_\perp$) and parallel ($\boldsymbol{\hat r}_\parallel$) to the unperturbed interface. 
For a flat interface located at ${r_\perp=X_0}$ and Cartesian coordinates this yields, 
\begin{align}
    \boldsymbol{X}(r_\parallel,t)
    =
    r_\parallel \, \boldsymbol{\hat{r}}_\parallel + 
    X(r_\parallel,t)\,\boldsymbol{\hat r}_\perp\,,
    \label{app:interfacePositionFlat}
\end{align}
where we assume small perturbations ${X(r_\parallel,t)=X_0+\delta X(r_\parallel,t)}$ and no overhangs.
For a spherical domain of radius $X_0$, the interface is parametrized as 
\begin{align}
    \boldsymbol{X}(r_\parallel,t)
    =
    X(r_\parallel,t) \,\boldsymbol{\hat r}_\perp\,,
    \label{app:interfacePositionCurved}
\end{align}
with $r_\perp$ and $r_\parallel$ denoting the radial and azimuthal directions, respectively, and ${X(r_\parallel,t)=X_0+\delta X(r_\parallel,t)}$.
With this choice, the interface profiles ${\boldsymbol{\Phi}=(\phi,\psi)^T}$ can be decomposed as
\begin{align}
    \boldsymbol{\Phi}(\boldsymbol{r},t) 
    = 
    \begin{pmatrix}
        \phi_I\left(r_\perp-\delta X(r_\parallel,t)\right) \\[2mm] \psi_I\left(r_\perp-\delta X(r_\parallel,t)\right)
    \end{pmatrix}
    + 
    \begin{pmatrix}
        \delta\tilde\phi(\boldsymbol{r},t) \\[2mm] \delta\tilde \psi(\boldsymbol{r},t)
    \end{pmatrix} \,,
\end{align}
where ${\boldsymbol{\Phi}_I=(\phi_I,\psi_I)^T}$ denotes the solutions to the stationary profile equations Eq.\@~\eqref{eq:stationaryProblem}, for a flat and curved interface, respectively.
As shown in Sec.\@~\ref{sec:stabilityAnalysis}, for weakly perturbed interfaces, this yields $\boldsymbol{\phi}=\boldsymbol{\phi}_I+\boldsymbol{\delta\phi}$, where 
\begin{align}
    \boldsymbol{\delta \Phi}(\boldsymbol{r},t)  = -\partial_\perp \boldsymbol{\Phi}_I(r_\perp) \,\delta X(r_\parallel,t) + \boldsymbol{\delta\tilde\Phi}(\boldsymbol{r},t) \,.
    \label{app:PerturbationAnsatz}
\end{align}
That is, the interface profile can be decomposed into a translation along the vertical direction and a small curvature-dependent shift in concentrations.
To proceed, we need an equation that relates the dynamics of the latter with that of the interface perturbations\@~$\delta X$.
To that end, we recall that the time evolution of the interface position can be locally decomposed as
\begin{align}
    \partial_t \boldsymbol{X}(r_\parallel,t) 
    = v_\parallel \boldsymbol{\hat{e}}_\parallel+ v_\perp \boldsymbol{\hat{e}}_\perp \, , 
    \label{app:InterfaceDecomposition}
\end{align}
where ${\boldsymbol{\hat{e}}_\parallel}$ and ${\boldsymbol{\hat{e}}_\perp}$ denote the local unit vectors parallel and perpendicular to the interface.
These can be expressed as
\begin{align}
    \boldsymbol{\hat{e}}_\parallel &= 
    \frac{G(X)\boldsymbol{\hat r}_\parallel + \left(\partial_\parallel X\right) \,\boldsymbol{\hat r}_\perp}{\sqrt{G(X)^2+\left(\partial_\parallel X\right)^2}} \\
    \boldsymbol{\hat{e}}_\perp &= 
    \frac{G(X)\boldsymbol{\hat r}_\perp-\left(\partial_\parallel X\right) \boldsymbol{\hat r}_\parallel}{\sqrt{G(X)^2+\left(\partial_\parallel X\right)^2}} \,,
\end{align}
where $\partial_\parallel$ denotes a derivative along $\boldsymbol{\hat r}_\parallel$ and ${G(X)=1}$ for a flat and ${G(X)=X}$ for a curved interface, respectively.
At the same time, taking a time derivative of the explicit parametrizations, Eq.\@~\eqref{app:interfacePositionFlat} and Eq.\@~\eqref{app:interfacePositionCurved}, yields 
\begin{align}
    \partial_t \boldsymbol{X}(r_\parallel,t) = 
    \partial_t X \, \boldsymbol{\hat r}_\perp\,.
    \label{app:timeDerInterface}
\end{align}
By equating Eq.\@~\eqref{app:InterfaceDecomposition} and Eq.\@~\eqref{app:timeDerInterface} and projecting onto $\boldsymbol{\hat r}_\parallel$, we obtain
\begin{align}
    v_\parallel \,G(X) -v_\perp\, \partial_\parallel X(r_\parallel,t) = 0 \,.
\end{align}
Projecting Eq.\@~\eqref{app:timeDerInterface} onto $\boldsymbol{\hat r}_\perp$ leads to the constitutive relation that governs the time evolution of the interface
\begin{align}
    \partial_t \delta X(r_\parallel,t) 
    = v_\perp\frac{\sqrt{G(X)^2+\left(\partial_\parallel X\right)^2}}{G(X)}  \, .
    \label{app:interfaceDerivativeI}
\end{align}
At the same time, mass conservation of the phase-separating density constrains the interface velocity via
\begin{align}
     v_\perp = 
     \frac{\boldsymbol{\hat e}_\perp \left (\boldsymbol{j}_\mathrm{in} - \boldsymbol{j}_\mathrm{out}\right)}{\Delta_\phi} \, ,
\end{align}
where $\boldsymbol{j}_\mathrm{in/out}$ denote the currents on either side of the interface which can be inferred from Eqs.\@~\eqref{eq:DenistyRescaled} and\@~\eqref{eq:ChemicalRescaled} and $\Delta_\phi$ is the difference in the plateau densities.
Together with Eq.\@~\eqref{app:interfaceDerivativeI}, this provides the desired relation linking the local interface geometry ($\delta X$) towards the densities at the interface.
For the steady state solution $\boldsymbol{\Phi}$, the net flux across the interface must vanish, implying that ${\boldsymbol{j}_\mathrm{in} - \boldsymbol{j}_\mathrm{out}}$ is at least of order $\mathcal{O}(\delta X)$.
That is, to leading order in the perturbation $\delta X$ one finds
\begin{align}
    \partial_t \delta X(r_\parallel,t) =
    \frac{\boldsymbol{\hat r}_\perp \left (\boldsymbol{j}_\mathrm{in} - \boldsymbol{j}_\mathrm{out}\right)}{\Delta_\phi} \, ,
    \label{app:LinearTimeEvolutionInterface}
\end{align}
for both, the flat as well as the curved interface.
To obtain a closed set of dynamical equations, it remains to derive expressions for the currents $j_\mathrm{in/out}$ to leading order in the interface perturbations $\delta X$.
\subsection{Sharp interface limit}
Within the sharp interface limit, and to linear order in\@~$\delta X$, we have shown in Eq.\@~\eqref{eq:linearizedAll}  that deviations from the stationary interface profiles $\boldsymbol{\delta\phi}$ [Eq.\@~\eqref{app:PerturbationAnsatz}] obey a time evolution of the form
\begin{subequations}
    \label{app:linearizedAll}
\begin{align}
    \partial_t \delta\phi(\boldsymbol{r},t) 
    &= M_\pm \boldsymbol{\nabla}^2 \delta\phi + K_\pm \boldsymbol{\nabla}^2 \delta\psi\,, \label{app:linearizedDensity} \\
    \partial_t \delta\psi(\boldsymbol{r},t) 
    &= \boldsymbol{\nabla}^2 \delta\psi + \frac{1}{l^2} \left(\delta\phi - \delta\psi\right)  \, ,\label{app:linearizedChemical}
\end{align}
\end{subequations}
where we have expanded around the corresponding minima of the free energy density [Eq.\@\eqref{eq:freeEnergyDensity}] in both the high- and low-density domains.
Strictly speaking, this involves a double expansion where not only the amplitude of the interface undulations but also the deviation between the equilibrium and non-equilibrium plateau densities is considered small.
Moreover, we have neglected terms of order $\mathcal{O}(\boldsymbol{\nabla}^4)$, assuming only weakly modulated perturbations inside both domains.
Under these approximations, the linearized currents take the form
\begin{align}
    \boldsymbol{j}_\pm= 
    - M_\pm \boldsymbol{\nabla}\delta\phi(\boldsymbol{r},t) 
    - K_\pm \boldsymbol{\nabla}\delta\psi(\boldsymbol{r},t)\,.
    \label{app:linearCurrents}
\end{align}
To determine their values at the interface, we need to solve equations Eq.\@~\eqref{app:linearizedDensity} and Eq.\@~\eqref{app:linearizedChemical}.
To that end, we employ an ansatz of the form 
\begin{align}
    \label{app:Ansatz}
    \begin{pmatrix}
        \delta \phi(\boldsymbol{r},t) \\[1mm]
        \delta \psi(\boldsymbol{r},t) \\[1mm]
        \delta X(r_\parallel,t)
    \end{pmatrix}
    = 
    \varepsilon_k \exp(\sigma_k t) \, p_k(r_\parallel)
    \begin{pmatrix}
        \delta \phi(r_\perp) \\[1mm]
        \delta \psi(r_\perp) \\[1mm]
        1
    \end{pmatrix}\, ,
\end{align}
where, to simplify notation, we use the same symbols $(\delta\phi,\delta\psi)$ for the full perturbation $(\delta\phi(\boldsymbol{r},t),\delta\psi(\boldsymbol{r},t))$ and the single argument function $(\delta\phi(r_\perp),\delta\phi(r_\perp))$.
Wherever necessary, we will explicitly distinguish between these two cases.
The basis functions $p_k$ are given by ${p_k=\cos(kr_\parallel)}$ for both the flat as well as the spherical case.
For a flat interface, periodicity in the parallel direction requires ${k=2\pi n/L_\parallel}$, ${n\in \mathbb{Z}\setminus\{0\}}$whereas for the spherical case, we use azimuthal functions ${p_k=\cos(k r_\parallel)}$ with $r_\parallel$ denoting the azimuthal angle and integer values ${k\geq 2}$.
Substituting this ansatz into the linearized dynamical equations, Eq.\@~\eqref{app:linearizedAll}, leaves us with a problem of the form
\begin{widetext}
\begin{align}
    0 =
    \begin{pmatrix}
          \sigma_k - M_\pm  L(k,\partial_\perp) & -K_\pm  L(k,\partial_\perp) \\[2mm]
         -l^{-2} &  \sigma_k - L(k,\partial_\perp) + l^{-2}
    \end{pmatrix} 
    \begin{pmatrix}
        \delta \phi(r_\perp) \\[2mm]
        \delta \psi(r_\perp) 
    \end{pmatrix}
    \, , \quad
    L(k, \partial_\perp) = 
    \begin{cases}
        -k^2 + \partial_\perp^2 , \, &\mathrm{flat} \\[2mm]
        -\left(\frac{k}{r_\perp}\right)^2 + \partial_\perp^2 + \frac{1}{r_\perp} \partial_\perp , \, &\mathrm{curved}
    \end{cases}\,,
    \label{app:linearProblem}
\end{align}
\begin{align}
    \begin{pmatrix}
        \delta \phi(r_\perp) \\[2mm]
        \delta \psi(r_\perp) 
    \end{pmatrix}
    =
   \begin{pmatrix}
        h_\pm c_1(k, q^1_\pm r_\perp) + i_\pm c_1(k, q^1_\pm r_\perp) + j_\pm c_2(k, q^2_\pm r_\perp) +k_\pm c_2(k, q^2_\pm r_\perp) \\[2mm]
        r_\pm c_1(k, q^1_\pm r_\perp) + m_\pm c_1(k, q^1_\pm r_\perp) + n_\pm c_2(k, q^2_\pm r_\perp) +o_\pm c_2(k, q^2_\pm r_\perp)
    \end{pmatrix}
    \, .
    \label{app:Solution}
\end{align}
\end{widetext}
The corresponding solution shown in Eq.\@~\eqref{app:Solution} is a superposition of eigenmodes to the linear problem, Eq.\@~\eqref{app:linearProblem}, where the functions ${c_{1/2}(k,r_\perp)}$ are eigenfunctions of the (linear) operator~${L(k,\partial_\perp)}$,
\begin{align}
    L(k,\partial_\perp)\, c_{1/2}(qr_\perp) = \lambda_k(q)\, c_{1/2}(qr_\perp) \, ,
\end{align}
with $\lambda_k(q)$ denoting the corresponding eigenvalue.
Depending on the interface curvature, one finds
\begin{align}
    c_{1/2}(k,r_\perp) =
    \begin{cases}
        \exp(\pm r_\perp) \, , \, &\mathrm{flat} \\[2mm]
        I_k(r_\perp), K_k(r_\perp) \, , \, &\mathrm{curved}\, 
    \end{cases}\,,
\end{align}
with $I_k$ and $K_k$ denoting modified Bessel functions of the first and second kind, respectively.
Substituting these expressions into Eq.\@~\eqref{app:linearProblem}, the coupled differential equations reduce to an algebraic problem
\begin{align}
     0 =
    \begin{pmatrix}
          \sigma_k - M_\pm  \lambda_k(q^\pm) & -K_\pm  \lambda_k(q^\pm) \\[2mm]
         -l^{-2} &  \sigma_k - \lambda_k(q^\pm) + l^{-2}
    \end{pmatrix} 
    \begin{pmatrix}
        v_1^\pm \\[2mm]
        v_2^\pm 
    \end{pmatrix} \,,
\end{align}
where $v_1^\pm$ ($\phi$ direction) and $v_2^\pm$ ($\psi$ direction) are the components of the eigenvector $\boldsymbol{v}$ with eigenvalue zero.
The inverse length scales $q^\pm_{1/2}$ appearing in  Eq.\@~\eqref{app:Solution} follow from the requirement that the determinant of the above matrix vanishes, with the subscripts $1/2$ indicating the two possible solutions.
Similarly, the parameters $r_\pm$, $m_\pm$, $n_\pm$, and $o_\pm$ must be chosen such that the solution remains a valid eigenvector of the problem.
Thus, in the high- (``$+$''), as well as in the low-density domain (``$-$'') four independent prefactors $h_\pm$, $i_\pm$, $j_\pm$, and $k_\pm$ remain, which must be determined by appropriate boundary conditions.
To this end,  we follow the same approach as in Sec.\@~\ref{app:stationaryInterfaces}:
For systems with finite extension along the direction perpendicular to the interface, we impose no flux boundary conditions at the system boundaries, whereas for infinitely sized systems it suffices to demand finite solutions at infinity.
At the interface itself, we employ the same type of boundary conditions as in the static case (see Sec.\@~\ref{app:stationaryInterfaces}), however, corrected for local deformations of the interface.

In Sec.\@~\ref{sec:stabilityAnalysis}, we established that the continuity of the chemical field across the sharp interface imposes the conditions
\begin{subequations}
\label{app:BCChemical}
    \begin{align}
    \psi \big\rvert_{r_\perp=X^-} 
    &= \psi \big\rvert_{r_\perp=X^+} 
    \, , \\
    \partial_\perp \psi \big\rvert_{r_\perp=X^-} 
    &= \partial_\perp \psi \big\rvert_{r_\perp=X^+}
    \, ,
    \end{align} 
\end{subequations}
which applies to the full interface profile ${\psi=\psi_I+\delta\psi}$; see Eq.\@~\eqref{app:PerturbationAnsatz}.
Recalling that ${X(r_\parallel,t)=X_0+\delta X(r_\parallel,t)}$ and expanding to leading order in the interface deformations $\delta X$, this reduces to 
\begin{subequations}
\label{app:BCChemicalExpanded}
\begin{align}  
    \delta \psi \big \rvert_{r_\perp=X_0^-} &= \delta \psi \big \rvert_{r_\perp=X_0^+} \, ,\\
    \left(\partial_\perp \delta \psi  + \partial_\perp^2 \psi_I\right)\big\rvert_{r_\perp=X_0^-}  &= \left(\partial_\perp \delta \psi  + \partial_\perp^2 \psi_I\right)\big\rvert_{r_\perp=X_0^+} \, ,
\end{align}
\end{subequations}
where we have used our ansatz, Eq.\@~\eqref{app:Ansatz}, to eliminate the dependence on $r_\parallel$ and $\delta X$.
That is, the above condition applies to the single argument function $\delta\psi(r_\perp)$.
For the density field $\phi$, the local Gibbs-Thomson relation, Eq.\@~\eqref{app:GibbsThomson}, requires
\begin{align}
    \phi \big\rvert_{r_\perp=X^\mp}
    = \phi_\pm + \frac{\gamma}{2\Delta\phi} H\,.
\end{align}
Expanding once more in the interface deformations yields
\begin{align}
     \partial_\perp \phi_I \delta X\big\rvert_{X_0^\pm} + \delta \phi \big\rvert_{X_0^\pm} = \frac{\gamma}{2\Delta\phi} \left(H -H_0\right)\, ,
\end{align}
where $H$ denotes the local interface curvature and $H_0$ is the curvature of the stationary interface.
At this stage, the dependence on $\delta X$ has not yet been eliminated.
Substituting our ansatz, Eq.\@~\eqref{app:Ansatz}, along with the expressions for the local curvature of a weakly perturbed interface, Eq.\@~\eqref{app:LocalGTFlat} and Eq.\@~\eqref{app:LocalGTSpherical}, the above relation simplifies to
\begin{align}
    \label{app:BCDensityExpanded}
    \partial_\perp \phi_I\big\rvert_{r_\perp=X_0^\pm} + \delta \phi \big\rvert_{r_\perp=X_0^\pm} = \frac{\gamma}{2\Delta\phi} \tilde H\, ,
\end{align}
where ${\tilde H = k^2}$ for a flat interface, ${\tilde H = (k^2-1)/X_0^2}$ for a spherical interface, and $\delta\phi$ denotes the single argument function $\delta\phi(r_\perp)$.
In total, the four boundary conditions at the interface, Eqs.\@~\eqref{app:BCChemicalExpanded},\@~\eqref{app:BCDensityExpanded}, together with the conditions at the system boundaries, comprise eight independent constraints for the eight unknown parameters $h_\pm$, $i_\pm$, $j_\pm$, and $k_\pm$.
This yields a unique time-dependent solution for the linearized dynamics, Eq.\@~\eqref{app:linearizedAll}, with only $\sigma_k$ remaining a free parameter.
\subsection{Self-consistency relation}
Having determined solutions for the fields $\delta\phi$ and $\delta\psi$, Eq.\@~\eqref{app:Ansatz} and Eq.\@~\eqref{app:Solution}, provides us with expressions for the linearized currents, Eq.\@~\eqref{app:linearCurrents}, which govern the time evolution of the interface via Eq.\@~\eqref{app:LinearTimeEvolutionInterface}.
Using our ansatz, Eq.\@~\eqref{app:Ansatz}, we, thus, obtain a self-consistency condition
\begin{align}
    \Delta_\phi \, \sigma_k =
    &- \left[M_+ \partial_\perp \delta \phi
    + K_+ \partial_\perp \delta \psi \right]_{r_\perp=X_0^-} \notag \\
    &+ \left[M_- \partial_\perp \delta \phi
    + K_- \partial_\perp \delta \psi\right]_{r_\perp=X_0^+} \, .
    \label{app:selfConsistencyRelation}
\end{align}
Since the right-hand side of this equation implicitly depends on $\sigma_k$, the above problem represents a highly non-linear equation, that can only be solved numerically.

In summary, the analysis of the stability of an initially flat or curved interface follows a three-step procedure:
(i) In the first step, we determine approximate stationary interface profiles $\boldsymbol{\Phi}_I$, whose derivative appears in the boundary conditions, Eq.\@~\eqref{app:BCChemicalExpanded} and Eq.\@~\eqref{app:BCDensityExpanded}. The details of this calculation are provided in Sec.\@~\ref{app:stationaryInterfaces}.
(ii) In the second step, we consider small undulations of the stationary interface, which result in deviations from the stationary interface profiles. By solving the linearized dynamics, Eq.\@~\eqref{app:linearizedAll}, we determine these deviations for a given configuration of the interface.
(iii) In the final step, using Eq.\@~\eqref{app:LinearTimeEvolutionInterface}, we link the shape of the interface profile to the dynamics of the interface position, resulting in the self-consistency relation Eq.\@~\eqref{app:selfConsistencyRelation}. Solving this equation for a given mode $k$ yields the growth rate $\sigma_k$, from which we may infer the stability of the undeformed interface against perturbations of the form $p_k$; see Eq.\@~\eqref{app:Ansatz}.

While the first two steps involve solving a set of linear algebraic equations, the third step can only be tackled numerically.
However, due to the complexity of the problem and the number of equations involved, in most cases, providing analytical solutions for steps (i) and (ii) is not only cumbersome but also offers only limited additional insight.
Therefore, we address all three steps using a combined Mathematica\@~\cite{Inc..2024, ZN} notebook.
To determine the dispersion relation $\sigma_k$ from the consistency condition Eq.\@~\eqref{app:selfConsistencyRelation}, we employ a Newton-Raphson algorithm, utilizing Mathematica's built-in``FindRoot'' function.
Note that the dispersion relation always has two branches from which we are only interested in the one with the larger real part, as it determines the onset of the instability.
Thus, to ensure convergence to the larger branch, we initialize the Newton-Raphson procedure with several different (imaginary) starting values, only retaining the largest solution.
\section{Stability of an equilibrium interface}
\label{app:modelBDispersion}
As noted above, for the theoretical data presented in this manuscript, we primarily analyze the stability of a flat or constantly curved interface using Mathematica and do not explicitly calculate the expressions for the stationary, as well as perturbed interface profiles. However, to demonstrate the procedure outlined above, in this section, we provide analytical solutions for each step in the limit of vanishing chemotactic interactions.

For simplicity, we consider an infinitely sized system with a flat interface positioned at ${X_0=0}$:
In the first step, we determine the stationary interface profile $\phi_I$, which, in this case, is particularly simple:
For vanishing chemotactic interactions, the stationarity condition, Eq.\@~\eqref{app:stationaryProblemDensity}, reduces to ${\partial_\perp \mu=0}$, where $\mu$ denotes the chemical potential [Eq.\@~\eqref{app:rescaledChemicalPotential}]. 
The exact solution to this equation is provided in Eq.\@~\eqref{app:1dInterface}, which, in our rescaled units, reads
\begin{align}
    \phi(r_\perp) = \phi_+ - \Delta\phi\left( \frac{1+\tanh(r_\perp)}{2}\right) \, .
\end{align}
In the sharp interface limit, this expression reduces to a step function where ${\phi=\phi_\pm}$ for ${r_\perp \lessgtr 0}$, respectively.
We could have obtained the same result solving the linearized problem, Eq.\@~\eqref{app:stationaryProblemLinearized}, which, in this limit, simplifies to ${\boldsymbol{\nabla}_\perp^2\delta\phi=0}$ in the high- as well as the low-density domain and only allows for a constant solution in each domain.

In the second step, we consider the deviations from the stationary interface profile for a slightly perturbed interface with undulations of the form ${\delta X= \varepsilon_k \cos(k r_\parallel)}$ as described in Eq.\@~\eqref{app:Ansatz}.
This requires solving the linearized problem in Eq.\@~\eqref{app:linearProblem}, which in our case simplifies to
\begin{align}
    \left[\sigma_k - M_\pm (-k^2+\partial_\perp^2) \right] \delta\phi(r_\perp)=0 \,.
\end{align}
Applying the boundary conditions specified in Eq.\@~\eqref{app:BCDensityExpanded} and requiring that the solution remains finite at infinity, one finds
\begin{align}
    \delta \phi(r_\perp) =
    \begin{cases}
        \dfrac{\gamma k^2}{2 \Delta\phi} \, \exp(q_+ r_\perp) \, , \quad &r_\perp < 0\,,\\[2mm]
        \dfrac{\gamma k^2}{2 \Delta\phi} \, \exp(-q_- r_\perp) \, , \quad &r_\perp > 0 \, ,
    \end{cases}
\end{align}
where we have introduced the inverse length scales ${q_\pm=(\sigma/M_\pm+k^2)^{1/2}}$.
Thus, inside the bulk phases, the curvature-induced shifts $\delta\phi$ relax exponentially.

In the final step, we use these solutions to solve the self-consistency relation, Eq.\@~\eqref{app:selfConsistencyRelation}, which, in the absence of chemotactic interactions, reads
\begin{align}
    \Delta \phi \, \sigma_k = -\dfrac{\gamma k^2}{2 \Delta\phi} \left(M_+ q_+ +M_- q_-\right)\,.
\end{align}
Using that for our system ${M_+=M_-=2M}$ (see Sec.\@~\ref{sec:stabilityAnalysis}) and introducing the auxiliary variable ${x_k=\sigma_k/2M}$, we obtain
\begin{align}
    x_k = - \frac{\gamma k^2}{\Delta \phi^2} \sqrt{x_k+k^2} \, .
\end{align}
Since the surface tension ${\gamma > 0}$, it follows that ${x_k<0}$ for all modes $k$.
Therefore, the only solution to the above equation is
\begin{align}
    x_k 
    &= \frac{\gamma^2k^4}{2\Delta \phi^4} \left(1-\sqrt{1+4\frac{\Delta\phi^4}{\gamma^2 k^2}}\right) \notag \\
    &= - \frac{\gamma}{\Delta\phi^2}k^3 + \mathcal{O}(k^4) \, ,
\end{align}
where in the second line we used an expansion to leading order in $k$.
In other words, for large wavelength ${k\rightarrow 0}$, a flat equilibrium interface has a dispersion relation given by
\begin{align}
    \sigma_k = - \frac{2\gamma M}{\Delta\phi^2} k^3 \, ,
\end{align}
which is strictly negative for all ${k>0}$.
Consequently, we find that in the absence of chemotactic interactions a weakly perturbed interface always relaxes back to the stationary (flat) solution, with a time evolution governed by
\begin{align}
    \partial_t \delta X(r_\parallel,t) = - \frac{2\gamma M}{\Delta\phi^2} k^3 \, \delta X(r_\parallel,t) \, .
\end{align}
This is nothing else but the classical capillary wave relation of an equilibrium phase-separating system (model B) \cite{Bray.2001}.
\begin{figure*}
    \centering
    \includegraphics[]{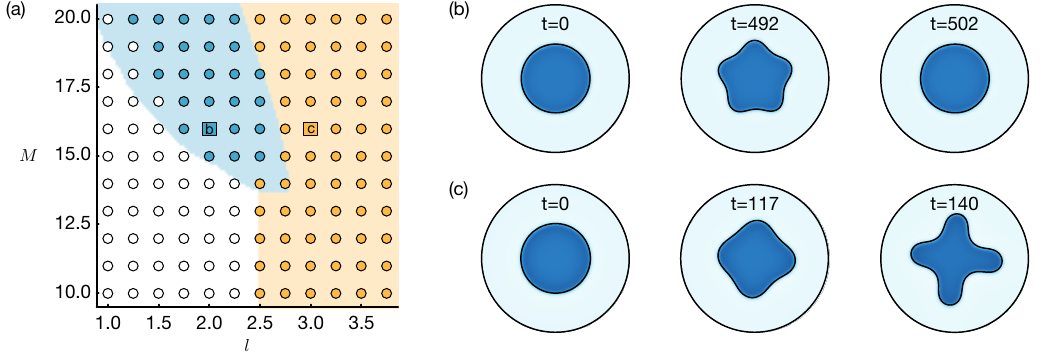}
    \caption{\textbf{Stability of a spherical interface} (a) Bifurcation diagram obtained from numerical simulations (bullets) and linear stability analysis of small undulations of a spherical interface (background color) in a two-dimensional spherical system of radius ${L_\perp=35}$, for a range of values of $M$ and $l$. 
    All simulations were initialized with the corresponding stationary profiles obtained from solving Eq.\@~\eqref{eq:stationaryProblem} and evolved using a finite element scheme with no-flux conditions at the system boundary (App.\@~\ref{app:Numerics}). 
    Depending on the observed behavior, each interface solution is categorized as stable (white bullets) if the droplet remains stationary or unstable (yellow/light gray bullets) if the droplet deforms (cf.\@~subpanel c). 
    If the interface exhibits traveling capillary waves or temporally modulated amplitudes (subpanel b), the perturbation is categorized as oscillatory (blue/gray bullets).
    The theoretical bifurcation diagram (background color) is constructed by evaluating the dispersion relation $\sigma_k$ obtained from numerically solving the self-consistency relation, Eq.\@~\eqref{eq:ConsistencyCondition}. 
    Analogously to the case of a flat interface (Fig.\@~\ref{fig:phaseDiagram}), the interface is classified as stable when Re${(\sigma_k)<0}$ (white), unstable when Re${(\sigma_k)>0}$ and ${k_\mathrm{osc}>k_\mathrm{max}}$ (yellow/light gray) or oscillatory when Re${(\sigma_k)>0}$ and ${k_\mathrm{osc}<k_\mathrm{max}}$ (blue/gray), cf.\@~\ref{sec:PhaseDiagram}. 
    (b--c) Time evolution of a spherical interface (black line), separating the high- (dark blue/gray) from the low-density (light blue/gray) domain for parameters corresponding to the square markers in sub-panel (a).
    The remaining parameters used for all panels are ${Pe=0.5}$, ${\phi_c=0.7}$, and ${\bar\phi=0.55}$.}
    \label{figAPP:phaseDiagramCurved}
\end{figure*}
\section{Bifurcation diagram of a spherical interface}
\label{app:CurvedInterfaces}
While the main part of this manuscript focuses on flat interfaces, in this section we briefly discuss the case of a spherical interface.
To this end, we follow the same approach as in Sec.\@~\ref{sec:PhaseDiagram}, constructing the bifurcation diagram for a curved interface and comparing the results with numerical simulations. 
The results of this analysis are shown in Fig.\@~\ref{figAPP:phaseDiagramCurved}(a), with two representative time evolutions of the initially spherical interface solution shown in Fig.\@~\ref{figAPP:phaseDiagramCurved}(b--c).

Analogous to the flat-interface case, the stationary interface profile obtained by solving the profile equations, Eq.\@~\eqref{eq:stationaryProblem}, is classified as stable if, in the numerical simulation, the droplet remains stationary and unstable if it deforms [Fig.\@~\ref{figAPP:phaseDiagramCurved}(c)].
If the interface exhibits traveling capillary waves [as in Fig.\@~\ref{fig:GearsAndBelts}(b)] or temporally modulated amplitudes [Fig.\@~\ref{figAPP:phaseDiagramCurved}(b)], the emerging perturbation is categorized as oscillatory. 
As before, we compare the observed behavior with the predictions from our theoretical stability analysis detailed in Sec.\@~\ref{app:StabilityAnalysis}.
Depending on the dispersion relation $\sigma_k$, obtained from solving the self-consistency condition, Eq.\@~\eqref{eq:ConsistencyCondition}, for a spherical interface, we distinguish three regimes:
The spherical solution is stable when Re${(\sigma_k)<0}$ (white), unstable when Re${(\sigma_k)>0}$ and ${k_\mathrm{osc}>k_\mathrm{max}}$ (orange) or oscillatory when Re${(\sigma_k)>0}$ and ${k_\mathrm{osc}<k_\mathrm{max}}$ (blue); see Sec.\@~\ref{sec:PhaseDiagram}.

We find that the bifurcation diagram [Fig.\@~\ref{figAPP:phaseDiagramCurved}(a)] obtained in this way is qualitatively similar to that of a flat interface [Fig.\@~\ref{fig:phaseDiagram}(e)]:
For ${M\leq13.5}$, the onset of the instability is independent of the mobility ratio $M$, and a stationary instability emerges when the diffusive length scale $l$ exceeds a critical threshold. 
For larger values of $M$, the unstable region extends beyond this threshold, leading to the emergence of an oscillatory instability.
As in the flat-interface case, the oscillatory instability requires fast signal propagation (${M\geq13.5}$) and is confined to short diffusive length scales ${l\leq 2.75}$.

A comparison of our theoretical predictions (background colors) with numerical simulations (bullets) shows good agreement; see Fig.\@~\ref{figAPP:phaseDiagramCurved}(a).
However, as for a flat interface, the stability analysis appears to slightly overestimate the onset of the oscillatory instability.
In summary, we find no qualitative difference between flat and constantly curved interfaces.
That is, the additional curvature of a spherical interface does not appear to influence the underlying mechanisms that destabilize a stationary interface.
\section{Fastest growing mode}
\label{app:fastestMode}
\begin{figure*}
    \centering
    \includegraphics[]{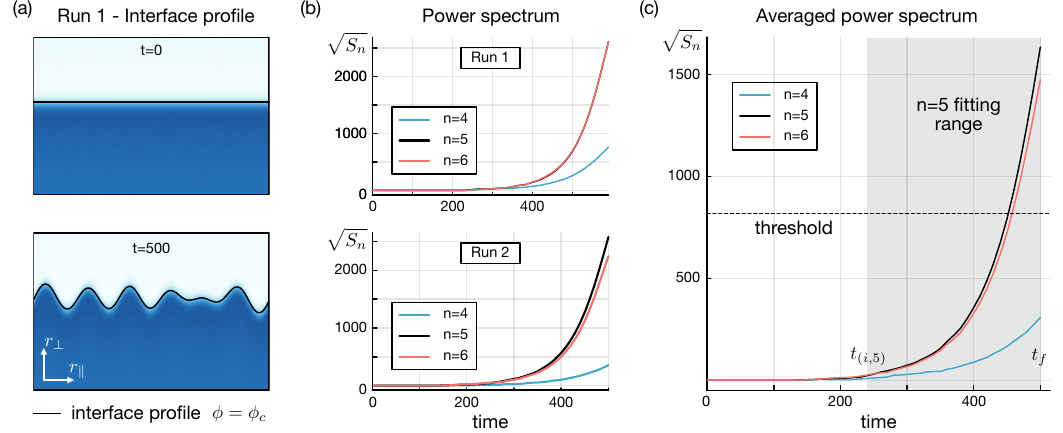}
    \caption{\textbf{Fastest growing mode} obtained from two statistically independent runs for ${Pe=0.45}$, ${l=7}$, and all remaining parameters chosen as in Fig.\@~\ref{fig:modeDiagram}. The simulations were initialized with the corresponding stationary profiles obtained from solving Eq.\@~\eqref{eq:stationaryProblem} for a system with an extension ${L_\perp=50}$ in the direction perpendicular to the interface and ${L_\parallel=200}$ in the parallel direction. After perturbing this state by uniform white noise ${\zeta\in[-0.01,0.01]}$, it was evolved using a finite element scheme for the dynamical equations Eqs.\@~\eqref{eq:DynamicsRescaled} with periodic boundary conditions in the direction parallel to the interface and no-flux boundary conditions in the perpendicular direction.
    (a) The simulations were terminated after a time ${T=50000}$ (${T=250000}$ for data points adjacent to the theoretical instability onset in Fig.\@~\ref{fig:modeDiagram}) or when the interface profile (${\phi=\phi_c}$ level set, black line), separating the high-density domain (dark blue/gray area) from the low-density domain (light blue/gray area), reached ${r_\perp=0.7L_\perp}$. The snapshots display a zoom-in on the interface in the range ${r_\perp \in [0.2L_\perp,0.76L_\perp]}$.
    (b) For each run, we calculated the power spectrum $S_n(t)$, Eq.\@~\eqref{eq:powerSpectrum}, of the interface profile, where ${n=kL_\parallel/(2\pi)}$ denotes the mode number of the interface modes $k$ defined in Eq.\@~\eqref{eq:LinearAnsatz}.
    (c) To extract the fastest growing mode (${n=5}$, black line), for each mode $n$ we took an ensemble average over the different runs. When at the terminal time-point $t_f$, the averaged trajectory $\sqrt{S_n(t)}$ exceeded a threshold $\text{argmax}_n(\sqrt{S_n(t_f)})/2$ (dashed black line), we fitted an exponential growth law to extract the average growth rate. The fitting range ${[t_{(i,n)},t_f]}$ (gray area for ${n=5}$) was chosen to cover 99\% of the curve, i.e.~${\sqrt{S_n(t_{(i,n)})}=0.01\sqrt{S_n(t_f)}}$.}
    \label{figAPP:modeAnalysis}
\end{figure*}
To extract the empirically fastest-growing mode from numerical simulations for the different parameter combinations displayed in Fig.\@~\ref{fig:modeDiagram}, we exploit that for small amplitudes of the interface perturbations our theory predicts an exponential growth of the different unstable modes; cf.\@~Eq.\@~\eqref{eq:LinearAnsatz}.
Under this assumption, the fastest-growing mode should out-compete all other unstable modes, and, thus, determine the emergent length scale of the interfacial perturbation.
However, for finite observation times, the emergent interface profile also crucially depends on the initial perturbation.
We account for this by applying a statistical average over three independent realizations of initial perturbations.

Our approach\@~\cite{ZN} is summarized in Fig.\@~\ref{figAPP:modeAnalysis}, which displays the different steps of our analysis for a single parameter combination (${Pe=0.45}$, ${l=7}$) and two independent runs:
For each set of parameters, the system is initialized with the stationary interface profiles obtained by solving Eq.\@~\eqref{eq:stationaryProblem} for a system of size ${L_\perp=50}$, ${L_\parallel=200}$.
While, in principle, the perturbations induced due to the finite mesh are sufficient to trigger an interface instability, the stationary solutions is perturbed by a uniformly distributed random variable ${\zeta\in[-0.01,0.01]}$
Thereby, we guarantee that different runs are statistically independent.
The perturbed initial states [Fig.\@~\ref{figAPP:modeAnalysis}(a) top-panel] are evolved by a finite element scheme for the dynamical equations Eqs.\@~\eqref{eq:DynamicsRescaled} with periodic boundary conditions in the direction parallel to the interface and no-flux boundary conditions in the perpendicular direction.

Each simulation is run until at least ${T=50000}$. 
Since close to the onset of the instability, the expected growth rates tend to zero, for data points adjacent to the theoretical instability onset displayed in Fig.\@~\ref{fig:modeDiagram} we employ an extended runtime of ${T=250000}$.
Further taking into account that the exponential growth of individual modes only holds for small amplitudes, we also terminate all simulations prematurely, if the interface position $X(r_\parallel,t)$ reaches a critical threshold ${r_\perp=0.7L_\perp}$. 
This threshold is sufficiently distant from the initial interface position around ${X_0\approx 0.5L_\perp}$ to observe significant interface deformations.

From the data obtained in this way, we extract the interface position and calculate its power spectrum $S_k(t)$ [see Eq.\@~\eqref{eq:powerSpectrum}], which is displayed for two independent runs in Fig.\@~\ref{figAPP:modeAnalysis}(b).
For convenience, instead of the wavevector $k$ we index the power spectrum by the mode numbers ${n=kL_\parallel/(2\pi)}$ and further choose to work with its square-root $\sqrt{S_n}$.
Since we are anticipating exponential growth of the individual modes, the latter only amounts to reducing the growth rates by a factor of two.
We proceed by averaging the individual runs over the smallest common time window ${t<t_f}$, where neither of the simulations has been terminated [see Fig.\@~\ref{figAPP:modeAnalysis}(c)].

To extract the fastest growing mode, we fit an exponential growth law ${S_n(t)=S_n(0)\exp(\omega_n t)}$.
For each mode $n$, the fitting window ${t\in[t_{(i,n)},t_f]}$ is chosen to cover 99\% of the curve, i.e., the initial timepoint $t_{(i,n)}$ for each mode fulfills ${\sqrt{S_n(t_{(i,n)})}=0.01\sqrt{S_n(t_f)}}$.
By this, we exclude data points, where the inferred profile is strongly affected by our discrete export-mesh, which has a mesh size of 0.25 in units of the equilibrium interface width $w$. 
Moreover, we only consider modes $n$ whose terminal spectral weight $\sqrt{S_n(t_f)}$ exceeds a threshold $\sqrt{S_n(t_f)}>\text{argmax}_n(\sqrt{S_n(t_f)})/2$.
The construction of the fitting range, as well as the threshold value, are illustrated in Fig.\@~\ref{figAPP:modeAnalysis}(c).
Repeating the above procedure for each combination of parameters, we obtain a phase diagram of fastest-growing modes which are displayed as bullets in Fig.\@~\ref{fig:modeDiagram}.
\section{Empirical phase velocities}
\label{app:phaseVelocities}
\begin{figure*}
    \centering
    \includegraphics[]{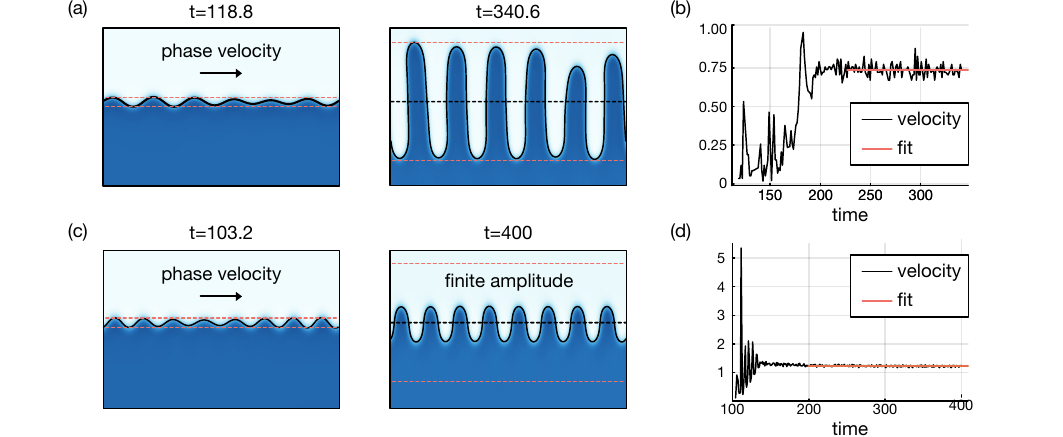}
    \caption{\textbf{Phase velocity of capillary waves} for parameter sets 12 (a,b) and 14 (c,d) in Tab.\@~\ref{tab:VelocityParams}. All simulations were initialized with the corresponding stationary profiles obtained from solving Eq.\@~\eqref{eq:stationaryProblem} and evolved using a finite element scheme for the dynamical equations Eqs.\@~\eqref{eq:DynamicsRescaled} with periodic boundary conditions in the direction parallel to the interface and no-flux boundary conditions in the perpendicular direction. In all simulations, the system size was chosen as ${L_\perp=40}$ in the direction perpendicular to the interface and ${L_\parallel=100}$ in the parallel direction. (a--b) The phase velocity was inferred by tracking and averaging the velocity of the extrema of the interface profile (${\phi=\phi_c}$ level-set, solid black line), which separates the high-density domain (dark blue/gray area) from the low-density domain (light blue/gray area). We started recording the phase velocity when the interface profile first deviated from the initial flat interface (dashed black line) by more than one equilibrium interface width $w$ and terminated our observation when it deviated by more than $15w$. The corresponding time points and intervals (red/gray dashed lines) are illustrated in the left and right panels of sub-figures (a) and (c), respectively. The instantaneous phase velocities during the tracking periods are shown in sub-figures (b+d). Once a constant phase velocity is established, we extract its value by fitting a horizontal line (solid red/gray line).}
    \label{figAPP:phaseVelocities}
\end{figure*}
In this section, we describe how to extract the phase velocity of capillary waves for the different parameter combinations (Tab.\@~\ref{tab:VelocityParams}) displayed in Fig.\@~\ref{fig:velocityDiagramm}(b) from the corresponding finite element simulations.
The procedure\@~\cite{ZN} is illustrated in Fig.\@~\ref{figAPP:phaseVelocities} for parameter sets 12 and 14 of Tab.\@~\ref{tab:VelocityParams}:
For each combination of parameters, the simulations are initialized with the corresponding stationary profiles obtained from solving Eq.\@~\eqref{eq:stationaryProblem} and evolved using a finite element scheme for the dynamical equations, Eqs.\@~\eqref{eq:DynamicsRescaled}, with periodic boundary conditions in the direction parallel to the interface and no-flux boundary conditions in the perpendicular direction.
Once a capillary wave has developed, its instantaneous velocity is determined by tracking the extrema of the interface profile $X(r_\parallel,t)$ which is approximated as the ${\phi=\phi_c}$ level-set.
The instantaneous phase velocity [Fig.\@~\ref{figAPP:phaseVelocities}(b,d)] is obtained by averaging over the velocities of the different extrema.
For the parameters shown in Tab.\@~\ref{tab:VelocityParams}, there are two different scenarios: Either the wave amplitude grows until it reaches the system boundary [Fig.\@~\ref{figAPP:phaseVelocities}(a)] or the system settles in a (transiently stable) finite amplitude capillary wave [Fig.\@~\ref{figAPP:phaseVelocities}(b)]. 
To account for this, we only track the phase velocities when the interface profile does not deviate more than fifteen times the equilibrium interface width from the position of the initial flat interface.
The corresponding tracking interval is indicated by dashed red lines in the right panels of Figs.\@~\ref{figAPP:phaseVelocities}(a,c).
To further exclude effects from our finite export mesh (mesh-size: $0.5$), we only start tracking the phase velocity, once the interface profile deviates by more than one equilibrium interface width from the flat one [red dashed lines in the left panels of Figs.\@~\ref{figAPP:phaseVelocities}(a,c)].
The instantaneous velocity profiles are shown in Figs.\@~\ref{figAPP:phaseVelocities}(b,d).
As can be seen, after some initial phase, a capillary wave forms and settles at a constant velocity, which can be extracted by fitting a horizontal line.
The results obtained in this way, together with the corresponding theoretically predicted velocities, are listed in Tab.\@~\ref{tab:VelocityParams} and shown in Fig.\@~\ref{fig:velocityDiagramm}(b).
\begin{table*}
    \caption{\textbf{Simulation parameters} ($Pe$, $M$, $l$) used to determine the empirical ($v_e$) and theoretical ($v_t)$ phase velocities of capillary waves for different combinations of the P\'eclet number $Pe$, the diffusion length $l$ and the relative mobility $M$ as displayed in Fig.\@~\ref{fig:velocityDiagramm}(b). The remaining parameters were chosen as ${\phi_c=\bar\phi=0.7}$, with a parameter-dependent runtime $T$ displayed in Tab.\@~\ref{tab:FigureParams}.}
    \begin{ruledtabular}
    \begin{tabular}{c|c|c|c|c|c|c|c|c|c|c|c|c|c|c|c}
         &1&2&3&4&5&6&7&8&9&10&11&12&13&14&15 \\
         \hline
         $Pe$ & 0.25 & 0.35 & 0.6 & 0.65 & 1 & 0.38 & 0.4 & 0.7 & 0.45 & 0.2 & 0.5 & 0.55 & 0.35 & 0.8 & 0.3\\[1mm]
         $l$ & 1.75 & 1.25 & 1 & 1.5 & 1 & 2 & 2.25 & 1.25 & 2.5 & 2  &1.5 & 1.75 & 2.75 & 1 & 3 \\[1mm]
         $M$ & 47 & 35 & 19 & 14 & 9.5 & 26 & 23 & 13.6 & 19 & 65 & 20 & 16.5 & 26 & 12.5 & 12.5 \\[1mm]
         $(T \, m)/10^4$ & 10 & 10 & 2.5 & 2 & 0.5 & 1 & 1.5 & 0.75 & 1 & 4 & 1 & 1 & 1.2 & 5 & 3 \\[1mm]
         $v_e$ & 1.25 & 1.66 & 1.73 & 0.69 & 1.32 & 0.65 & 0.5 & 1.08 & 0.42 & 1.1 & 0.95 & 0.73 & 0.46 & 1.23 & 0.35\\[1mm]
         $v_t$ & 1.29 & 1.78 & 1.97 & 0.87 & 1.31 & 0.72 & 0.56 & 1.23 & 0.38 & 1.1 & 0.99 & 0.69 & 0.35 & 1.7 & 0.39
    \end{tabular}
    \end{ruledtabular}
    \label{tab:VelocityParams}
\end{table*}
\section{Conveyor belts}
\label{app:conveyorTransport}
\begin{figure}
    \centering
    \includegraphics[]{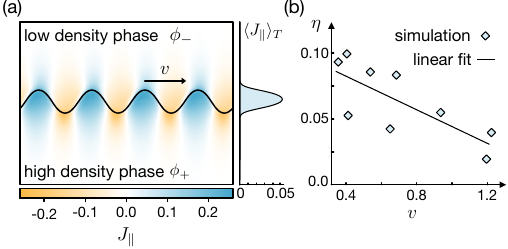}
    \caption{\textbf{Localized transport by a liquid conveyor belt.} in a system of size ${L_\perp}=40$ and ${L_\parallel=100}$ for ${\phi_c=\bar\phi=0.7}$. (a) Lateral currents $J_\parallel$ drive the propagation (black arrow) of a capillary wave with phase velocity $v$. The difference in magnitude of positive and negative current domains leads to a non-vanishing time-averaged current ${\langle J_\parallel\rangle_T}$ (black line). 
    (b) Transport efficiency $\eta$\@~\eqref{eq:efficiency} for nine different parameter combinations (cf.\@~Tab.\@~\ref{tab:conveyorParams}). The best linear fit (black line) reads $\eta = 0.11 - 0.06 v$. All results were obtained from finite element simulation where the system was initialized with the corresponding flat interface solution to Eqs.\@~\eqref{eq:stationaryProblem} and evolved using the full dynamical model Eq.\@~\eqref{eq:DynamicsRescaled} with periodic boundary conditions in the lateral and no-flux boundary conditions in the vertical direction. The data shown in (a) corresponds to parameter set 6 in Tab.\@~\ref{tab:conveyorParams}.}
    \label{figApp:interfacialTransport}
\end{figure}
\begin{table}[b]
    \caption{\textbf{Simulation parameters} ($Pe$, $M$, $l$) used to determine the phase velocity $v$, time-averaged lateral current $\langle J_\parallel\rangle_T$ and transport efficiency $\eta$ of a liquid conveyor belt for different combinations of the P\'eclet number $Pe$, the diffusion length $l$ and the relative mobility $M$ as displayed in Fig.\@~\ref{fig:gearData}(b). The remaining parameters were chosen as ${\phi_c=\bar\phi=0.7}$, with a parameter-dependent runtime $T=30000/M$.}
    \begin{ruledtabular}
    \begin{tabular}{c|c|c|c|c|c|c|c|c|c}
         & 1 & 2 & 3 & 4 & 5 & 6 & 7 & 8 & 9 \\[1mm]
         \hline
         $Pe$ & 0.4 & 0.6 & 0.28 & 0.35 & 0.7 & 0.25 & 0.53 & 0.25 & 0.35\\[1mm]
         $l$ & 2 & 1& 2& 2.25& 1.25& 1.75& 1.75& 2.25& 1.25\\[1mm]
         $M$ & 23 &19 &40 &25 &12 &50 &15.5 &45 &35 \\[1mm]
         $v$ & 0.56& 1.73& 0.96& 0.57& 0.91& 1.31& 0.49& 0.75& 1.69\\[1mm]
         $\langle J_\parallel\rangle_T/10^{-2}$& 3.9& 4.8& 5.6& 2.1& 2.7& 5.1& 3.2& 4.5& 2.3\\[1mm]
         $\eta/10^{-1}$ & 1& 0.40& 0.83& 0.53& 0.42& 0.51& 0.32& 0.45& 0.23\\[1mm]
    \end{tabular}
    \end{ruledtabular}  
    \label{tab:conveyorParams}
\end{table}
Figure.\@~\ref{fig:GearsAndBelts} demonstrates that capillary waves propagating along the interface between high- and low-density phases generate a net-current ${\langle J_\parallel\rangle_T}$, parallel to the interface.
This current is sharply localized at the interface and rapidly decays to zero within the high- and low-density phases [Fig.~\ref{figApp:interfacialTransport}(a)].
As such, liquid conveyor belts offer a compelling mechanism for facilitating highly localized, directional mass transport.
To evaluate the efficiency of this transport, we compare it to an idealized scenario in which the wave pattern moves as a rigid body.
In general, the most efficient form of transport arises when all molecules move coherently at a uniform velocity.
Although such rigid-body-like motion cannot occur in systems subject to diffusion, it nonetheless serves as a useful benchmark against which active transport processes can be probed.

For a liquid conveyor belt, the dynamics deviate significantly from the idealized rigid-body scenario.
As shown in Fig.\@~\ref{figApp:interfacialTransport}(a), for a phase-motion towards the right-hand side, the instantaneous lateral current ${J_\parallel}$ exhibits alternating regions of positive and negative flux:
While in the high-density peaks mass is transported to the right, in the low-density valleys mass is transported to the left.
That is, the propagation of capillary waves does not arise from a coherent chemotactic flux but instead emerges from continuous mass redistribution between different segments of the wave pattern.
Nonetheless, because the magnitudes of the positive and negative currents are typically unequal, a net current in the propagation direction of the wave is maintained on average.

To evaluate the efficiency of the emergent transport mechanism, we compare the maximum of the time-averaged current at the interface  [cf.\@~Fig.~\ref{figApp:interfacialTransport}(a)]
\begin{align}
    J = \mathrm{argmax}_{r_\perp}\langle J_\parallel\rangle_T
\end{align}
with the corresponding mean current $J_\mathrm{r}$ of a rigidly moving capillary wave with phase velocity $v$.
Averaging $J_\mathrm{r}$ over one period of the wave pattern yields ${J_\mathrm{r}\approx v \phi_c}$ 
The transport efficiency is then defined as
\begin{align}
    \eta = \mathrm{argmax}_{z}\left(\dfrac{\langle J_\parallel\rangle_T}{v \phi_c}\right) ,
    \label{eq:efficiency}
\end{align}
with a (hypothetical) optimal value of ${\eta=1}$.
To build intuition for the efficiency of liquid conveyor belts, we determined $\eta$ using nine different parameter combinations listed in Tab.\@~\ref{tab:conveyorParams}.
To that end, we employed finite element simulations where the system was initialized with the corresponding flat interface solution to the stationary profile equations, Eqs.\@~\eqref{eq:stationaryProblem}, and evolved according to the full dynamical model, Eq.\@~\eqref{eq:DynamicsRescaled}.
The parameters were chosen similarly to the ones in Fig.\@~\ref{fig:velocityDiagramm}, with the requirement, that the system settles in a (transiently stable) finite amplitude capillary wave pattern.

The numerically determined efficiency\@~\cite{ZN} is presented in Fig.\@~\ref{figApp:interfacialTransport}(b):
As already stated in Sec.\@~\ref{sec:liquidMechanics}, lateral transport through a liquid conveyor belt is rather inefficient with typical values of $\eta$ ranging from 0.02 to 0.1.
Additionally, the data reveals a trend of decreasing efficiency for faster waves.
Fitting a linear function $\eta=\eta_{\mathrm{max}}+m v$ and extrapolating it to ${v=0}$, for slow capillary waves, we expect an optimal efficiency of roughly ${\eta_{max}\approx 0.11}$.
For increasing phase velocities, the efficiency $\eta$ decreases with a slope of roughly ${m=-0.06}$.
\section{Interface spectrum}
\begin{figure}
    \centering
    \includegraphics[]{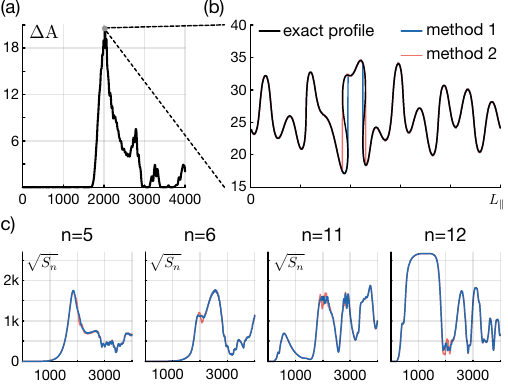}
    \caption{\textbf{Interface overhangs} Comparison of the interface profiles obtained using the two closure schemes ``method\@~1'' (blue/dark gray lines) and ``method\@~2'' (red/gray lines). In method\@~1 (pruning), the high-density overhang is removed, whereas in method\@~2 (filling), the low-density region beneath the overhang is filled up. (a) The difference between the two methods is quantified by the total area $\Delta \text{A}$ enclosed between the resulting profiles. (b) Illustration of the exact interface profile (black) and the corresponding approximations obtained using method\@~1 or method\@~2 at ${\mathrm{t}=2000}$, corresponding to the middle inset of Fig.\@~\ref{fig:superposition}(b). c) Time-dependent power spectra $S_n$ of the dominant modes (${n={5,6,11,12}}$) obtained using both methods. For better illustration, the curves were smoothed using a Gaussian-weighted moving average with a window size of ${\Delta t=60}$ and a standard deviation of ${\sigma=12}$. All parameters match those used in Fig.\@~\ref{fig:superposition}(b).}
    \label{figAPP:overhangs}
\end{figure}
To obtain the data shown in Fig.\@~\ref{fig:modeDiagram} and Fig.\@~\ref{fig:superposition}(a--b), we computed the power spectrum, Eq.\@~\eqref{eq:powerSpectrum}, of the interface profile $X(r_\parallel,t)$, which requires taking its Fourier transform\@~\cite{ZN}.
While this can be done straightforwardly, the ${\mathrm{t}=2000}$ inset in Fig.\@~\ref{fig:superposition}(b) illustrates that, in some instances, the interface profile develops small overhangs.
In such cases, the profile can no longer be described as a single-valued function $X(r_\parallel,t)$, rendering its power spectrum ill-defined.
To circumvent this issue, we apply two alternative closure schemes, illustrated in Fig.\@~\ref{figAPP:overhangs}(b).
Both approaches modify the interface profile by either reducing or extending the high-density domain, i.e., the area below the black profile in Fig.\@~\ref{figAPP:overhangs}(b). 

In the first approach (method 1), overhanging segments of the high-density domain are removed (``pruned''), whereas in the second approach (method 2), the low-density region beneath the overhang is ``filled in''. 
The discrepancy between these two methods can be quantified by the area enclosed between the resulting profiles.

For the parameters used in Fig.\@~\ref{fig:superposition}, we find that this discrepancy remains small, as illustrated in Fig.\@~\ref{figAPP:overhangs}(a), with the largest possible deviation shown in Fig.\@~\ref{figAPP:overhangs}(b).
To assess whether the power spectrum is significantly affected, Fig.\@~\ref{figAPP:overhangs}(c) compares the time-dependent power spectra of the dominant modes [${n=\{5,6,11,12\}}$, see Fig.\@~\ref{fig:superposition}(b)] obtained using both methods.
The results indicate that the two approaches yield nearly identical curves, with only minor differences around ${t\approx 2000}$, coinciding with the maximal deviation in  Fig.\@~\ref{figAPP:overhangs}(a).
We conclude that, for our purposes, both closure schemes provide equivalent results. 
Therefore, throughout this manuscript, we consistently apply method 1.

Additionally, as pointed out in the figure captions, for Fig.\@~\ref{fig:superposition}(a--b) and Fig.\@~\ref{figAPP:overhangs}, the time-dependent power spectra $\sqrt{S_n}$ were smoothed using a short-ranged moving average.
This was done to mitigate apparent high-frequency fluctuations [see Fig.\@~\ref{fig:superposition}(a--b), Fig.\@~\ref{figAPP:overhangs}(c)] that arise due to the discrete nature of our export mesh (grid size\@~0.25).
Specifically, we applied a weighted Gaussian moving average, using a window size of $\Delta t=60$ and a standard deviation of ${\sigma=12}$. 
In this approach, at each time point $t$, the corresponding data point was replaced by a locally weighted sum of its neighboring values within a time frame ${[t-60,t]}$.
The weights follow a Gaussian distribution with zero mean and a standard deviation ${\sigma=12}$ to ensure that closer points contributed more strongly to the smoothed value. 
This method preserves the trend of the power spectra while effectively reducing high-frequency noise.
\section{Rotation and translation of active Gears}
\label{app:translationGears}
\begin{figure}[b]
    \centering
    \includegraphics[]{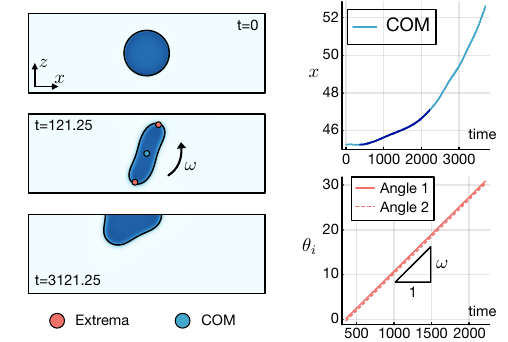}
    \caption{\textbf{Center-of-mass ($v$) and angular ($\omega$) velocity of an active gear} rotating between two parallel walls with no-flux boundary conditions for parameter set 6 of Tab.\@~\ref{tab:GearParameters}. Although the system is fully symmetric, the gear does not remain centered between the two horizontal walls but eventually drifts and crashes into the upper wall. The center-of-mass (COM, top-right) and angular trajectories (bottom-right)  of the extremal points (red/gray dots) are inferred from finite-element simulations as described in Fig.\@~\ref{fig:gearVelocity}, with a fitting range indicated by the dark blue/gray line in the COM plot.}
    \label{figAPP:gearVelocityException}
\end{figure}
\begin{table*}[]
    \caption{\textbf{Simulation parameters} ($Pe$, $M$, $l$, $\bar\phi$) used to determine the center-of-mass ($v$) and angular ($\omega$) velocity of an active gear rotating between two solid walls of length ${L=90}$ and distance $H$ [see Fig.\@~\ref{fig:gearsBC}] for ${\phi_c=0.7}$. For each set of parameters, we conducted finite element simulations with symmetric and asymmetric boundary conditions (cf.\@~Fig.\@~\ref{fig:gearsBC}) and extracted the corresponding velocities ${(v_s,\omega_s)}$ and ${(v_{as},\omega_{as})}$, which are shown in Fig.\@~\ref{fig:gearData}. In the case of asymmetric boundary interactions, we kept the product ${M\, g_G=0.002}$ fixed. For parameter set 9 and symmetric boundary conditions, the active gear turned out unstable and we could not measure its phase and angular velocity.}
    \centering
    \begin{ruledtabular}
    \begin{tabular}{c|c|c|c|c|c|c|c|c|c|c}
         Parameterset & 1 & 2 & 3 & 4 & 5 & 6 & 7 & 8 & 9 & 10  \\[1mm]
         \hline
         $Pe$ & 0.42 & 0.43 & 0.38 & 0.475 & 0.45  & 0.4 & 0.52 & 0.48 & 0.39 & 0.36 \\[1mm]
         $M$ & 20 & 23 & 25 & 27 & 24 & 27 & 24 & 22 & 23 & 28\\[1mm]
         $l$ & 2 & 1.5 & 1.75 & 1 & 1.25 & 1.5 & 1 & 1.25 & 2 & 1.75 \\[1mm]
         $H$ & 35 & 40 & 45 & 40& 50 & 30 & 25 & 35 & 50 & 60 \\[1mm]
         $\bar\phi$ & 0.465 & 0.484 & 0.51 & 0.511 & 0.54 & 0.472 & 0.481 & 0.507 & 0.546 & 0.552 \\[1mm]
         $v_s/10^{-5}$ & 44.17 & 1.13 & -6.15 & 3.09 & -4.15 & -93.53 & -9.08 & 8.84 & & -6.59\\[1mm]
         $-\omega_s$ & 0.068 & 0.0908 & 0.0609 & 0.1704 & 0.0846 & 0.1039 & 0.3252 & 0.1152 & & 0.0417 \\[1mm]
         $v_{as}/{10^{-3}}$ & 8.27 & 5.43 & 3.65 & 3.01 & 0.79 & 8.04 & 6.76 & 3.42 & 6.39 & 6.91\\[1mm]
         $-\omega_{as}$ & 0.0645 & 0.0918 & 0.0604 & 0.1680 & 0.0831 & 0.104 & 0.2936 & 0.1140 & 0.0306 & 0.0438\\[1mm]
    \end{tabular}
    \end{ruledtabular}
    \label{tab:GearParameters}
\end{table*}
In Sec.\@~\ref{sec:liquidMechanics}, we analyzed a single active gear positioned between two horizontal walls. 
When no-flux boundary conditions were applied at both walls, the gear remained stationary, whereas it acquired a finite translational velocity when the symmetry between the walls was broken. 
Specifically, we introduced a weak gravity-like interaction, Eq.\@~\eqref{eq:gravity}, counteracted by a repulsive bottom wall, implemented via Dirichlet boundary conditions, ${\psi\big\rvert_{z=0}=1}$. 
The emergent angular and center-of-mass velocities for ten different parameter combinations (see Tab.\@~\ref{tab:GearParameters}) are shown in Fig.~\ref{fig:gearData}(a).

We found that for symmetric gear-wall interactions, the gear’s translational velocity $v$ was essentially zero (also see Tab.\@~\ref{tab:GearParameters}).
The only exception to this rule were the two dumbbell-shaped gears (parameter sets 1,6 in Tab.\@~\ref{tab:GearParameters}) which also showed a non-zero velocity even in the symmetric setup.
In this section, we demonstrate that this anomaly arises from an instability where the gear’s center of mass deviates from the system's mid-plane.
To this end, we consider the time evolution obtained from finite element simulations for parameter set 6 of Tab.\@~\ref{tab:GearParameters} and symmetric boundary conditions; see Fig.\@~\ref{figAPP:gearVelocityException}.

As for the other parameter sets (cf.\@~Fig.\@~\ref{fig:gearVelocity}), the system is initialized with a spherical droplet.
After a brief period, the spherical domain undergoes an interfacial instability and settles into a dumbbell-shaped rotating gear with a constant angular velocity (bottom right of Fig.\@~\ref{figAPP:gearVelocityException}).
Subsequently, the gear experiences a slight drift that eventually leads to it colliding with the top wall (top right of Fig.\@~\ref{figAPP:gearVelocityException}).
The observed behavior differs from the one regularly observed for asymmetric gear-wall interactions in two ways:
First, the $x$-component of the gear's center of mass does not exhibit a constant velocity but shows some non-linear time evolution.
Second, the gear's translational and angular velocity have the same sign, whereas, for asymmetric gear-wall interactions, we always found their signs to be anti-correlated.
We conclude that the apparent translation of the gear’s COM in this symmetric setup is due to an instability rather than constant propulsion, as observed for asymmetric gear-wall interactions. 
We speculate that dumbbell-shaped gears are particularly prone to this instability due to their elongated shape, which distinguishes them from other observed gear shapes.
To nevertheless correctly account for the observed phenomena, when determining the gear's velocity $v$, we fit the COM trajectory over a range indicated by the dark blue line in Fig.\@~\ref{figAPP:gearVelocityException}.

\section{COMSOL parameters}
\label{app:Numerics}
\begin{table*}[]
    \caption{\textbf{Simulation parameters} used for figures presenting numerical results inferred from finite element simulations using COMSOL Multiphysics Version 6.1 \cite{AB.2023}.}
    \begin{ruledtabular}
    \begin{tabular}{c|c|c|c|c|c|c}
        Figure &  mesh size & element order & rel.\@~tolerance  & runtime T & stop criterion & initial conditions\\[1mm]
        \hline 
        Fig.\@~\ref{fig:overview} & 2/3 & linear & 1e-5  & [150k,150k,1500,1500,666,55] & runtime & homogeneous + noise \\[1mm]
        Fig.\@~\ref{fig:categorization} & 1/2 & linear & 1e-5 & [250,250,500] & runtime & flat interface\\[1mm]
        Fig.\@~\ref{fig:GearsAndBelts} & 1/2 & linear & 2e-5 & [3488, 3021] & runtime & flat/spherical interface\\[1mm]
        Fig.\@~\ref{fig:phaseDiagram} & 1/3 & linear & 1e-5 & 40000/M, 100000/M & 0.6$L_\perp$ & flat interface \\[1mm]
        Fig.\@~\ref{fig:modeDiagram},\ref{figAPP:modeAnalysis} & 1/3 & linear & 4e-5 & 50000/M, 250000/M & 0.7$L_\perp$ & flat interface + noise\\[1mm]
        Figs.\@~\ref{fig:velocityDiagramm}, \ref{figAPP:phaseVelocities} & 1/2 & linear & 1e-5 & cf.\@~Tab.\@~\ref{tab:VelocityParams} & runtime & flat interface \\[1mm]
        Fig.\@~\ref{fig:superposition}, \ref{figAPP:overhangs}& 1/3 & linear & 1e-5 & 4000 & runtime & flat interface \\[1mm]
        Figs.\@~\ref{fig:gearVelocity}, \ref{fig:gearData}, \ref{figAPP:gearVelocityException} & 1/2 & linear & 2e-5& 50000/M - 75000/M & runtime & curved interface \\[1mm] 
        Fig.\@~\ref{fig:dropletBreakage} & 1 & linear & 1e-5 & 2500 & runtime & homogeneous + noise \\[1mm] 
        Figs.\@~\ref{figAPP:phaseDiagram1} & 2/3 & linear & 2e-5 & 100000 & runtime & homogeneous + noise \\[1mm]
        Fig.\@~\ref{figAPP:phaseDiagram2} & 2/3 & linear & 2e-5 & 6000/100000 & runtime & homogeneous + noise \\[1mm]
        Fig.\@~\ref{figAPP:phaseDiagramCurved} & 1/3 & linear & 1e-5 & 25000/M, 100000/M &0.65$L_\perp$ & spherical interface \\[1mm]
        Fig.\@~\ref{figApp:interfacialTransport} & 1 & quadratic & 5e-5 & 30000/M & runtime & flat interface
    \end{tabular}    
    \end{ruledtabular}
    \label{tab:FigureParams}
\end{table*}
All finite element simulations presented in this manuscript were performed using COMSOL Multiphysics Version 6.1 \cite{AB.2023}. 
A collection of COMSOL simulation files covering flat and curved interfaces, homogeneous initial conditions, and rotating gear configurations is available in Ref.\@~\cite{ZN}.
For each simulation, we employed the PARDISO solver with adaptive time stepping, utilizing an implicit 5th-order backward differentiation formula (BDF) and a relative tolerance as specified in Table~\ref{tab:FigureParams}. 
Simulations were carried out on a two-dimensional mesh with triangular Delaunay tessellation, where the maximum element size is detailed in Tab.\@~\ref{tab:FigureParams}. 
The system was discretized using Lagrangian shape functions of linear or quadratic order, as indicated for each case in Table~\ref{tab:FigureParams}. 
For all other settings, the COMSOL default values were applied. 
Simulation files with the specified settings can be found in reference\@~\cite{SM}.

Unless stated otherwise, all simulations were terminated after reaching a prescribed runtime T (Tab.\@~\ref{tab:FigureParams}). 
For the data presented in Figs.\@~\ref{fig:phaseDiagram}, \ref{fig:modeDiagram}, \ref{figAPP:modeAnalysis}, and \ref{figAPP:phaseDiagramCurved}, the simulations were stopped when the interface profile (${\phi=\phi_c}$) exceeded a predefined threshold value $r_T$, also specified in Tab.\@~\ref{tab:FigureParams}.
Specifically, we terminated simulations when $\mathrm{max}_{r_\perp>r_T}(\phi)>\phi_c$.

Table\@~\ref{tab:FigureParams} also provides details on the initial conditions for each simulation. 
Simulations initialized with a flat/spherical solution to the stationary profile equations, Eqs.\@~\eqref{eq:stationaryProblem}, are labeled as ``flat/spherical interface''.
The respective stationary profiles were obtained by initializing a one-dimensional system with a step function at values $\phi_\pm$ with the length of the domains chosen to reproduce the desired average density $\bar\phi$. 
Subsequently, the initial state was evolved until ${T=50000}$, with the same settings specified above, using a relative tolerance of $10^{-5}$ and a maximal mesh size of $1/4$.
If the system was initialized with a uniform density, the corresponding entries in Tab.\@~\ref{tab:FigureParams} are labeled as ``homogeneous'', while the keyword ``noise'' indicates that the initial state included additional random white noise perturbations.
For Figs.\@~\ref{fig:velocityDiagramm}(b), \ref{fig:gearData}(a), and \ref{figApp:interfacialTransport}, the simulation parameters corresponding to each data point are summarized in Tables\@~\ref{tab:VelocityParams}-\ref{tab:GearParameters}, respectively.
%
\end{document}